\documentclass[onecolumn]{svjour3}         
\smartqed  
\usepackage{graphicx}
\usepackage{fix-cm}
\usepackage{journals}
\usepackage{txfonts}
\usepackage{natbib}
\usepackage{amsbsy}
\usepackage{fix-cm}
\graphicspath{{.}{Figures/}}
\newcommand{\argmin}{\mathop{\operator@font arg\,min}\limits}

\begin{document}

\title{Imaging the heart of astrophysical objects with optical
  long-baseline interferometry 
}
\subtitle{}


\author{ 
J.-P.~Berger$^{1,2}$  	           \and
  F.~Malbet$^{1}$  	        \and
  F.~Baron$^{3,4}$              \and 
  A.~Chiavassa$^{5,19}$	                      	     \and
  G.~Duvert$^{1,6}$	                  \and
  M.~Elitzur$^{7}$	        	     \and
  B.~Freytag$^{8}$	         	     \and
  F.~Gueth$^{9}$	         	     \and
  S.~H\"onig$^{10,11}$  	           	     \and
  J.~Hron$^{12}$	        	     \and
  H.~Jang-Condell$^{13}$		     \and
  J.-B. Le Bouquin$^{2,1}$	      	     \and
  J.-L.~Monin$^{1}$	 	 	     \and
  J.D.~Monnier$^{3}$	        	     \and
  G.~Perrin$^{14}$	          	     \and
  B.~Plez$^{15}$    	          	     \and
  T.~Ratzka$^{16}$  	            	     \and
  S.~Renard$^{1}$  	         	     \and
  S.~Stefl$^{2}$	        	     \and
  E.~Thi\'ebaut$^{8}$	         	     \and
  K.~Tristram$^{10}$	          	     \and
  T.~Verhoelst$^{17}$	        	     \and
  S.~Wolf$^{18}$	          	     \and
  J.~Young$^{4}$	        
}

\authorrunning{Berger et al.} 

\institute{
{\bf 1:} Universit\'e J.~Fourier (Grenoble 1)/CNRS, UMR\,5571, Institut de Plan\'etologie et d'Astrophysique de Grenoble, B.P.~53, F-38041 Grenoble cedex 9, France 
\and
{\bf 2:} European Southern Observatory, Santiago, Chile
\and
{\bf 3:} University of Michigan, Ann Arbor, USA
\and
{\bf 4:}  Cavendish Laboratory, University of Cambridge, Cambridge, United Kingdom
\and
{\bf 5:} Max-Planck-Institut f{\"ur} Astrophysik, Garching bei M{\"u}nchen, Germany
\and
{\bf 6:}  Jean Marie Mariotti Center, Grenoble, France  
{\bf 7:}  University of Kentucky, Lexington, USA
\and
{\bf 8:}  Centre de Recherche en Astrophysique de Lyon, UMR\,5574 
Universit\'e de Lyon/\'Ecole Normale Sup\'erieure de Lyon/CNRS, Lyon, France
\and
{\bf 9:} Institut de Radioastronomie Millim\'etrique, Grenoble, France
\and
{\bf 10:} Max-Planck-Institut f{\"u}r Radioastronomie, Bonn, Germany
\and
{\bf 11:} University of California in Santa Barbara, Department of Physics,
Santa Barbara, USA
\and
{\bf 12:} Institut f{\"ur} Astronomie, Universit{\"a}t Wien, Wien, Austria
\and
{\bf 13:} Department of Physics \& Astronomy
1000 E. University of Wyoming 3905 Laramie, WY, 82071
\and
{\bf 14:} Laboratoire d'Etudes Spatiales et d'Instrumentation en Astrophysique,
UMR\,8109 Observatoire de Paris/CNRS, Meudon, France
\and
{\bf 15:} Laboratoire Univers et Particules de Montpellier, UMR 5299, CNRS, Universit\'e Montpellier 2, 34095 Montpellier, France
\and
{\bf 16:} Universit\"ats-Sternwarte M\"unchen, M\"unchen , Germany
\and
{\bf 17:} Institute of Astronomy, KULeuven, Belgium     
\and
{\bf 18:} University of Kiel, Kiel, Germany
\and
{\bf 19:} Institut d'Astronomie et d'Astrophysique, Universit\'e Libre
de Bruxelles, Bruxelles, Belgium
}

\date{Received: date / Accepted: date}

\maketitle

\begin{abstract}
  The number of publications of aperture-synthesis images based on optical
  long-baseline interferometry measurements has recently increased
  due to easier access to visible and infrared interferometers. The
  interferometry technique has now reached a technical maturity level
  that opens new avenues for numerous astrophysical topics requiring
  milli-arcsecond model-independent imaging. In writing this paper our
  motivation was twofold: 1) review and publicize emblematic excerpts of the impressive
  corpus accumulated in the field of optical interferometry image
  reconstruction;
2) discuss future prospects for this technique  by selecting four representative
astrophysical science cases in order to review the potential benefits
of using optical long baseline interferometers.

For this second goal we have simulated interferometric data from those
selected astrophysical environments and used state-of-the-art codes to
provide the reconstructed images that are reachable with current or
soon-to-be facilities. The image reconstruction process was ``blind'' in the
sense that reconstructors had no knowledge of the input brightness
distributions. We discuss the impact of optical interferometry in
those four astrophysical fields. We show that
image reconstruction software successfully provides accurate
morphological information on a variety of astrophysical topics and
review the current strengths and weaknesses of such reconstructions.

We investigate how to improve image reconstruction and the quality of
the image possibly by upgrading the current facilities.  We finally
argue that optical interferometers and their corresponding
instrumentation, existing or to come, with 6 to 10 telescopes, should
be well suited to provide images of complex sceneries.

\end{abstract}

\section{Introduction}

The recent years have seen a significant increase in scientific
publications making use of images based on homodyne optical
long-baseline interferometry
\citep[e.g.][]{2007Sci...317..342M,2009A&A...496L...1L,Schmitt:2009,2009ApJ...707..632L,2009A&A...497..195K,2009A&A...508..923H,2010Natur.464..870K,Millour:2011}.
While model fitting of visibilities remains today the most accurate way to
characterize an object brightness distribution, the technique has now
reached a technical maturity level that opens new avenues for numerous
astrophysical topics requiring milli-arcsecond model-independent
imaging.

These results are the culmination of years of efforts that started
with speckle interferometry, aperture masking and long-baseline
interferometry. For the latter active or defunct prototype facilities
such as COAST, IOTA, I2T, GI2T, Mark III, NPOI, PTI, SUSI\footnote{ISI
belongs to the same fraternity but operates in heterodyne mode and
will therefore not be discussed.}, have
progressively paved the way for large diameter telescope arrays such
as CHARA, Keck Interferometer, and VLTI.

In order to assess the main astrophysical questions that visible and
infrared interferometry can actually answer, we first review the status of
optical interferometry in terms of image synthesis achievements
and then follow up with a discussion on the important issues from four main
astrophysical fields namely planetary formation, stellar atmospheres,
circumstellar shells, and active galactic nuclei\footnote{This choice, obviously not exhaustive, should not hide the
  wealth of astronomical topics requesting milli-arcsecond resolution
  imaging: Cepheids, magnetic, Be, O, supermassive stars, stellar mass
  loss, jet formation, dynamics of close stellar clusters, SMBH
  galaxies etc.}. Then, we simulate interferometric data from those
environments and provide the reconstructed images that are achievable
with current or near-future facilities, in order to discuss the impact
of such techniques in those fields.

The paper is organized as follows. The principle of interferometric
imaging as well as a brief description of present imaging
interferometers is given in Sect.~\ref{sec:imaging}. The four
astrophysical topics are briefly described in Section
\ref{sec:astro-simus} and for each one the parameters of the
simulations are presented. The imaging facilities are reviewed in
Section~\ref{sec:facilities}. In Sect.~\ref{sec:results} we describe the
process which allows us to produce realistic interferometric data and
then reconstructed images. Finally, Section ~\ref{sec:discussion} is
devoted to a discussion of the results and the future perspectives for
optical long-baseline interferometry.


\section{Imaging with a long baseline interferometer}
\label{sec:imaging}
\subsection{Principles of optical long baseline interferometry}

Principles of long baseline interferometry are well explained in
numerous articles, conference proceedings and monographies. The reader
is referred to \cite{Thompson:2001} for a detailed description of the
technique which lays out nicely most of the fundamental principles,
albeit mostly oriented towards radio astronomy. More recently there
have been interesting collective works, review articles and
monographies dedicated to the specifics of optical long baseline
interferometry
\citep[e.g.][]{2000plbs.conf.....L,Monnier:2003,2007NewAR..51..583H,
  2007NewAR..51..563M, Glindemann:2011}.  In our definition, optical
interferometry covers the wavelength regime from the visible
$0.4\,\mu\rm{m}$ to the mid infrared $10-20\,\mu\rm{m}$.

An interferometric array measures the spatio-temporal coherence of the
electromagnetic field through two or more telescopes \citep[e.g.][]{Goodman:1985}. The core observable of an interferometer is the
complex visibility $V_{kl}$:

\begin{equation}
V_{kl} = |V_{kl}| \exp^{j\Phi_{kl}}
\end{equation}

Where $|V_{kl}|$ is the amplitude of the visibility and $\Phi_{kl}$
its phase. $V_{kl}$ is extracted from the measurement of the contrast
and phase of the interferogram formed between two telescopes $k$ and
$l$. Unlike radio interferometers, where the detection of the
electromagnetic field is done at the telescope level (heterodyne
interferometry), most optical interferometers require the actual
formation of interferometric fringes on a detector.  

The fundamental theorem of long baseline interferometry is the
Van-Cittert Zernike theorem \citep{Goodman:1985,Thompson:2001}. It
relates the complex visibility $V$ measured at a spatial frequency
$\vec{B}/\lambda$ ($\vec{B}$ being the telescope baseline vector
and $\lambda$ the wavelength) to the actual object brightness
$I(\vec{x})$ distribution through a Fourier transform relation.

\begin{equation}
V(\vec{B}/\lambda) = \frac{\int_{-\mathbf{\infty}}^{\mathbf{\infty}}
I(\vec{x}) \exp^{-2\pi j
  \vec{x} \,\vec{B} / \lambda}\,  \rm{d} \vec{x}}{\int_{-\mathbf{\infty}}^{\mathbf{\infty}}
I(\vec{x}) \rm{d} \vec{x}}
\end{equation}

The components $(u,v)$ of $\vec{B} / \lambda$ form the spatial frequency
coordinate system. $\vec{x}$ represents the angular position
vector. Therefore, interferometric observations provide information
about 
spatial frequency components of the brightness distribution. Similarly
to the ``Rayleigh criterion'' used in single pupil telescopes, one can
define the resolution of a two-telescope interferometer of baseline
$B$ as $\lambda / 2B$ (in radians) which corresponds
approximately to the milli-arcsecond level in the near infrared with
typical hectometric baselines.  
The amplitude of the visibility is related to the projected brightness
angular size while the phase provides information on the brightness photocenter location.

In practice, visible and infrared interferometry require mixing the
light received from an astronomical source and collected by several
independent telescopes separated from each other by tens or even
hundreds of meters\footnote{With the notable exception of the
  heterodyne ISI interferometer \citep{Townes:2008}}. The light beams
are then overlapped and form an interference pattern if the optical
path difference between the different arms of the interferometer
---taking into account paths from the source up to the detector--- is
smaller than the coherence length of the incident wave (typically of
the order of several microns).  This interference pattern is composed
of fringes, i.e.\ a succession of stripes of faint (destructive
interferences) and bright (constructive interferences) intensity. By
measuring (i) the contrast of these fringes, i.e.\ the normalized flux
difference between the maximum and minimum intensity, also called the
visibility amplitude, and (ii) their phase, i.e. their position, one
can construct an estimation of the so-called complex visibility.

\subsection{Obstacles to imaging with an optical interferometer}

In principle, measuring the complex visibilities at all spatial
frequencies up to a maximum value $\vec{uv}_{\rm max}$, should
permit the retrieval of the synthesized image of the spatial
intensity distribution of the object with a spatial resolution
$1/\vec{uv}_{\rm max}$ by inverse Fourier transforming the 2-D
visibility map. However there are two main obstacles to this simple
inversion.

First, in all practical cases, the $(u,v)$ plane cannot be fully
covered, because the spatial frequencies sampled by an interferometer
are limited by the number of pairs of telescopes. With the Earth
rotation these spatial frequencies follow $(u,v)$ tracks in shape of
arcs of ellipses \citep{Thompson:2001,2007NewAR..51..597S} which
provides \emph{super-synthesis} (also called ``Earth rotation
synthesis'') and therefore increases the coverage. In the particular
cases where no strong wavelength dependence of the object brightness
distribution is expected, one can use measurements at different
wavelengths to help further extend the $(u,v)$ coverage, but only
radially.

Secondly, the phases of the visibilities are frequently lost because
they have been scrambled by atmospheric blurring effects
\citep{Quirrenbach:2000}. Instrumental effects also contribute but in
a more static way. In the optical domain (as in the radio but on a
wilder scale), rapid fluctuations in the atmospheric optical index
induce random optical path variations. Typical atmospheric coherence
times at these wavelengths are in the range of 1 to 10
milliseconds. Consequently, measuring direct phase information is
challenging. Several techniques have been explored to circumvent this
difficulty, all of which have inherited from experience in radio
interferometry,  and are well detailed in \citet{Monnier:2007}.
\begin{itemize}
\item The phase of the product of the 3 complex visibilities measured
  with 3 telescopes forming a closed triangle, called the closure phase, was first introduced by
  \citet{Jennison:1956} in order to cancel out instrumental and
  atmosphere-induced phase errors and recover partial phase
  information. It was later pioneered in the optical domain by
  \citet{Baldwin:1986} and subsequent aperture masking experiments at
  Cambridge and is now a routine observable of optical
  interferometers. Parallel to that, triple-correlation techniques
  used in speckle interferometry \citep[e.g][]{Weigelt:1991} also make use of closure phase although the
  formal relation between the two techniques was established by
  \cite{Roddier:1986}. In practice, the closure phase is estimated from
  the phase of the so-called ``bi-spectrum'' $B_{ijk}$ which
  is constructed from the product of complex visibilities estimated in
  a closed triangle
  $B_{ijk}=V_{ij}\,V_{jk}\,V_{ki}$.
\item The \emph{self-calibration} method as described by
  \citet{Cornwell:1981} aims at retrieving phase information on a
  maximum number of baselines from closure phase estimators. It
  integrates the phase recovery procedure in the image reconstruction
  process. For that purpose it uses an iterative scheme that starts
  from a trial image and seeks, through modeling and fitting of atmospheric and
  telescope phase errors, to find the best phase estimation consistent
  with closure phases.
\item \emph{Phase referencing} encompasses various techniques that aim
  at measuring, as directly as possible, the phases. The idea is to
  calibrate the phase measurement with a phase reference, either
  internal or external. \emph{Differential phases}, which are obtained
  by comparing phases at different wavelengths with a reference
  channel, provide direct means of phase estimation. This is true provided one can
  constrain the morphology at the reference wavelength (e.g an
  unresolved emission in a continuum channel) and model the
  atmospheric and instrumental differential phase. The reader is
  referred  to the work of \cite{Vakili:1997}, for example, for
  illustration of the differential phase extraction, and
  \citet{Schmitt:2009} and previous NPOI team publications for
a detailed demonstration of  spectral differential
  imaging. \emph{Astrometric phase referencing} uses the capability of
  observing two objects within the atmospheric isoplanatic patch to
  provide an astrometric reference to the phase (with the motivation
  of authorizing longer integration times if a nearby bright source
  can be used as a reference). This requires a specific infrastructure
  allowing for dual-star observation. It has been implemented at PTI,
  NPOI, and is being implemented at Keck and VLTI \cite[e.g][]{Delplancke:2008} but has not led yet to notable mapping results.
\end{itemize}

\subsection{Principles of image reconstruction}

Because of the voids in the $(u,v)$ plane coverage and the
lack of complete phase information, a given set of data can be fitted
by several different brightness distributions. The image power spectra
and bispectra of all these distributions fit the
data within the error bars in the least squared sense. Hence,
additional prior information is required to discriminate between those
images and to further constrain the solutions. The best image is
defined as the most probable image given the data and prior
information.  The bias toward the prior is the price to pay to select
an image from such sparse data. Reconstructing an image consists then
in minimizing a quantity connected to the data and to the prior with
respect to the pixel values of the image \citep{2009arXiv0909.2228T}:
\begin{equation}
  \label{eq:imaging-problem}
  \boldsymbol{x}_{\mathsf{best}} = {\rm arg min}_{\boldsymbol{x}\in\boldsymbol{\Omega}}
  \left[ f_{\mathsf{data}}(\boldsymbol{x};\boldsymbol{y})
  + \mu\,f_{\mathsf{prior}}(\boldsymbol{x}) \right]
\end{equation}
\noindent where $\boldsymbol{x}$ are the image parameters (\textit{e.g.} the
\emph{pixel} values) constrained to belong to the feasible set
$\boldsymbol{\Omega}$ of, \textit{e.g.}, non-negative and normalized images,
$f_{\mathsf{data}}$ measures the discrepancy between the image model and the
data $\boldsymbol{y}$, $f_{\mathsf{prior}}$ is the \emph{regularization} term
which enforces the priors and the hyperparameter $\mu$ is a weight factor that
tunes the balance between these two terms. 

The term $f_{\mathsf{data}}$ is usually derived from an analytical
model of the data and the statistics of the noise, although slightly
different approximations \citep[\textit{e.g.},][]{2005JOSAA..22.2348M}
are made by current image synthesis algorithms.  Note that
$f_{\mathsf{data}}$ may be itself the sum of different terms to
account for different kinds of data such as power spectra and phase
closures.

The term $f_{\mathsf{prior}}$ is used to inject in the minimization
process a priori knowledge on the object. In addition to fundamental
constraints such as positivity (i.e. intensity can not be negative) and
limited field of view there are numerous possible priors that can be
included in the regularization. Although automation of the
image reconstruction process is highly desirable, most of current
$(u,v)$ coverages in optical interferometry leave room for multiple
possible solutions. Therefore the observer cannot avoid introducing,
through the regularization, his knowledge of the expected brightness
distribution.  The interested reader is referred to the recent work by
\cite{Renard:2011} for a benchmarking of regularizations.

There are several popular regularizations.
\begin{enumerate}
\item Maximum Entropy Method \citep[MEM,][]{Gull:1984,Skilling:1984,1986ARA&A..24..127N}. This is
  one of the most employed techniques. It tries
  to find the smoothest image compatible with the data while keeping
  high frequency information. For that purpose it uses a smoothness
  scalar estimator called \emph{entropy} which has several possible
  definitions, one of the most popular being the Gull-Skilling entropy
  \citep{1984MNRAS.211..111S}. MEM will find the image that best fits
  the data and keeps entropy maximum. Because of the smoothness
  requirements MEM will introduce \emph{super-resolution} to the
  image, i.e. it will put information at spatial frequencies higher than the
  usual optical Rayleigh criterion.
\item \emph{Compactness} allows the definition of a spatial zone outside of
  which no flux is expected.
\item \emph{Total Variation} tries to minimize the total gradient of
  the image. It favors parts of the image with uniform zones and fast
  intensity changes (like a stellar surface). It has been found by
  \citet{Renard:2011} as one the best ``all-terrain'' regularization
  methods when
  applied to different object morphologies.
\item \emph{Smoothness} regularization smooths the image but can be
  adjusted to preserve some sharp variations at edges or point-like
  structures, favor some regions in the image etc. There are various
  ways to force and control smoothness. 
\item \emph{Sparseness} aims at restoring an image with the
  smallest number of components (\textit{e.g.}  CLEAN method in
  radio-astronomy or Building Block Method in optical interferometry,
  see below).
\end{enumerate}

Figure \ref{fig:priors} illustrates visually the effect of
different prior regularizations on different types of brightness
distributions. The first column is a doughnut like structure surrounding a
central star. The middle column is a photosphere with limb-darkening. The
last column is a galaxy composed of extended and point-like emission.
The prior regularizations, from top to bottom, are: the original image, smoothness, compactness, total
variation, smoothness with sharp boundaries (e.g  $\ell_2-\ell_1$ norm),
$\ell_p$ norm with $p=0.5$ and $p=2$, squared-root MEM and MEM with
gaussian a priori image.

\begin{figure}[p]
  \centering
\includegraphics[width = 0.6\textwidth]{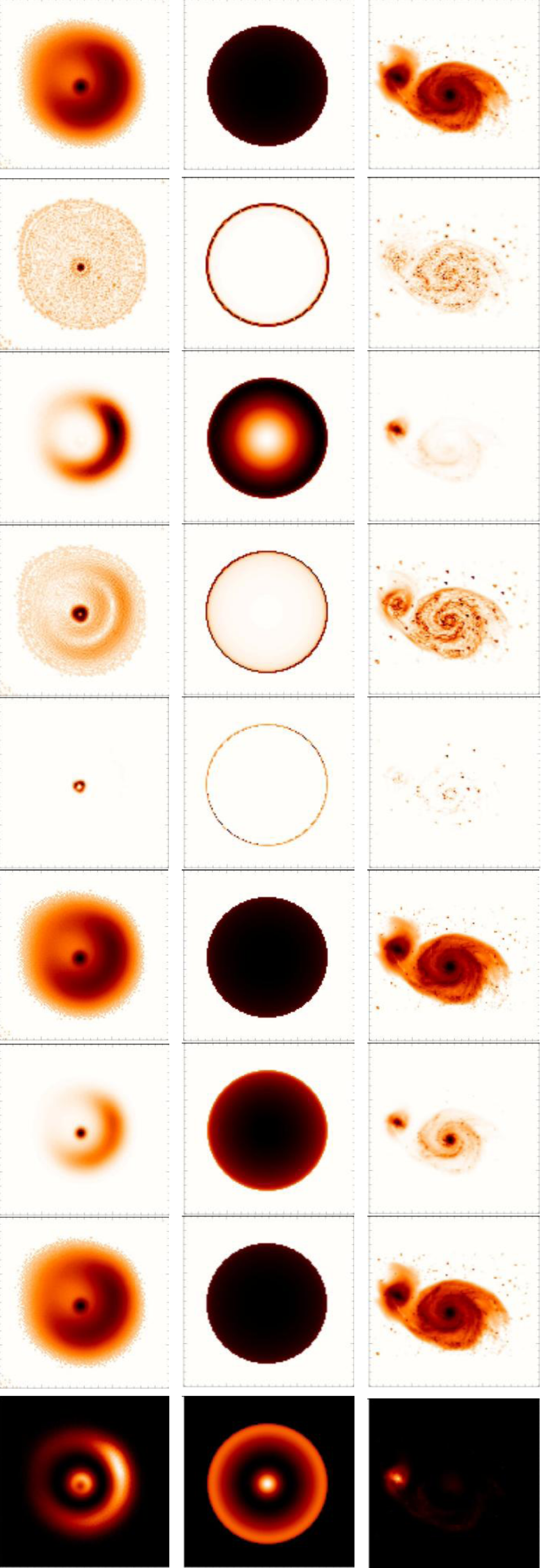}
 \caption{Effects of different priors. From left to right:
   reconstructed images of a doughnut
   like structure surrounding a central star, a photosphere with limb-darkening and a galaxy composed of extended and point-like
   emission. From top to bottom: original image, smoothness, compactness, total
variation, smoothness with sharp boundaries (e.g $\ell_2-\ell_1$ norm),
$\ell_p$ norm with $p=0.5$ and $p=2$, squared-root MEM and MEM with
gaussian a priori image.}
\label{fig:priors}
\end{figure}

\subsection{Imaging algorithms and their application}

In the radio domain the CLEAN algorithm \citep{1974A&AS...15..417H}
and its different flavors associated or not with self-calibration
\citep{Cornwell:1981} have been the dominant tools for image
reconstruction. This is not the case in optical long baseline
interferometry where mapping techniques are still the subject of active
research. It is essentially the absence of direct phase information that has
triggered new developments with respect to radio-astronomy. Although
this effort started with the rise of aperture masking
\citep{Baldwin:1986,Haniff:1987,Readhead:1988} in single-dish
telescopes, the similarities are so strong that both techniques can use
the same tools. For our purposes we can safely ignore the
specificities. Here we discuss the methods that are being
actively used in actual astrophysical programs and are sometimes
tailored to deal with different source structures and the number of
telescopes used.

\paragraph{The Building Block Method (BBM)} has been proposed by
\citet{1993A&A...278..328H} and offers many similarities with the CLEAN
method except that it can handle bispectra (and hence closure
phases). The minimization is carried through a matching pursuit
algorithm that imposes the sparsest solution. 

\paragraph{The BiSpectrum Maximum Entropy Method (BSMEM)} uses the MEM
method to regularize the image reconstruction process using visibility
amplitudes and bispectra. It supposes independent gaussian statistics
for amplitude and closure phase and uses the Gull-Skilling
\citep{1984MNRAS.211..111S} entropy as regularization criterion. Image
entropy provides a metric to measure smoothness in the image. This
criterion will minimize the regularization weight based on the
comparison with a reference image that can be updated through an
iterative process. It uses a specific numerical optimizer MEMSYS that
implements the strategy established by \citet{Skilling:1984} and can
handle any type of data.

\paragraph{MEM and self-calibration} Here, the maximum entropy method
is embedded in an interactive process that starts with a guess on the
instrument phases compliant with the closure phase measurements. The
iteration includes fitting telescope atmospheric and instrumental
phase errors and remapping of the image while respecting the closure
phases. It therefore iterates in order to converge to the most
probable image using the best guessed phases.

\paragraph{The MACIM} algorithm \citep[MArkov Chain
Imager][]{2006SPIE.6268E..58I}, unlike the preceding tries to find
globally the images that are optimal in the Bayesian sense (and not the image
that apparently fits best the data). For that purpose it tries to maximize 
the \emph{a posteriori} probability:
\begin{equation}
\mathrm{Pr}(\vec{x},\vec{y}) = \exp\left[-\frac{1}{2}(f_{\mathsf{data}}+\mu f_{\mathsf{prior}})\right]
\end{equation}
and therefore can provide a joint probability density of the probable
images. It can use or not use regularizations (currently MEM and dark zone
connectivity) and handles any kind of data.

\paragraph{The MIRA} algorithm \citep[Multi-aperture Image
Reconstruction Algorithm, ][]{2008SPIE.7013E..43T} minimizes 
Eq. ~(\ref{eq:imaging-problem}) by using a non linear optimization
algorithm. Because the method does not allow a global optimization, it
will depend on using the initial image as a constraint. It does not try
to reconstruct the phases directly and therefore can handle closure
phases data and any other observable. It encodes many types of
regularization which offers the possibility of testing the reliability of
the reconstructed images.

\paragraph{The WISARD} algorithm \citep[Weak-phase Interferometric
Sample Alternating Reconstruction Device, ][]{Meimon:2008}, 
is an application of the self-calibration idea to optical
interferometry. It reconstructs the phases from the
closure phases. The phase inversion process has multiple minima and
therefore the global minimization is a multi-modal problem and can
depend on the initial image.

\begin{table*}
  \centering
  \caption{Comparison of the image synthesis algorithms used in
    optical interferometry.} 
\label{tab:imaging-algorithms}
\begin{tabular}{p{14mm}p{28mm}p{32mm}p{32mm}}
    \hline
    Code\par name 
    &Data
    &Regularization 
    &Optimization Strategy\\
    \hline                                                                             
    \hline                                                                             
    MiRA      
    &any                                                                     
    &positivity, total variation, $\ell_2$, $\ell_2{-}\ell_1$, Gull-Skilling entropy with floating or given prior image 
    &limited memory quasi-Newton with bound (positivity) and normalization constraints \\
    \hline
    BSMEM     
    &bispectra, power spectra, \par complex visibilities                   
    &Gull-Skilling entropy with given prior image, multi-scale entropy                                      
    &non-linear conjugate gradients with unsupervised hyper parameter control           \\
    \hline
    WISARD    
    &pseudo-complex visibilities formed from phase closures and power spectra 
    &positivity, $\ell_2$, $\ell_2{-}\ell_1$                                                               
    &alternative minimization and self calibration                                     \\
    \hline
    BBM       
    &bispectra                                                               
    &positivity, sparseness                                                                                
    &matching pursuit                                                                  \\ 
    \hline
    MACIM     
    &any                                                                     
    &any                                                                                                   
    &global optimization by simulated annealing                                        \\
    \hline
  \end{tabular}
\end{table*}

Table~\ref{tab:imaging-algorithms} summarizes the features of the
different algorithms that were specifically designed to cope with
optical interferometry data.  These algorithms are the main challengers of
the \emph{Interferometric Imaging Beauty Contests} which, every two
years since 2004, quantitatively compares the results of various image
synthesis methods on simulated optical interferometric data
\citep{2004SPIE.5491..886L, 2006SPIE.6268E..59L, 2008SPIE.7013E..48C, 2010SPIE.7734E..83M}.
Among these algorithms, BSMEM \citep{1994IAUS..158...91B,
  2008SPIE.7013E.121B} and MiRA \citep{2008SPIE.7013E..43T} have
repeatedly won the contest for data sets typical of next generation
instruments. These are the ones that were used to produce the images from simulated
data sets presented in this paper, together with  \textsc{MACIM}
\citep{2006SPIE.6268E..58I}.  

\subsection{Examples}

\begin{figure*}[!t]
  \centering
\includegraphics[width=0.73\textwidth]{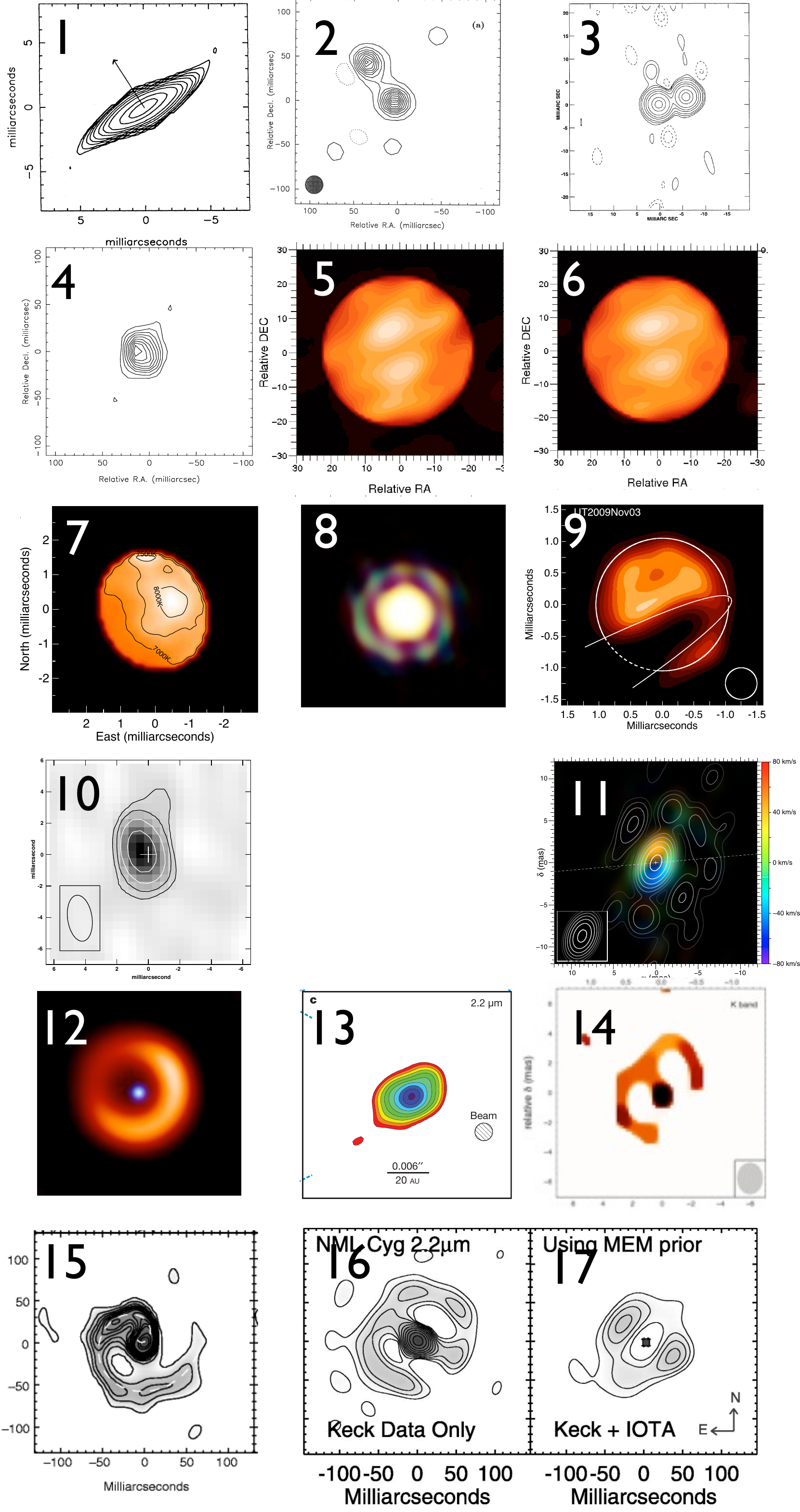}
\caption{The wall: examples of published reconstructed images (see text for
  details). 1:\citet{Quirrenbach:1994} 2:\citet{Baldwin:1996}
  3:\citet{Benson:1997} 4:\cite{2000MNRAS.315..635Y} 5\&6
  \citet{2009A&A...508..923H} 7:\citet{2007Sci...317..342M},8:
  \citet{2009A&A...496L...1L},
  9:\citet{2010Natur.464..870K},10:\citet{Schmitt:2009}, 11:
  \cite{Millour:2011},12:\citet{Tuthill:2002}, 13:\citet{Kraus:2010},
  14:\citet{Benisty:2011}, 15 \citet{Tuthill:1999}, 16\&17
  \citet{Monnier:2004}. Credit for figure 8: ESO/J.-B. Le Bouquin et al.}
  \label{fig:archives}
\end{figure*}

In this section, we describe some selected significant astrophysical
publications making use of reconstructed imaging
(Fig.~\ref{fig:archives}). While we acknowledge that these choices
suffer from selection effects, (for example we have limited the amount
of binary images) we believe that these examples illustrate well the
state of the art.

Image 1 in Fig.~\ref{fig:archives} shows one of the early attempts
by \citet{Quirrenbach:1994} to reconstruct images out of visibility
amplitude data using the Mark III interferometer. Even without phases
they were able, using MEM-inspired methods, to map the H$_\alpha$
emission around $\zeta$ Tauri showing clearly its extension and
flattening (related to its inclination). The absence of phases appears
in the centro-symmetry of the image.

Images 2 and 3 in Fig. \ref{fig:archives} show the pioneering images of binaries
obtained at COAST \citep{Baldwin:1996} and NPOI (now NOI) \citep{Benson:1997}
using standard radio packages associated with
self-calibration (for the NOI reconstruction).

\citet{2000MNRAS.315..635Y} used aperture masking and the COAST to map
the surface of Betelgeuse. In image 4, the aperture masking image
obtained with the William Herschel Telescope is shown. The surface is
marginally resolved since the angular resolution is barely
sufficient. The image was obtained with the combination of the MEM
method and self-calibration.  More recently
\citet{2009A&A...508..923H} imaged the spotty surface of Betelgeuse
using IOTA (images 5 and 6 of Fig. \ref{fig:archives}).  Both WISARD
and MIRA reconstructions lead to the same result confirming what had
already revealed by \citet{2000MNRAS.315..635Y}. In this case, both
the dynamic range and the ability to characterize the spots are
limited by the instrument accuracy and the difficulty for an
interferometer to lock on low visibilities. This constraint has to be
taken into account in the design of any future ``imaging instrument'',
as we will discuss later.

Images 7 to 9 of Fig. \ref{fig:archives} are reconstructed images of
respectively Altair, T Leporis and Epsilon Aurigae.  Images 7
and 9 were obtained with CHARA using MACIM and BSMEM \citep[respectively][]{2007Sci...317..342M,2010Natur.464..870K}.  Image 8 was taken with VLTI using
MIRA \citep{2009A&A...496L...1L}.  These images are an illustration of how
image reconstruction can retrieve moderately complex morphologies with
a limited $(u,v)$ coverage that might have escaped traditional model
fitting strategies.

Spectral phase referencing has proven to be a powerful tool to image line
emitting regions. For this purpose, a sufficient spectral resolution is
required in order to compare purely continuum emission to line
emission interferometric observables. Results using this technique are
illustrated in images 10 and 11 of Fig. \ref{fig:archives} where the
$H_{\alpha}$ emission around the star $\beta$ Lyrae has been mapped
using the AIPS package \citep{Schmitt:2009}. This was possible
because each baseline phase corresponding to the line emission was
isolated from the continuum. In a similar spirit, \citet{Millour:2011}
have used AMBER with high spectral resolution, MIRA, and a self-calibration
approach to retrieve absolute phases in the Br$_{\gamma}$ line of
the supergiant HD 62623. Since
the line was both spectrally and spatially resolved it was possible to
reveal the kinematics of a disk around the central star.

Mapping the infrared emission of protoplanetary disks has always been
one of the key science drivers for optical arrays and their focal
instrumentation. Images 12 to 14 of Fig. \ref{fig:archives} are an
illustration of the best disks maps achieved so far in this
field. Images 12,13,14 have been obtained, respectively, by Aperture
Masking at Keck \citep{Tuthill:2002} and long baseline interferometry
at VLTI \citep{Kraus:2010,Benisty:2011}.  Images 12 and 14 clearly
show the inner rim of the circumstellar disk caused by dust
sublimation. Maps 13 (using the BBM method) and 14 (using MIRA)
display limited features in the image because of the relatively sparse
$(u,v)$ coverage obtained with the VLTI.

The images obtained with aperture masking at Keck by
\citet[][Wolf-Rayet 104]{Tuthill:1999} and \citet[][Evolved star NML
Cyg]{Monnier:2004} show the most complex features of all the images
selected here (images 15-17 of  Fig. \ref{fig:archives}). This
complexity is a direct
illustration of the importance of a sufficient $(u,v)$ coverage since both
used masks with 21 apertures. MEM priors were chosen in the
reconstruction process. The case of the NML Cyg map is also very
illustrative of the requirement to strategically pave the $(u,v)$
plane. Therefore, one should always consider complementing long-baseline
interferometry with single-dish aperture masking if possible or
vice-versa. In this particular case, the Keck aperture masking data
lacked high spatial frequency constraints on the central source (left
image). The addition of IOTA data allowed  the
central source extension to be characterized and improved the quality of the
reconstruction. Conversely, extended fluxes might alter interferometric
data and should therefore be at least calibrated and ideally measured
with short spatial frequency instruments such as aperture masking
instruments.

\section{The future of aperture synthesis:
 astrophysical objects at the milli-arcsecond scale}
\label{sec:astro-simus}

While optical long baseline interferometers have made tremendous
technical and scientific progress, we wanted to evaluate what
improvements could be brought to the field of aperture synthesis
imaging. Following the recommendations of the SOC supervising the
\emph{Workshop on Interferometric Imaging 2009} (see
acknowledgements), we selected four main astrophysical topics for
which visible and infrared interferometers are expected to bring
breakthroughs and for which milli-arcsecond scale imaging will
certainly improve our understanding of the physical phenomena. We
restricted ourselves to the study of the continuum emission of various
object classes. The following sections detail those topics.

\subsection{Planetary signatures in protoplanetary disks}

\subsubsection{Context and major issues}

Currently\footnote{as of December 2011}, more than 700
exoplanets\footnote{http://exoplanet.eu} together with several thousand of
planet candidates have been detected so far and
although the detection techniques used may cause strong
selection effects, it is now obvious that the solar-system planets
cover a small region in the parameter space of possible planetary
characteristics. To understand the observed diversity of planets, one
has to consider the history of the planetary systems by studying
their formation and evolution. Triggered and supported by the rapidly
growing field of the search for extrasolar planets and the
observational material already collected, our theoretical
understanding about the planet formation process has been improved
significantly.  Nevertheless, the link between theoretical models and
observations is still rather weak. This link is provided by
circumstellar disks in which planets are expected to form.  These
disks are considered as the natural outcome of the protostellar
evolution, at least in the case of low and intermediate mass stars
\citep[e.g.][]{1987ApJ...312..788A, 1993ARA&A..31..129L}.  These disks
have masses
\citep[$10^{-3}-1\,\mbox{M}_\odot$;][]{1990AJ.....99..924B} and sizes
\citep[$50-1000\,\mathrm{AU}$;][]{1996AJ....111.1977M,
  1996A&A...309..493D} comparable to those expected for the primitive
solar nebula.

Theoretical investigations show that the planet-disk interaction
creates structures in young circumstellar disks which are usually much
larger in size than the planet itself and thus more easily detectable
\citep{2003ApJ...588..494M, 2004ApJ...612.1152V, 2007MNRAS.377.1324C}.
The specific result of the planet-disk interaction depends on the
evolutionary stage of the disk.  Typical signatures of planets
embedded in disks are gaps and spiral density waves in the case of
young, gas-rich protoplanetary disks, and characteristic asymmetric
density patterns in older debris disks.  Numerical simulations
demonstrate that high-resolution imaging performed with observational
facilities (already existing or soon becoming available) will allow
these ``fingerprints'' of planets to be traced in protoplanetary and
debris disks \citep[e.g.][]{2005ApJ...619.1114W} but the prospect of
such detections lies mostly in the mm domain.  These observations will
provide a deep insight into specific phases of the formation and early
evolution of planets in circumstellar disks.  For example,
\citet{2002ApJ...578L..79W} showed that the Atacama Large Millimeter
Array (ALMA) will allow tracing of hyper-dense vortices which are
supposed to represent an early stage of planet formation. Some authors
have tested the detectability of planet-induced features, such as
temperature increase due to planetary accretion, in the
near-to-mid-infrared domain with interferometers
\citep[e.g.][]{2008PhST..130a4025W} and have already pointed out that
this represents a major challenge. These studies have to be pursued.

For this work we wanted to test the limit of the planet-induced
structures that could be detected. We have therefore not considered
other crucial observational diagnostics that could also (and already
do) allow the properties of protoplanetary environments at the
astronomical unit scale to be further constrained. These observable
diagnostics include the disk inner rim vertical structure, the dust
content and its vertical distribution, kinematics, asymmetries etc.
Since these are already the subject of active research we instead
chose to investigate more challenging observables which could become
accessible in the future.

\subsubsection{Simulated images of disks perturbed by planets}
\label{sec:disks-simul}

The model for radiative transfer in the vicinity of an embedded
protoplanet used in this work is detailed in
\citet{2008ApJ...679..797J}.  It is based on a one-dimensional
analytical solution for radiative transfer in an optically thick
protoplanetary disk \citep{1998ApJ...500..411D} extended to three
dimensions.  The calculation is carried out iteratively for
self-consistency between temperature and density perturbations.  The
generation of simulated images is described in
\citet{2009ApJ...700..820J}.

The synthetic image produced for the image reconstruction is of a
face-on disk at a distance of 140\,pc.  Since the simulation
boundaries are limited in radius at both the inner and outer edges,
the disk brightness is smoothly extrapolated as a power-law interior
and exterior to the simulated region in order to eliminate image edge
artifacts.  The star is modeled as a black body of radius
$R_*=2.6\,R_{\odot}$ and effective temperature $T=4280\,\mbox{K}$,
corresponding to a 1\,Myr old star of mass $M_*=1\,M_\odot$
\citep{2000A&A...358..593S}.

\begin{figure*}[t]
  \centering
  \begin{tabular}{cc}
\includegraphics[height=0.45\textwidth]{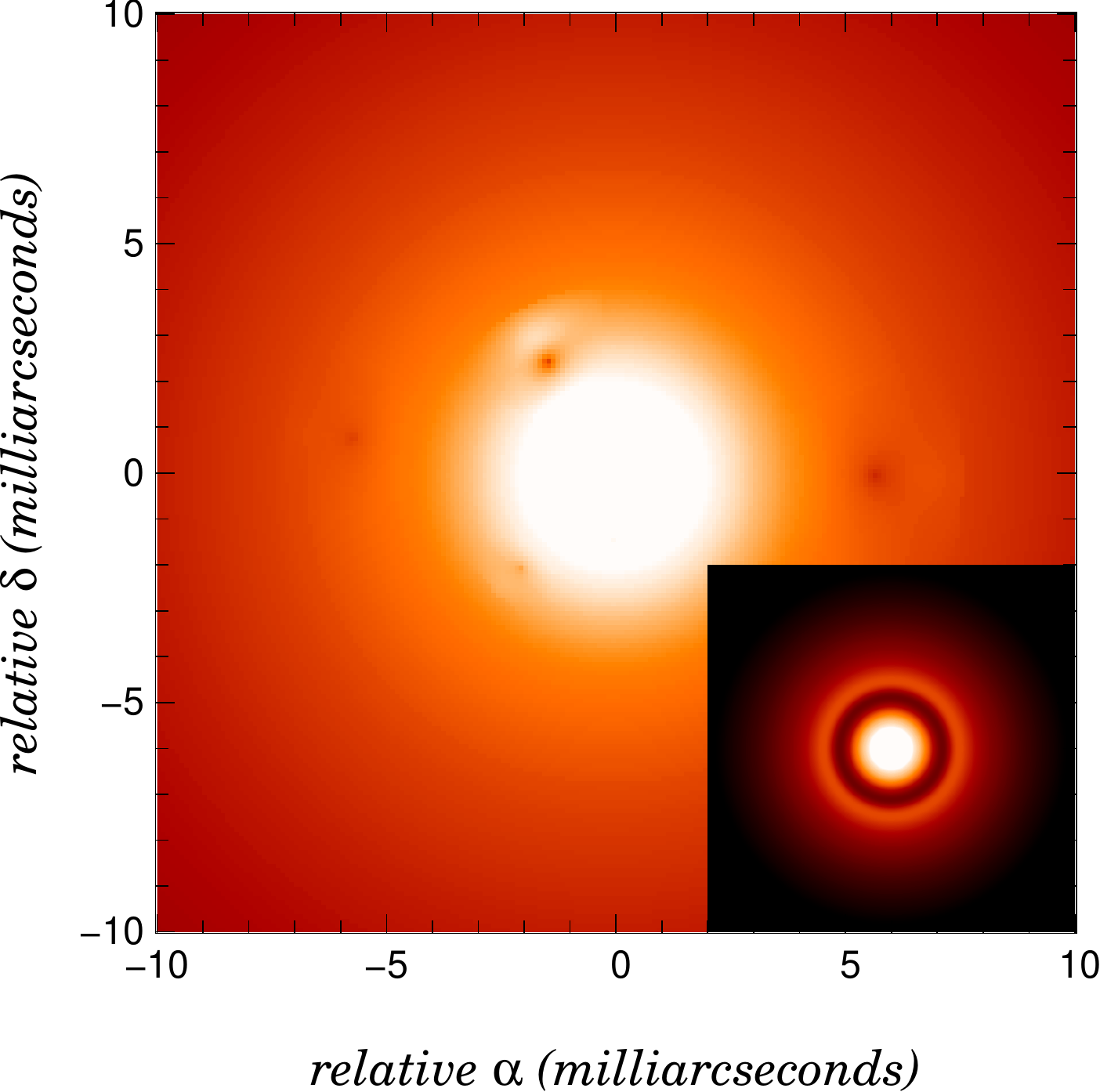} &
\includegraphics[height=0.45\textwidth]{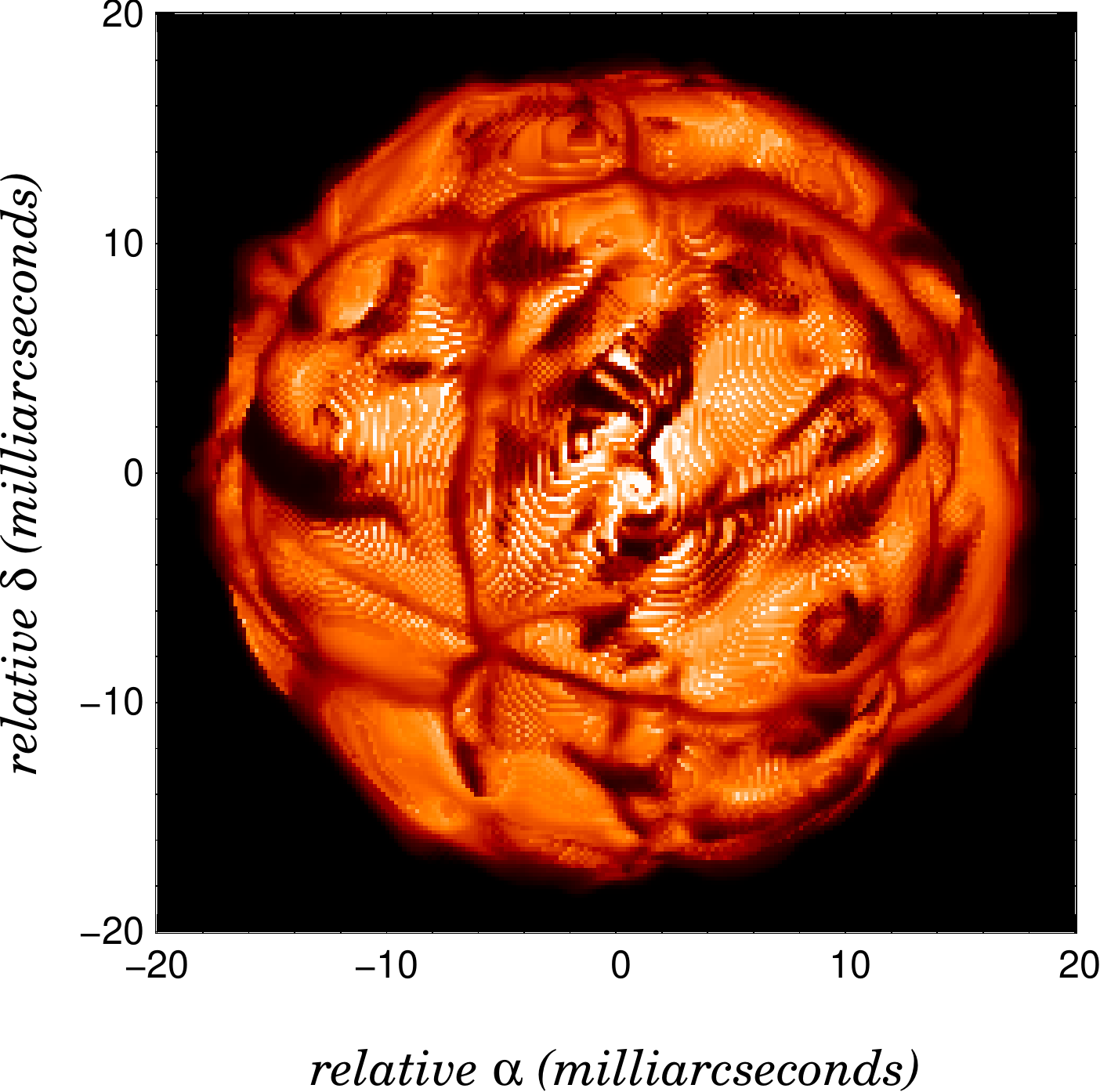}\\
\includegraphics[height=0.45\textwidth]{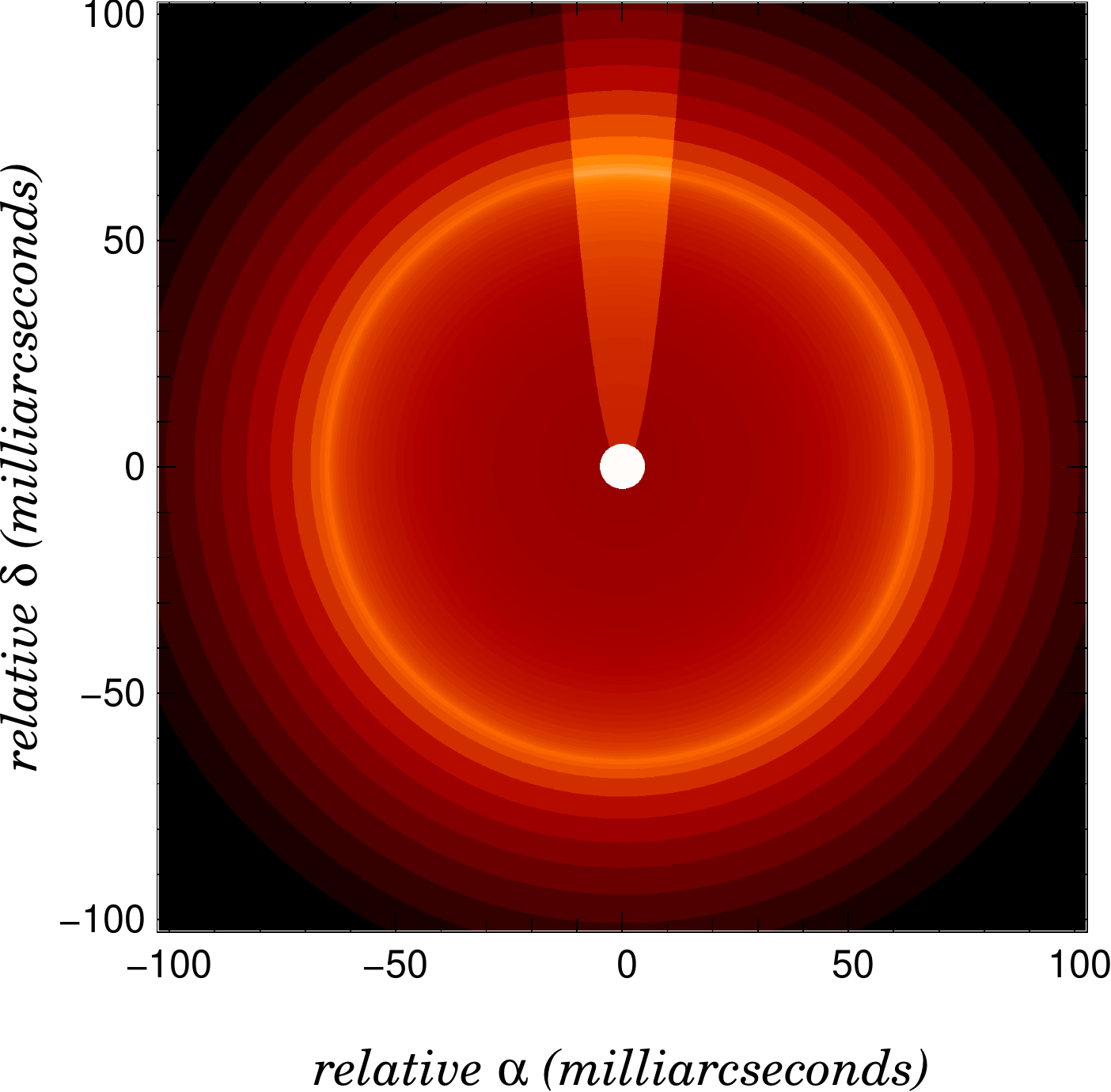} &
\includegraphics[height=0.45\textwidth]{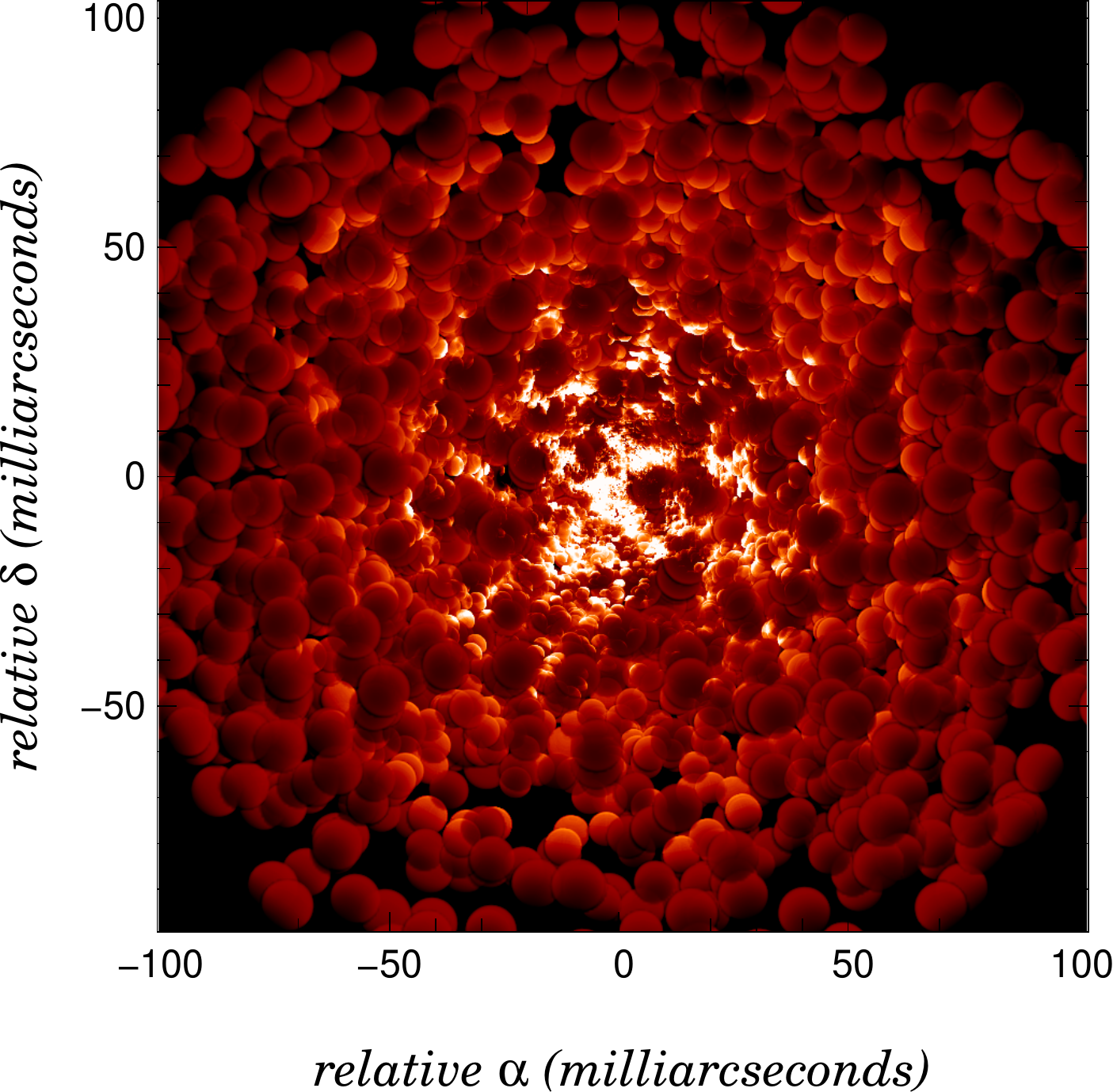}\\
\end{tabular}
\caption{Simulated images of our 4 science cases. Top, left to right:
  (a) disk in the $N$ band with shadows created by low-mass embedded
  planets (Sect.~\ref{sec:disks-simul}), inset shows disk with gap in
  the $N$ band; (b) stellar surface of a supergiant star with
  convective cells in the $K$ band
  (Sect.~\ref{sec:star-simul}). Bottom, left to right: (c) evolved
  star with surrounding multi-layer envelope and plume emission in the
  $K$ band (Sect.~\ref{sec:molecular-layers}); (d) a clumpy dust
  torus in the active galactic nuclei NGC1068 in the $K$ band (Sect.~\ref{sec:agn-simul}). The simulated images are presented in
  their largest field of view when available. }
  \label{fig:imageswide}
\end{figure*}

The first set of images consisted of planets (top left image of Fig.~\ref{fig:imageswide}) creating localized
shadows \citep{2009ApJ...700..820J}: (i) 10 and
$20\,\mbox{M}_{\oplus}$ planets at $8\,\mathrm{AU}$, (ii) 10 and
$50\,\mbox{M}_{\oplus}$ planets at $4\,\mathrm{AU}$, (iii) a
$20\,\mbox{M}_{\oplus}$ planet at $2\,\mathrm{AU}$, and (iv) a
$50\,\mbox{M}_{\oplus}$ planet at $1\,\mathrm{AU}$. All these planets are present
in the image. The second set of images (inset of the top left
image of Fig.~\ref{fig:imageswide}) were of gaps created by tidal
forces between the planet and disks, similar to those shown in
hydrodynamic simulations by \citet{2003MNRAS.341..213B}.  These gaps
were modeled as \emph{ad-hoc} azimuthally-symmetric perturbations.  If
the unperturbed surface density profile is defined as $\Sigma_0(r) =
\int_{-\infty}^{\infty} \rho(r,z) \, dz$ then the surface density
profile of a disk with a gap whose center is located at radial
distance $r_{g}$ of depth $d$ and width $w$ is $\Sigma = \Sigma_0
[1-d\exp(-(r-r_{g})^2/2w^2)] $.  Holding $\Sigma$ fixed, the
vertical density and temperature structures were recalculated using
the radiative transfer model described above.  Two images of a gap at
$r_{g}=10 \,\mathrm{AU}$ was calculated: (1) ``big gap'' scenario with $d=0.9$
and $w=1.8 \,\mathrm{AU}$, corresponding to a 110 $M_{\oplus}$ (0.35 $M_J$)
planet; and (2) ``small gap'' scenario with $d=0.5$ and $w=1.0 \,\mathrm{AU}$,
corresponding to a 37 $M_{\oplus}$ (0.12 $M_J$) planet.



\subsection{Granulation in late-type stars due to large convective cells}

\subsubsection{Context and major issues}

Outside the spots, the solar surface is covered by bright granules
with warm upflowing material surrounded by dark, cool inter-granular
lanes where matter falls downward.  Local multi-dimensional radiation
hydrodynamics (RHD) simulations of stellar surface convection, solving
the coupled equations of hydrodynamics and non-local radiation
transport including constant gravity, ionization effects, and
realistic (binned) opacities, first successfully reproduced the
properties of solar granulation \citep{1982A&A...107....1N}. This
includes pattern and dynamics (incl.\ size and timescales) of the
granulation, as well as center-to-limb variation, and line profiles
\citep{1981A&A....96..345D}.  The excellent agreement between
observations and models allowed these methods to be extended with
some confidence to other stars, replacing classical 1D stellar atmospheres
for high-accuracy abundance determinations \citep{2001ApJ...556L..63A,
  2008A&A...488.1031C}.

The size and also the contrast of the surface structures can be
directly linked to the efficiency of convection, to the formation of molecular
lines (CO on the Sun: \citealt{2005A&A...438.1043W}; water on Arcturus:
\citealt{2002ApJ...580..447R}; water on Betelgeuse:
\citealt{2006ApJ...637.1040R}), to the micro-variability \citep[limiting the
detectability of planets][]{2006A&A...445..661L} and to the vertical
extent of ``overshoot'' of convective influence
\citep{1996A&A...313..497F} into the outer layers (molsphere, wind: Sect.~\ref{sec:molecular-layers}; and dust formation:
\citealt{2008A&A...483..571F}).

Simulations of solar-type stars show the granulation properties known
from the Sun, with a horizontal size scaled with the surface pressure
scale height and slightly different contrast
\citep[e.g.][]{2004MNRAS.347.1208R}. However, the situation changes
for objects near the ``granulation boundary'' (A-type stars, cepheids,
red supergiants): \citet{2008AJ....135.1450G} observes reversed
C-shaped line bisectors on Betelgeuse and relates them to granulation
and giant convection cells accompanied by short-lived oscillations .
Also \citet{1971A&A....10..290S} and \citet{1975ApJ...195..137S}
attribute the observed photometric variability in red supergiants to
convection cells larger than one would expect from scaled solar
granulation, which is confirmed later by interferometric observations
of global surface structures on Betelgeuse \citep{1990MNRAS.245P...7B,
  2000MNRAS.315..635Y}. \citet{2007A&A...469..671J} find large line
shifts and asymmetries in RSG spectra and attribute them to the
existence and motion of a few large granules. Local RHD models of
surface patches of A-type stars and cepheids
\citep{2008PhST..133a4005F} and global models of red supergiants
(using a spherical potential and including the entire outer convective
envelope and the atmosphere, \citealt{2002AN....323..213F}) also show
similarly large cells and granulation with properties deviating from
the solar case.

For a large group of stars near the granulation boundary -- from the
main-sequence to red supergiants -- our benchmark star, the Sun, is
therefore no longer useful.  There is a clear need for new references, and
interferometric imaging will be capable of providing one for the
important case of red supergiants, shedding light on this particular
kind of surface granulation and its relation e.g.\ to the outer layers
and wind formation. Recent images of Betelgeuse obtained with IOTA
in the $H$ band \citep{2009A&A...508..923H} display two low contrast spots, one
of them probably marginally resolved. The spots are consistent with
signatures of large convective cells \citep{2009A&A...506.1351C}.

\subsubsection{Stellar surface simulation}
\label{sec:star-simul}

\citet{2009A&A...506.1351C} computed intensity maps at different
wavelengths (corresponding to broadband filters $H$ and $K$) with the
radiative transfer code OPTIM3D from snapshots of the 3D
hydrodynamical simulation (code CO$^5$BOLD) of a red supergiant star
presented above. They used a model with stellar
parameters close to those of Betelgeuse \citep{2005ApJ...628..973L}: a
$12\,\mathrm{M}_{\odot}$ stellar mass, a luminosity averaged over time of
$L=93000\pm1300\,\mathrm{L}_{\odot}$, an effective temperature of
$T_{\rm{eff}}=3490\mathrm{K}$, a radius of $R=832\mathrm{R}_{\odot}$,
and surface gravity $\log(g)=-0.337$. The numerical
resolution is $235^3$ with a grid spacing of $8.6\mathrm{R}_{\odot}$.

The granulation pattern has a significant impact on interferometric
observables. In order to derive their characteristic signature,
\citet{2009A&A...506.1351C} computed visibilities and closure phases
from these intensity maps. They found that convection-related surface
structures cause fluctuations in interferometric observables which
manifest themselves mostly at high spatial frequencies after the first
null point in the visibility function. In general, the visibilities
and closure phases deviate greatly from the uniform disk or
limb-darkened cases, due to the small scale structure on the model
stellar disk.  These observables have been compared to existing
Betelgeuse observations \citep{2000MNRAS.315..635Y,
  2004UKNAM.........Y, 2009A&A...508..923H}. The 3D simulation gives
good fits to the observed visibility points and closure phases
providing a consistent solid detection and characterization of the
granulation pattern from the optical to the near-infrared
\citep{2010arXiv1003.1407C}.

The granulation pattern can be characterized with today's
interferometers by searching for angular or temporal visibility
fluctuations or by looking at different spectral regions corresponding
to spectral features and continuum. High-angular resolution images
should verify or contradict the 3D simulations in terms of surface
intensity contrast, granulation size and temporal variation.  The
figure at the top right side of Fig.~\ref{fig:imageswide} shows the
image, computed in a narrow $K$ band channel, that has been used in
this study.


\subsection{Molecular layers around late-type stars}
\label{sec:molecular-layers}

\subsubsection{Context and major issues}

The late stages of stellar evolution are well studied because they are
the source of most of the dust produced in the universe
\citep{1994LNP...428..163S}. Dust is detected at large distances and
stars are known to produce molecules which are the building blocks of
dust particles. A molecular shell containing water vapor and carbon
monoxide was first suspected around red supergiants by
\citet{2000ApJ...540L..99T} who coined the name molsphere to describe
it. This explanation was shown to be consistent with interferometric
data collected in the near and mid-infrared
\citep{2004A&A...418..675P, 2004ApJ...611L..37W, 2004A&A...421.1149O}
and subsequent works showed the list of molecules could be further
extended to include SiO and Al$_2$O$_3$, the latter being a dust species
that can survive the high temperature range of the molsphere
(1500--2000\,K) and provide nucleation sites for SiO
\citep{2007A&A...474..599P, 2006A&A...447..311V}.

Molecules in the atmosphere of Mira stars are probably lifted by
large-amplitude pulsations. This is possibly a different process from
the one at play in supergiants, yet similarities have been found
between the two classes of objects. Diameters of Mira stars are known
to vary by factors of a few in and out of TiO bands
\citep[e.g.][]{1977ApJ...218L..75L} as well as in other bands although
with smaller amplitude. Measured interferometric diameters have been
difficult to explain with models partly because of pulsations. Some
dynamic models have been compared to interferometric data
\citep[e.g.][]{2004A&A...421..703W, 2005A&A...431.1019F}. Using the
\citet{1987A&A...186..200S} model and removing the contribution of the
continuum in the $K$ band, \citet{1999A&A...345..221P} have shown a
good agreement with R~Leo interferometric data raising the hypothesis
that molecules are a major contributor to the infrared surface
brightness distribution of these objects.  This was confirmed by
fitting $K$ and $L$ band data using a model incorporating a molecular
shell a stellar radius above the photosphere
\citep{2002ApJ...579..446M, 2004A&A...426..279P}. Recent works have
provided more detailed studies of the shell
(e.g. \citet{2005A&A...429.1057O, 2007ApJ...654L..77E,
  2007A&A...470..191W, 2009A&A...503..183O}). In parallel models have
been made more sophisticated and in particular now provide a physical
explanation for molspheres \citep{1999A&A...348L..17W}.

All studies had initially hypothesized circular symmetry and
homogeneity for the layer. Images would yield a refinement of the
structure of the layer and of the star, as well as the layer
brightness distribution. The images of T~Leporis in narrow spectral
channels across the $H$ and $K$ bands taken with VLTI
\citep{2009A&A...496L...1L} have confirmed the presence of the
circumstellar layer. The ring appears circularly uniform to within the
precision of image reconstruction with sparse apertures, consistent
with the type of shell model proposed by
\citet{2004A&A...426..279P}. Images taken with IOTA of $\chi$~Cygni in
the $H$ band do not disclose such a bright shell, although it is
detected by model-fitting to visibilities. The star itself shows
varying asymmetries as a function of the pulsation phase
\citep{2009ApJ...707..632L}. However insufficient dynamic range in the
reconstructed image does not allow any departure from circular
symmetry of the shell to be measured.

\subsubsection{Modeling  molecular layers}
\label{sec:mol-layers-simul}


Interferometric observations (and thus also reconstructed images) of
molecular layers can be compared either to self-consistent
(hydrodynamical) atmosphere models, or to flexible analytical models
which aim at a reproduction of the observed intensity distribution
with a minimum of physics. 

The first approach works reasonably well for Mira stars. For
example, \citet{2008A&A...479L..21W} use the models by
\citet{2004MNRAS.352..318I}, and \citet{2009ApJ...691.1328W} use
the models by \citet{2006ApJ...642..834K}, both on O-rich AGB
stars. Theoretical preparatory work on C-rich AGB stars has been
published by \citet{2009A&A...501.1073P}, based on models by
\citet{2003A&A...399..589H}, and a confrontation with observations is in
progress \citep{2011A&A...525A..42S}. 

In some cases, a more ad-hoc approach is required. This usually
implies a parametrized distribution of matter and opacity, and some
radiative transfer calculation. These models should then be checked
against thermal and chemical equilibrium.  Such an approach has also
been used with AGB star observations
\citep[e.g. ][]{2002ApJ...579..446M, 2004A&A...426..279P,
  2009A&A...496L...1L}, but it is hitherto the only option for red
supergiant stars \citep[e.g.][]{2006A&A...447..311V,
  2008A&A...484..371O}, or very asymmetrical circumstellar
environments \citep{2009A&A...504..115K}.

The synthetic images used in this work consist of an RSG photosphere
model, computed with the {\sc marcs} code \citep[][and further
  updates]{1975A&A....42..407G}, surrounded by two layers of molecular
material: a layer of CO is located at 1.15\,R$_{\star}$, with a
temperature of 2350\,K, and a column density of
10$^{23}$cm$^{-2}$. The second layer contains 10$^{21}$cm$^{-2}$ of
H$_2$O and is located at 1.7\,R$_{\star}$, with a temperature of
1850\,K. Around this object, we placed a dust shell which reproduces the
observed silicate features in $\alpha$\,Orionis ($\dot{M} \sim 3
\times 10^{-8}$\,M$_{\odot}$\,yr$^{-1}$). This shell starts at
13\,R$_{\star}$, and has a typical $r^{-2}$ outflow density
profile. To mimic the recent results of \citet{2009A&A...504..115K},
we included a plume as a characteristic of inhomogeneous mass loss,
possibly related to convection, as recently observed in
Betelgeuse. The chosen plume shape is a simplification of a structure
originating at the photosphere, experiencing dilution as it travels
outwards. It represents a local increase of 50\% of the circumstellar
intensity.  Images were computed across the $K$\,band, but only the
one near 2.4\,$\mu$m is presented here (bottom left side of
Fig.~\ref{fig:imageswide} and zoomed image at top centre of
Fig.~\ref{fig:images}), as it shows both the CO and the H$_2$O layer.
  

\subsection{Dusty tori in active galactic nuclei}
\label{sect:AGN}

\subsubsection{Context and major issues}

Although there are numerous classes of active galactic nuclei (AGN),
they have been unified into a single scheme
\citep[e.g.,][]{1993ARA&A..31..473A, 2002apsp.conf..151A,
  1995PASP..107..803U}. The basic premise of unification is that every
AGN is intrinsically the same object: a super-massive black hole whose
activity is powered by accretion through a surrounding disk. This
central engine is further surrounded by a dusty toroidal structure,
which provides anisotropic obscuration of the central region. The
observed diversity is simply the result of viewing this axisymmetric
geometry from different angles. Directions with clear sight of the
central engine yield ``type 1'' objects, where clouds in close
proximity to the black hole, moving at high velocities because of the
strong central gravity and ionized by the strong central continuum
radiation, give rise to broad line emission in high ionization
lines. AGN whose central engine and broad-line region (BLR) are
blocked from view are ``type 2''. 

The primary evidence for the torus arose from spectropolarimetric
observations of type 2 sources, which reveal hidden type-1 emission
via reflection off material situated above the torus opening. While
compelling, this evidence is only indirect in that it involves
obscuration, not direct emission by the torus itself. Direct evidence
comes from IR observations --- an obscuring dusty torus should
reradiate in the IR the fraction of nuclear luminosity it absorbs, and
the continua from most AGN indeed show significant IR emission. Recent
progress in IR high-angular resolution techniques achieved direct
imaging \citep[e.g.][]{2004A&A...425...77W} and interferometric
measurements \citep{2004Natur.429...47J, 2007A&A...471..453M,
  2007A&A...474..837T} of the torus in a few AGN. These observations
show that the torus is rather compact. The size scale is set by the
dust sublimation radius $R_{\rm d} \simeq 0.4\,(L/10^{45}\,\rm
erg\,s^{-1})^{1/2}\,\mbox{pc}$, which determines the torus inner
radius. All observations are consistent with $R_{\rm out}/R_{\rm d}$
no larger than $\sim$ 20--30, and perhaps even only $\sim$ 5--10. The
observations also indicate close proximity between dusty regions with
widely different temperatures. These findings can only be understood
if the dust distribution is clumpy because in that case the dust
temperature is much higher on the illuminated face of an optically
thick cloud than any other part of its surface, resulting in
co-location of different dust temperatures. The torus clumpiness,
predicted early on by \citet{1988ApJ...329..702K}, naturally explains
many features of the IR observations that cannot be understood in
smooth-density models \citep{2002ApJ...570L...9N, 2008NewAR..52..274E,
  2008ApJ...685..147N, 2008ApJ...685..160N}.

\subsubsection{Clumpy tori simulations}
\label{sec:agn-simul}

We simulated model images based on 3D clumpy torus models
\citep{2006A&A...452..459H, 2009arXiv0909.4539H}. The main parameters of the
models are (1) the radial dust distribution, parametrized as a
power-law $\eta(r)\propto r^a$, (2) the mean number of clouds $N_0$
along an equatorial line-of-sight, and (3) the half-opening angle
$\theta_0$ (defined as $\sigma$ in a Gaussian distribution along the
altitudinal direction). The inner radius of the torus $R_{\rm d}$ is
set by dust sublimation at temperature
$T_\mathrm{sub}=1500\,\mathrm{K}$ and scales with the luminosity of
the AGN, $R_{\rm d}\propto L^{1/2}$. While the model spectral energy
distribution (SED) has a negligible dependence on the actual size of
the dust clouds \citep[see][Sect.~2.5]{2009arXiv0909.4539H}, the clouds
might show up in the interferometric signal as small-scale emission or
obscuration regions. The cloud sizes are parametrized by a radial
power-law.

Starting from pre-calculated databases, clouds are randomly
distributed within the torus. Each cloud is associated with a
corresponding cloud in the database of directly- or
indirectly-illuminated clouds, the cloud image is placed in the model
space, and images at all spectral channels are computed using the
obscuration along the observer line-of-sight. The simulations are
performed for a torus which is seen edge-on by the observer, as we
would expect in a type 2 AGN, located at the distance of NGC~1068, the
only AGN with published interferometric data in both near- and mid-IR
\citep{2004A&A...425...77W, 2004A&A...418L..39W, 2004Natur.429...47J,
  2009MNRAS.394.1325R} which is close enough so that clumpiness in the
torus can be observed (see the right image of
Fig.~\ref{fig:imageswide} to get a wide view of the simulated AGN and the
top right image of Fig.~\ref{fig:images} to get a zoom on the central
part). For that simulation we did not take into account the incoherent
flux contribution from the surrounding galaxy which will bias the
measurements. We discuss this issue further in Section
\ref{sec:discussion-agn}.

\section{Available imaging facilities}
\label{sec:facilities}


\begin{table}[t]
  \renewcommand{\arraystretch}{1.25} 
  \centering 
  \caption{Imaging interferometers in use or in construction.}
  \label{tab:interferometers}
  \begin{tabular}{lcccc}
    \hline
    Facility &Wavelength &Numbers of &Aperture    &Baseline\\
             &(microns)  &apertures  &(m)         &(m)\\
    \hline\hline
    NOI    &$0.45-8.5$  &6       &0.12           &$2-99(437)$\\
    \hline
    CHARA    &$0.5-2.5$  &6       &1           &$50-330$\\
    \hline
    VLTI     &$1-13$    &4      &8.2           &$40-130$\\
             &$1-13$    &4      &1.8           &$8-200$\\
    \hline
    MROI$^\ast$     &$1-2.5$   &10       &1.4         &$7.5-350$\\
    \hline
    \multicolumn{5}{l}{\footnotesize $^\ast$ in construction}\\
  \end{tabular}
\end{table}

NOI, CHARA, VLTI, and MROI are the present and near-future
interferometers which can perform image synthesis (see
Table~\ref{tab:interferometers}). We used their configurations to
produce realistic simulated reconstructed images for the astrophysical
topics described in Sect.~\ref{sec:astro-simus}. We describe below the
main characteristics of these instruments and their most recent
imaging results.

\subsection{CHARA}
\label{sect:CHARA}

The Center for High-Angular Resolution Astronomy (CHARA) Array is a
6 telescope interferometer operated by Georgia State University on the
Mt.\ Wilson site outside Los Angeles, California
\citep{2005ApJ...628..453T}.  CHARA has the longest baselines of any
current optical interferometer (maximum 330m baseline), allowing
sub-milli-arcsecond angular resolution in the visible and infrared.
While most of the 30 CHARA refereed papers to date present infrared observations using only single baseline data
\citep[e.g.,][]{2005ApJ...628..439M}, we focus here on progress in
{\em imaging} which requires at least 3 telescopes to be used.

Currently, the only combiner at CHARA to have published imaging
results is the Michigan Infrared Combiner
\citep[MIRC,][]{2006SPIE.6268E..55M}.  MIRC now combines
up to 6~CHARA 
telescopes simultaneously and can produce snapshot images of stellar
surfaces and interacting binaries, as long as the surface features are
not too complicated.  \citet{2007Sci...317..342M} presented a detailed
image of the star Altair, the first resolved image of a main-sequence
star, confirming the strong gravity darkening and oblateness expected
for this rapid rotator.  The rapid rotators Alderamin and Rasalhague
have now also been imaged and modeled by CHARA
\citep{2009ApJ...701..209Z}.  Another breakthrough made possible by
imaging was the first ``movie'' of the interacting binary
$\beta$~Lyrae \citep{2008ApJ...684L..95Z}, imaging the pair of stars
through half of the $\sim$13 day orbital period. More recently
impressive images of the well known eclipsing system $\epsilon$
Aurigae revealed the transiting shadow of a flat disk
\citep{2010Natur.464..870K}. 
New imaging targets include the gas disks of Be stars and rotating
spots on active stars.

Imaging efforts at CHARA are rapidly intensifying with the recent
commissioning of the PAVO and VEGA combiners, both visible combiners
capable of closure phase measurements and interferometric imaging. A
6-telescope fringe tracker (CHAMP) has been delivered in Summer 2009
and the 2-telescope CLASSIC combiner has been upgraded to use three
telescopes (CLIMB). MIRC is starting to use all six CHARA telescopes
simultaneously allowing the imaging of planet-forming
disks in young stellar objects for the first time.

\subsection{VLTI}
\label{sect:VLTI}

The Very Large Telescope Interferometer
\citep[VLTI,][]{2008SPIE.7013E..11H} is located at Cerro Paranal,
Chile, and is operated by the European Southern Observatory. The
infrastructure is composed of four relocatable 1.8m Auxiliary
Telescopes (ATs) equipped with visible-wavelength tip/tilt correction,
four fixed 8.2m Unit Telescopes (UTs) equipped with visible-wavelength
adaptive optics, and 6 delay lines of 100\,m stroke. The minimum
baseline is 8m while the maximum one is about 200m. The focal
laboratory is equipped with a $JHK$-band 3-beam combiner AMBER
\citep{2007A&A...464....1P} allowing precise measurements of
closure-phases, and with an $N$-band 2-beam combiner MIDI
\citep{2003SPIE.4838..893L}. The VLTI instruments are opened to the
astronomical community thanks to dedicated observatory support,
leading to a total of more than 191 refereed papers in 8 years
(2004--2011).

The first reconstructed image produced by the VLTI is of the binary
$\theta^1$~Ori~C \citep{2009A&A...497..195K}, using three
configurations of 3 ATs. The spectral coverage of AMBER across the
$K$-band was used to enhance the coverage of the $(u,v)$
plane. However because observations were not close enough in time, the
authors had to artificially ``rotate'' the $(u,v)$ plane between each
configuration to account for the binary rotation.  Almost at the same
time, \citet{2009A&A...496L...1L} reconstructed a spectro-image of the
Mira variable T~Lep at a spectral resolution $R\sim35$ using AMBER and
four 3-AT configurations. Using the MIDI instrument with the 6
possible UT baselines, \citet{2009MNRAS.394.1325R} reconstructed an
image of the torus at the heart of NGC~1068. The image is
point-symmetric because of the lack of phase information. Other
imaging works are in progress concerning B[e] stars
\citep{2009A&A...507..317M}, young stellar objects and supergiants.

ESO is currently commissioning PRIMA \citep{2006SPIE.6268E..27D}, an
instrument which will be able to simultaneously observe two stars
separated by up to an arc minute, using 2 telescopes. Beyond the
astrometric capability which is the primary focus of PRIMA, it will
allow differential phase measurement and possibly an improvement in
absolute phase measurements for imaging. The fringe sensors that come
with it are also expected to be more sensitive. Since October 2011 a
visitor instrument PIONIER allows four VLTI telescopes (ATs or UTs) to
be combined and has provided first images of interacting binaries
\citep{2011A&A...535A..67L, 2011A&A...536A..55B, 2010SPIE.7734E..99B}. Around 2013, the VLTI is expected to
receive the first second-generation general user instruments making
use of 4 telescopes simultaneously. These are Matisse working in the
$L$, $M$, $N$ bands \citep{2008SPIE.7013E..70L} and Gravity with dual
beam operation and an astrometric capability
\citep{2008SPIE.7013E..69E}.

Later, it is planned that the VSI instrument will eventually combine 6
or even 8 telescopes simultaneously, providing realistic snapshot
imaging capability \citep{2008SPIE.7013E..68M}.  In parallel, phase A
studies of a second generation fringe-tracking instrument, capable of
cophasing an array of 4 and potential more telescopes are being
conducted. This instrument will be essential to improve the
sensitivity of second generation science combiners
\citep{Meisner:2010,Blind:2011}.

\subsection{MROI}
\label{sect:MROI}

Located in New Mexico (USA), the Magdalena Ridge Obs\-ervatory
Interferometer \citep[MROI, ][]{2008SPIE.7013E..26C} is a visible and
infrared telescope array under construction. MROI will comprise ten
1.4-m diameter unit telescopes laid out in a Y-shaped array. The
telescopes will be relocated approximately three times per year. Four
scaled array configurations will be available, the most compact
configuration giving \emph{shortest} baselines of 7.8\,m for
comparison with single-dish measurements, and the most extended
configuration having a longest baseline of 340\,m.

MROI has been designed for model-independent imaging of faint targets
with a layout optimized for baseline bootstrapping employing equal
spacings between adjacent telescopes on each array arm, essentially
100\% sky coverage down to 60 deg from zenith, a beam train that
minimizes the number of optical surfaces encountered by the light
beams, and a group-delay fringe tracking capability, with limiting
point-source sensitivity of $H=14$.

First light is expected in 2012, with first fringes in 2013. Phase A
deployment (6 telescopes and the $JHK$ science capability) could be
completed as soon as 2015. Phase B (10 telescopes and optical science)
will begin when more funding for unit telescopes, associated delay
lines and an optical beam combiner is obtained.

\subsection{NOI}

NOI (ex NPOI\footnote{Navy Prototype Optical Interferometer}) has pioneered aperture synthesis in the optical domain
and is still operational today with active plans for future
developments \citep{Armstrong:1998,Hutter:2008}. The array has two
subarrays dedicated to imaging and astrometry. While the imaging array
uses 12\,cm beam-stopped relocatable siderostats it will eventually use
six siderostats of 35\,cm aperture. The astrometric subarray is composed
of six 50\,cm siderostats with shorter baselines (19 to 38\,m). The focal
instrumentation operates in the 450\,nm to 850\,nm band. The current
imaging capability is obtained through the combination of two
``imaging'' siderostats with four of the astrometric ones. NPOI offers
the unique feature of being able to maintain active group delay fringe
tracking for its 6 siderostat operation therefore offering a unique
snapshot imaging capability. The current maximum operational baseline
for imaging is 99\,m but the array will have the capability to span
baselines from 2 to 437 meters. The integration of the former
Keck-interferometer 1.8\,m telescope outriggers to the imaging array is
under study.  In parallel a study to install 1.4\,m composite telescopes
equipped with adaptive optics is underway with the first step being
construction of a prototype. Since 1998 NPOI has been particularly
successful at improving its automation (in particular in the field of
array-phasing), studying binary systems of various types, and
demonstrating a spectral phase referencing capability \citep[e.g.][]{Benson:1997,Zavala:2010,Schmitt:2009}.


\section{Results from image reconstruction simulations}
\label{sec:results}

In this section we first describe the method for generating
realistic interferometric data from the
MROI and VLTI, utilizing the  telescope relocation capability of these facilities. Then we show the
images that were reconstructed by our colleagues without prior knowledge of the
scientific content. Finally we compare the reconstructed images with the original ones and discuss our findings.

\subsection{Simulating realistic interferometry data}

\begin{figure*}
  \centering
 \includegraphics[width=0.95\hsize]{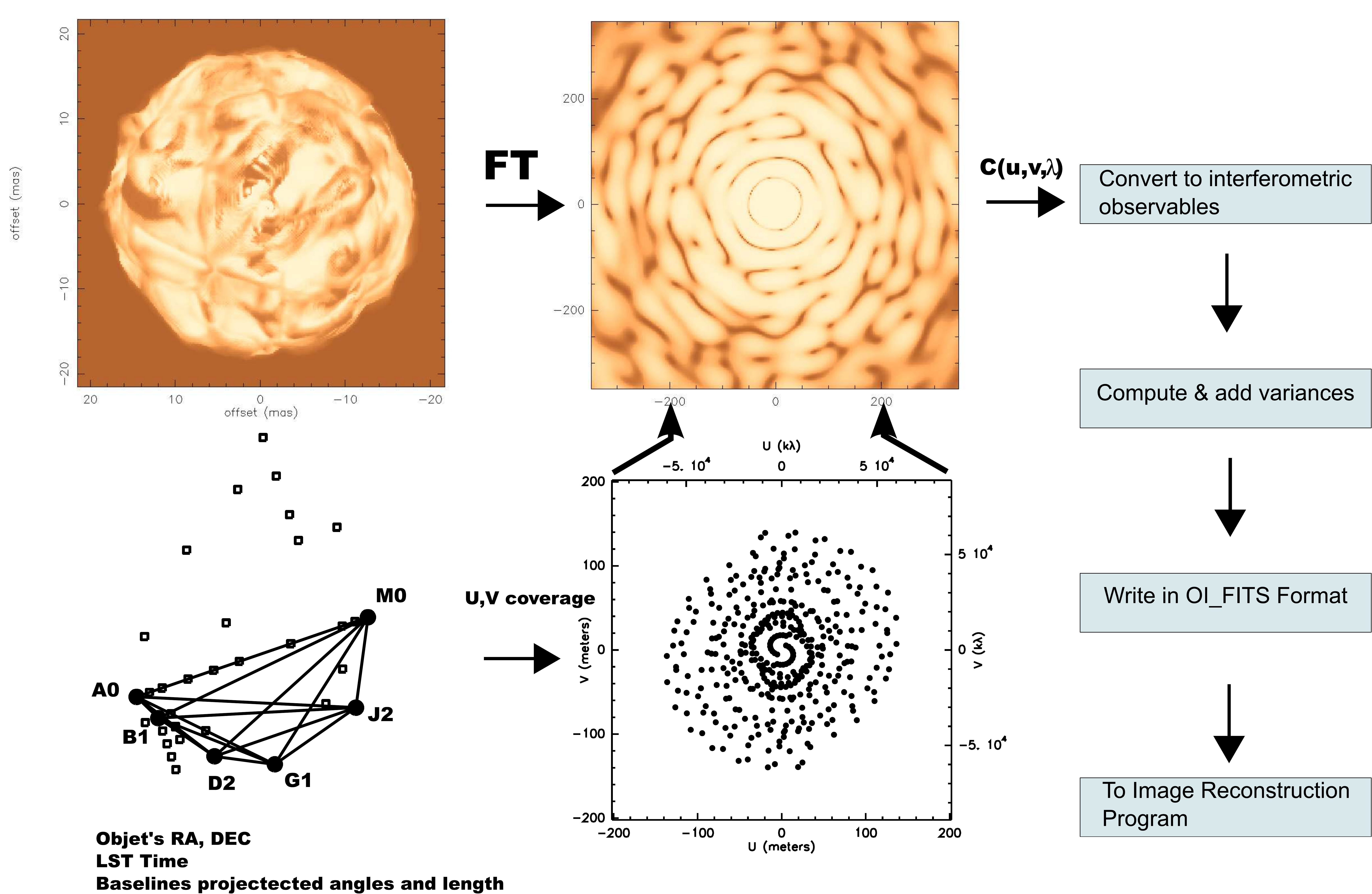}
 \caption{From an image to visibilities: Modeling interferometric
   observables with \texttt{ASPRO}.}
\label{fig:asproflowchart}
\end{figure*}

The simulations used here have been made with the \texttt{ASPRO}
package \citep{2002SPIE.4844..295D}. \texttt{ASPRO} is a
general-purpose observation preparation tool developed by the
\emph{Jean-Marie Mariotti Center}\footnote{The JMMC is a network of
  French laboratories specialized in optical interferometry techniques.}
for optical interferometers, with a particular emphasis on the VLTI
instruments (Sect.~\ref{sect:VLTI}). \texttt{ASPRO} and its little
brother \texttt{ASPRO2} are available
as a web-based utility, with a client-server architecture and a Java
user interface, at \texttt{http://jmmc.fr}, and also as a standalone
program. \texttt{ASPRO} simulates the observation by an interferometer
of an astronomical object, at one or several times, and delivers
simulated interferometric observables in the OIFITS format
\citep{2005PASP..117.1255P}. 

The interferometric chain is modeled as the combination of an
interferometer infrastructure and focal instruments.  The
interferometer infrastructure comprises the telescopes, delay lines,
tip-tilt correctors, adaptive optics, and fringe trackers. It adds geometrical
requirements such as the positions and sizes of the telescopes
apertures at the time of observation with respect to the position of
the science object and the geometrical delays thus induced between each
pair of telescopes. In addition, it includes environmental constraints
such as the atmospheric seeing, the different horizons seen by
telescopes and technical limitations (limits on the delay-line
strokes, flux dependence on the performance of active optical elements...). For each
interferometer simulated in \texttt{ASPRO} these requirements are
known and tabulated; only the atmospheric seeing (and its associated
coherence time) is a variable factor that can be adjusted by the user.

The focal instrument is described by its spectral capability
(bandwidth, resolution, number of spectral channels, central
wavelength, photometric band...) and its sensitivity and throughput.
The former are, for most instruments, fixed as a list of instrumental
``modes''. For example, for AMBER, there are 18 such modes, one giving
simultaneous $J$, $H$, $K$ observations at $35$ resolution, 3 for medium
($1500$) resolution in $H$ or $K$, and 14 for high ($12000$) resolution in
K. The sensitivity and throughput, i.e., the limiting magnitudes, the
visibility losses, the biases and the variance of the values of the
various interferometric observables are computed with a
dedicated noise model (Appendix \ref{sec:noise-model}).

The science object is described by its position on the sky and its
angular and spectral energy distributions. Those can be either a
series of images, one for each wavelength, grouped in a data cube, or a
combination of simple analytic models (binary, uniform disk, etc...).

\subsection{Simulation methodology}
\label{sec:simulation_methodology}

\begin{table}[p]
\centering
\begin{tabular}{cc}
\hline
Array & Configuration\\
\hline
MROI & N2 W2 W4 S2\\
&N4 W2 W4 S4\\
&N2 N4 S2 S4\\
\hline
&N4 W4 W6 S4\\
&N6 W4 W6 S6\\
&N4 N6 S4 S6\\
\hline
\hline
VLTI & A0 B1 D2 G1 J2 M0\\
&A1 B3 D2 G1 H0 I1\\
& UT1-UT2-UT3-UT4\\
\hline
\end{tabular}
\caption{Array configurations for both MROI and VLTI used in the simulations}
\label{tab:configs}
\end{table}

\begin{figure*}[p]
  \centering
  \begin{tabular}{cc}
\includegraphics[height=0.30\textwidth]{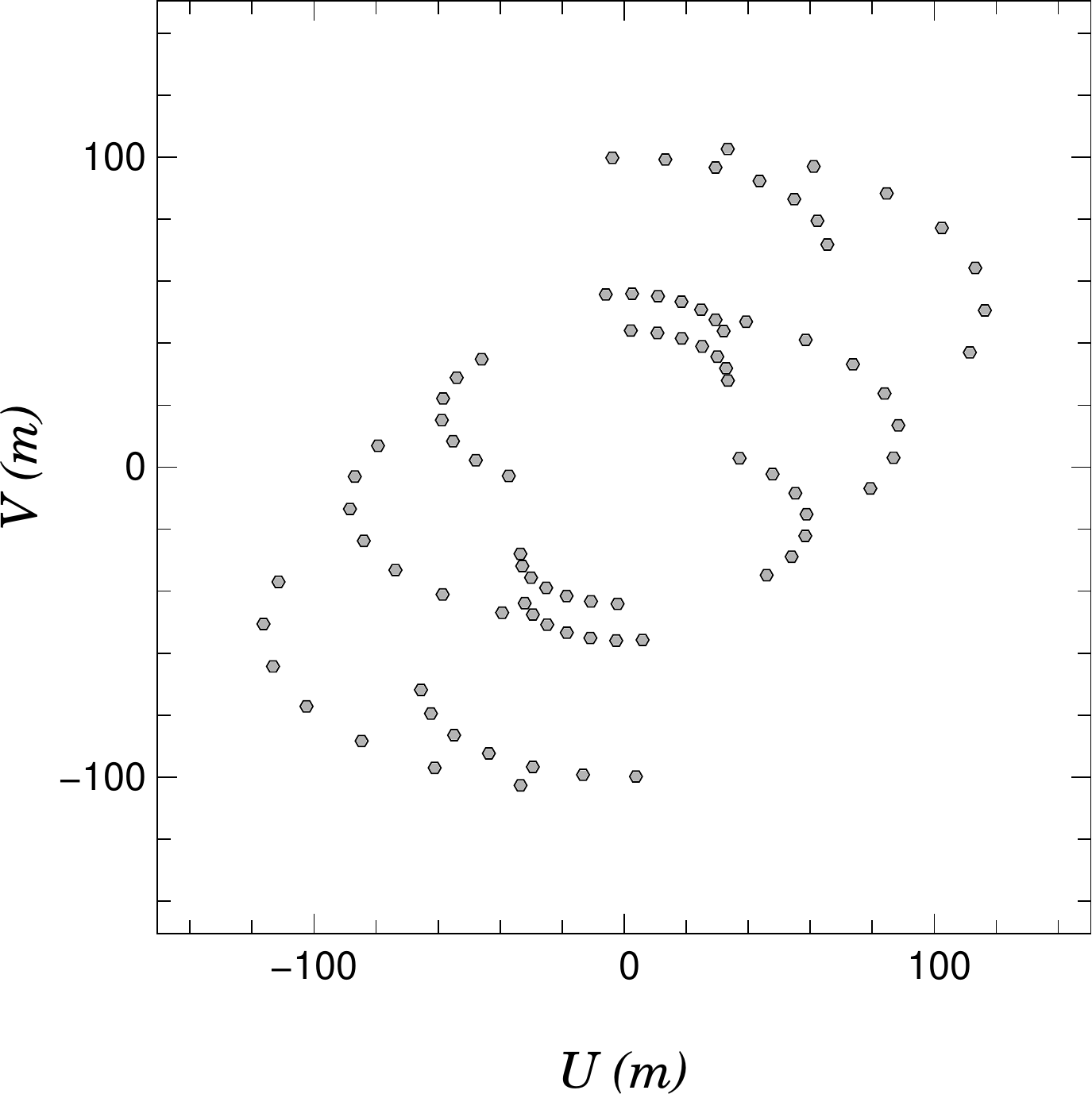} &
\includegraphics[height=0.30\textwidth]{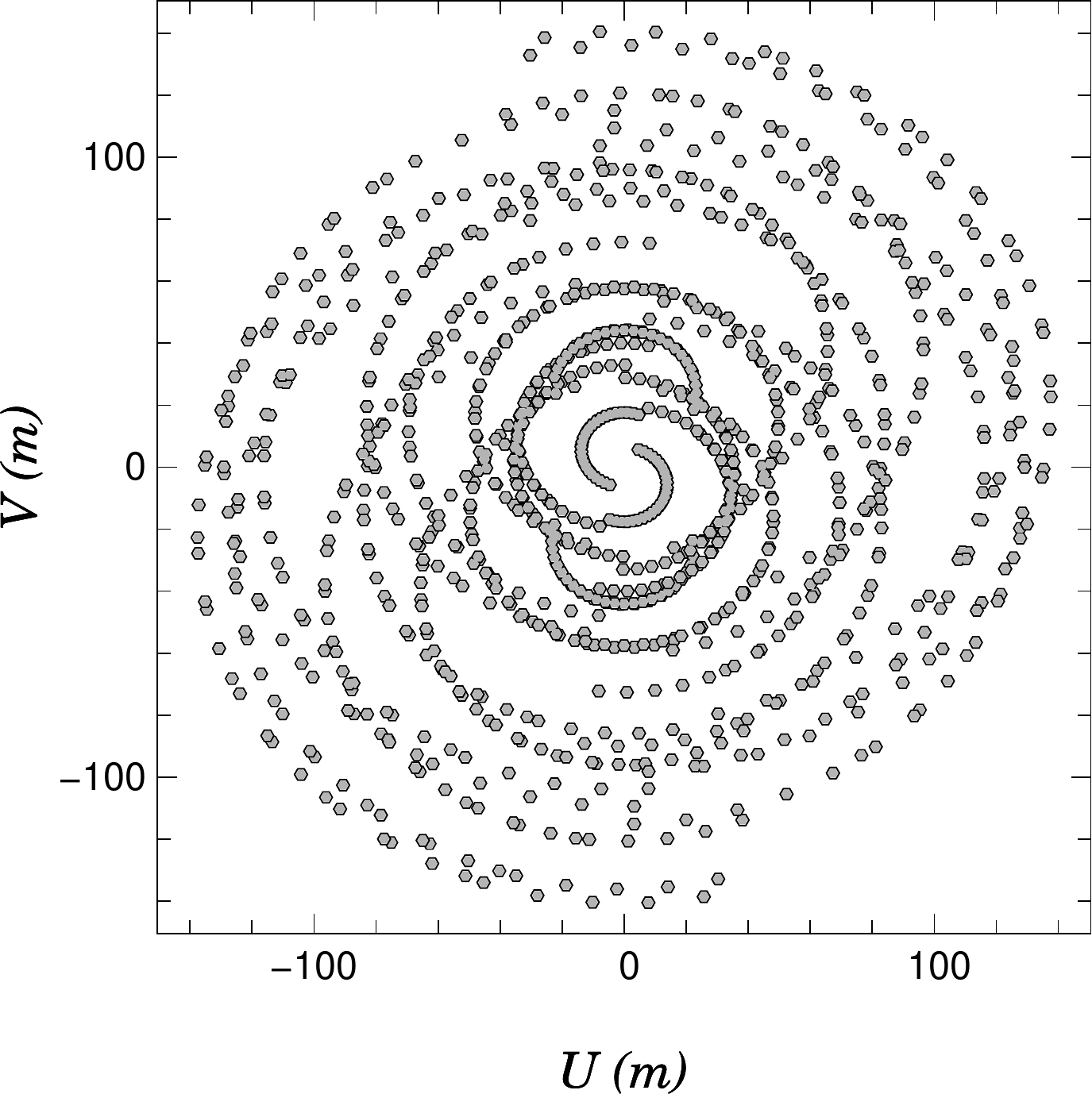} \\
\includegraphics[height=0.30\textwidth]{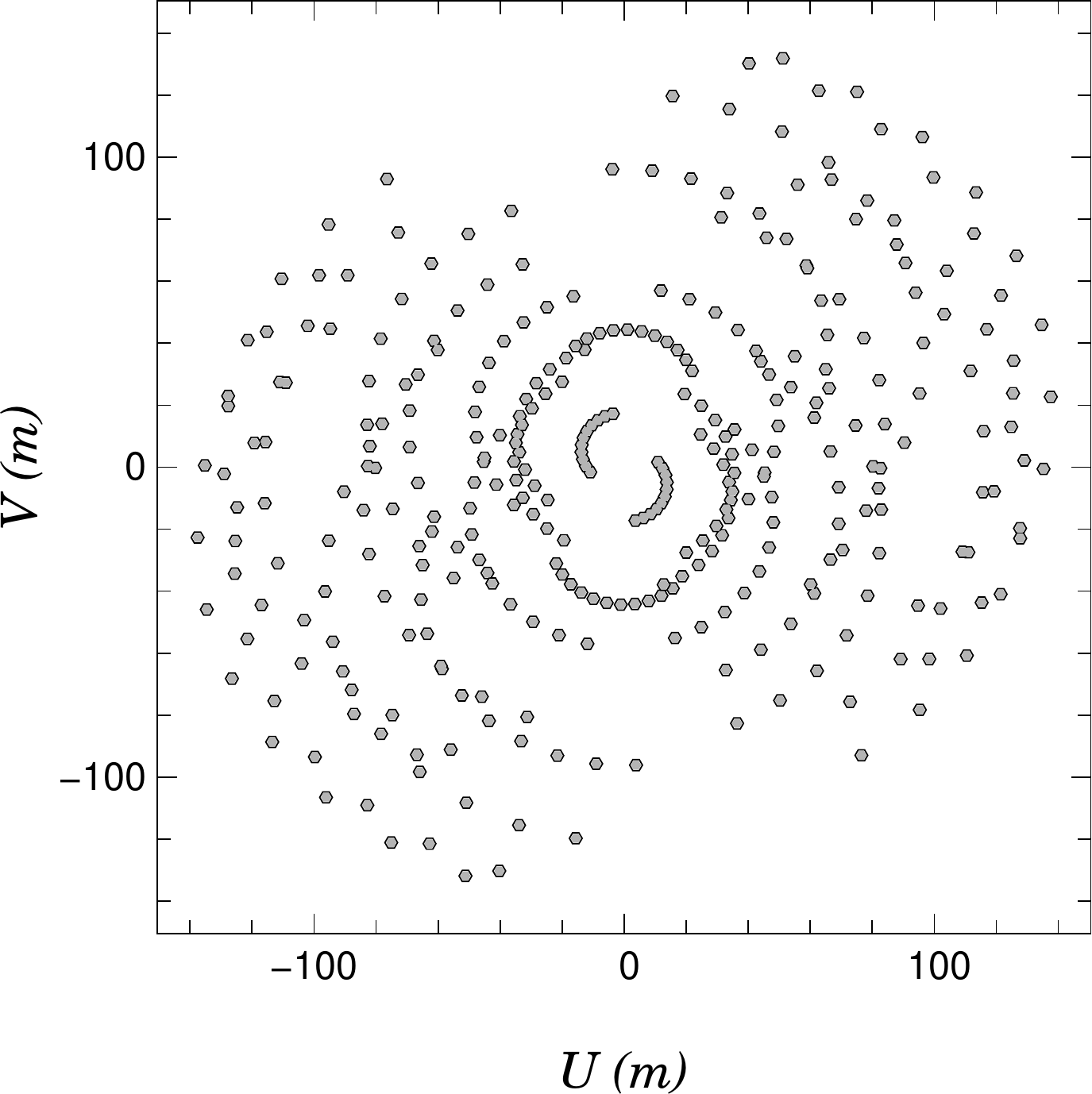} &
\includegraphics[height=0.30\textwidth]{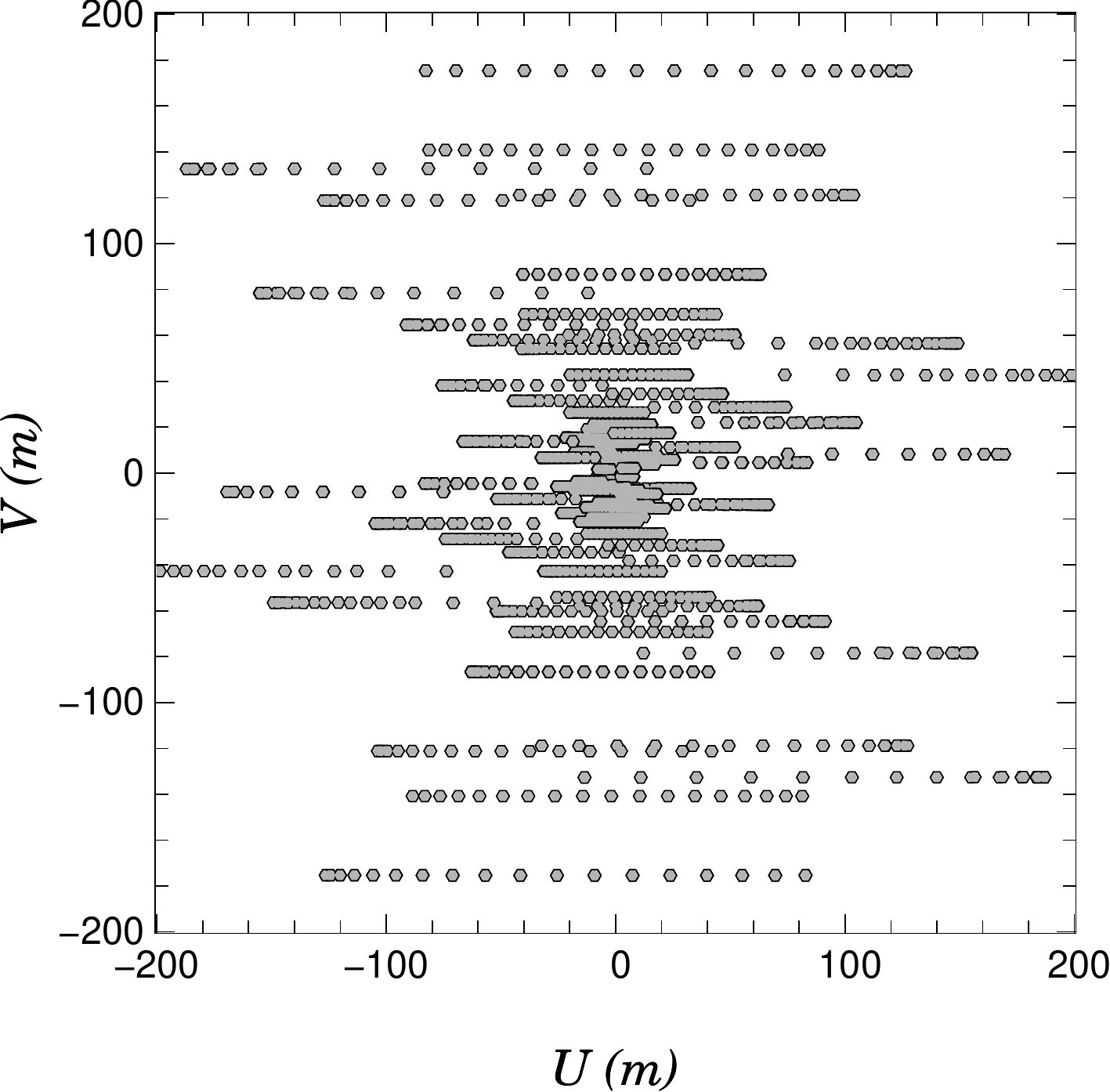}\\
\end{tabular}
\caption{The $(u,v)$ plane coverage used for the image
  reconstructions. Top from left to right: Matisse/VLTI-UT coverage
  for the protoplanetary disk model; VSI/VLTI-AT coverage for the
  late-type supergiant star model for Betelgeuse; VSI/VLTI-AT coverage
  for the molecular layer model around a late type star; I4/MROI
  coverage for the dusty tori model for the active galactic nuclei  NGC~1068.}
  \label{fig:uv}
\end{figure*}

\begin{table}[p]
  \caption{Description of the parameters of the different science cases}
  \label{tab:science_cases}
  \begin{center}
    \begin{tabular}{lllll}
      \hline
      Science cases &Interf. &Instrum. &Wavel. &Algorithm\\
      \hline\hline
      Protoplanetary disks & VLTI-UT & Matisse & $N$ band & BSMEM \\
      Stellar surface & VLTI-UT & VSI & $K$ band & MIRA \\
      Molecular layers & VLTI-UT & VSI & $K$ band & MIRA\\
      AGN & MROI & I4 & $K$ band & BSMEM \\
      \hline
  \end{tabular}
  \vspace*{-2em}
\end{center}
\end{table}
\begin{table}[p]
  \centering
  \caption{Number of squared visibilities ($V^2$), closure phases (CP)
    and their corresponding average errors over the $N_{V^2},N_{CP^2}$
    measurements. To simulate a potential calibration limit in the minimum visibility and
    closure phase that could be measured we introduced a threshold of
    0.002 in visibility and 0.1 degree in closure phase.}
  \label{tab:uv+vis}
 \begin{tabular}[c]{lcccc}
\hline
& Disk & Supergiant & Evolved star & AGN \\
\hline
\hline 
$N_{V^2}$ &42&375&165&972\\
$\sigma_{V^2}$&0.02&0.005&0.01&0.02\\
$N_{\rm CP}$ &21 &250&110&486\\
$\sigma_{\rm CP}$ &$2^\circ$&$0.2^\circ$&$0.3^\circ$&$1^\circ$\\
\hline
 \end{tabular}
\end{table} 

The simulation of observations, depicted in
Fig.~\ref{fig:asproflowchart}, consists of: 
\begin{enumerate}
\item building the list of $(u,v)$ positions of the baselines at each
  wavelength of the required focal instrument mode and at the
  various times (for example, an observation every 20 minutes for 6
  hours around the meridian) intended by the observer;
\item computing the Fourier Transform (FT) of the
  flux spatial distribution of the object at each wavelength at each
  $(u,v)$ point defined above (in the present case, where the user provides a model of the
  object as a cube of images);
\item from the complex visibilities obtained, computing all the
  interferometric observables, V$^2$, amplitude and phase of the
  bispectrum, and, if the instrument has spectral capabilities, the
  amplitude and phase of the differential visibility;
\item for each of these interferometric observables, applying the
  noise model (Appendix \ref{sec:noise-model}) and a lower-limit
  accuracy threshold to get the expected
  variance of each value; 
\item randomize the interferometric
  observables according to their
  expected variance;
\item export these simulated observations in the OIFITS format. 
\end{enumerate}

The $(u,v)$ coverage for both MROI and VLTI was derived from the
allowed 6 telescope configurations (Table \ref{tab:configs} and
Fig. \ref{fig:uv} ). In the case of MROI this was obtained by
switching rapidly 6 beams in front of a four beam combiner. This
explains why three$\times$four telescope combinations are used to get all 15 baselines.

Undoubtedly, because we do not take into account a detailed model of
fringe tracking performances (outside the scope of such a paper), we
set ourselves on the optimistic side in terms of calibration
biases. We can therefore consider that the images obtained here are
the best one can expect from the interferometers and that actual data
may contain additional artifacts.

By simulating the expected results of the actual observation and
providing OIFITS files, \texttt{ASPRO} allows assessment of not only the
feasibility of the observations but also the science potential of the
planned observations. The OIFITS file obtained can be used to check
whether image reconstruction techniques retrieve the desired
information and to compare the results to what a model-fitting software can
achieve. The interferometric observables 
can also be compared to the model observables in terms of chi-square
distance as a first estimate on the feasibility of the planned
observation. Table \ref{tab:uv+vis} contains a summary of the
computed errors on the visibility and closure phase for the different
input images.

\subsection{Results}

\begin{figure*}[p]
  \centering
  \begin{tabular}{cc} 
   \includegraphics[height=0.42\textwidth]{wii09-giantK} &
    \includegraphics[height=0.42\textwidth]{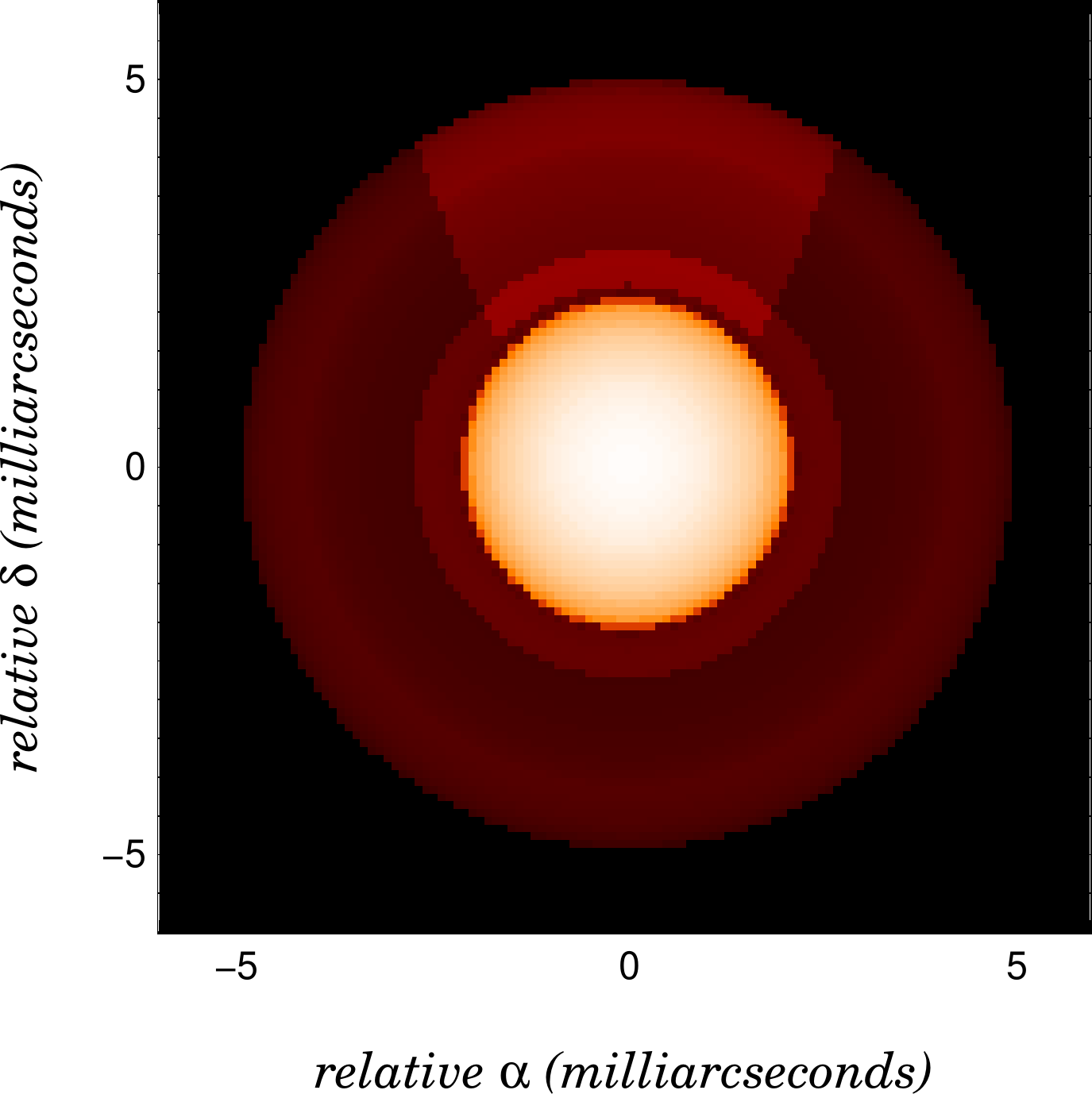}\\
    \includegraphics[height=0.42\textwidth]{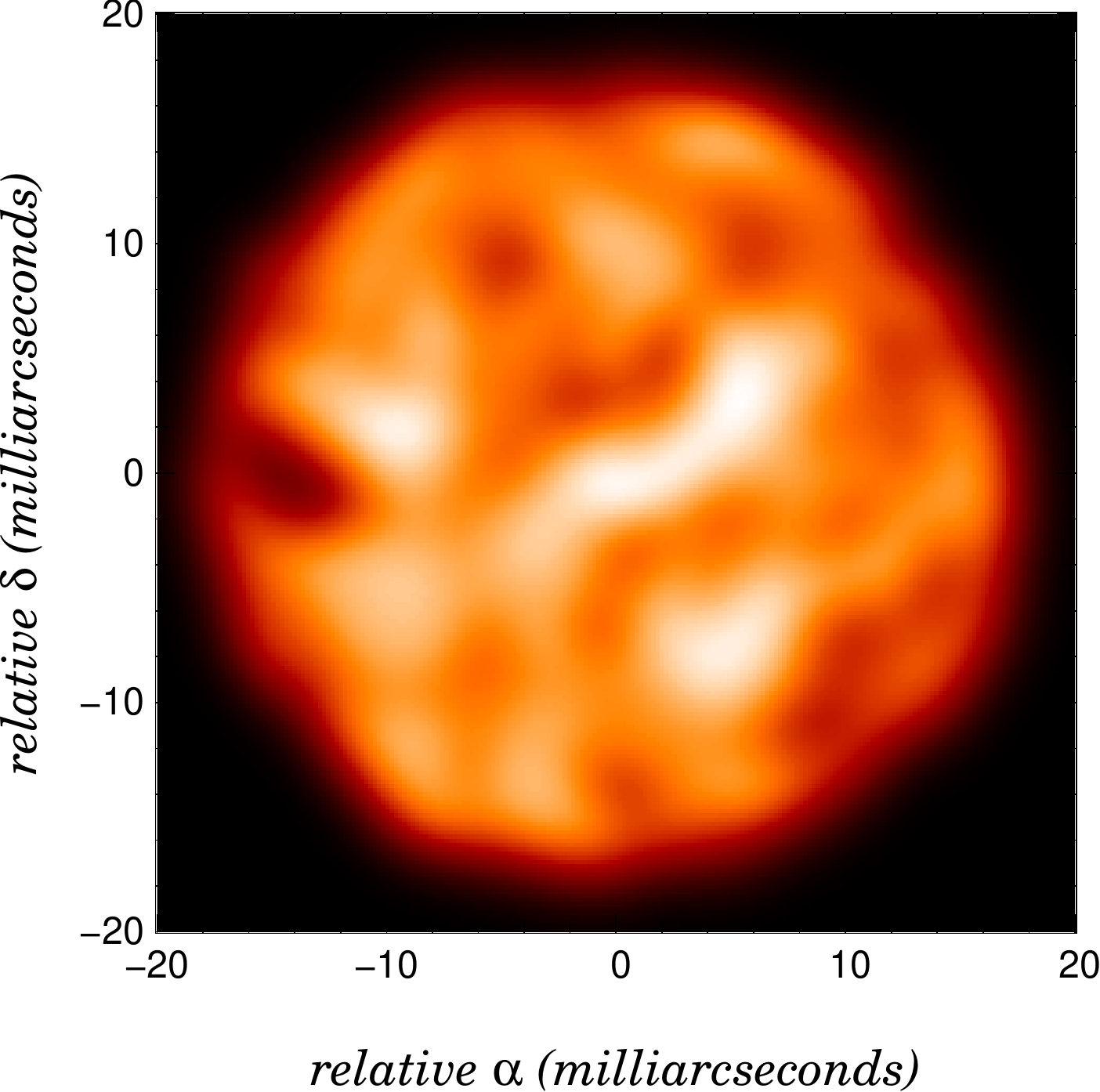} &
    \includegraphics[height=0.42\textwidth]{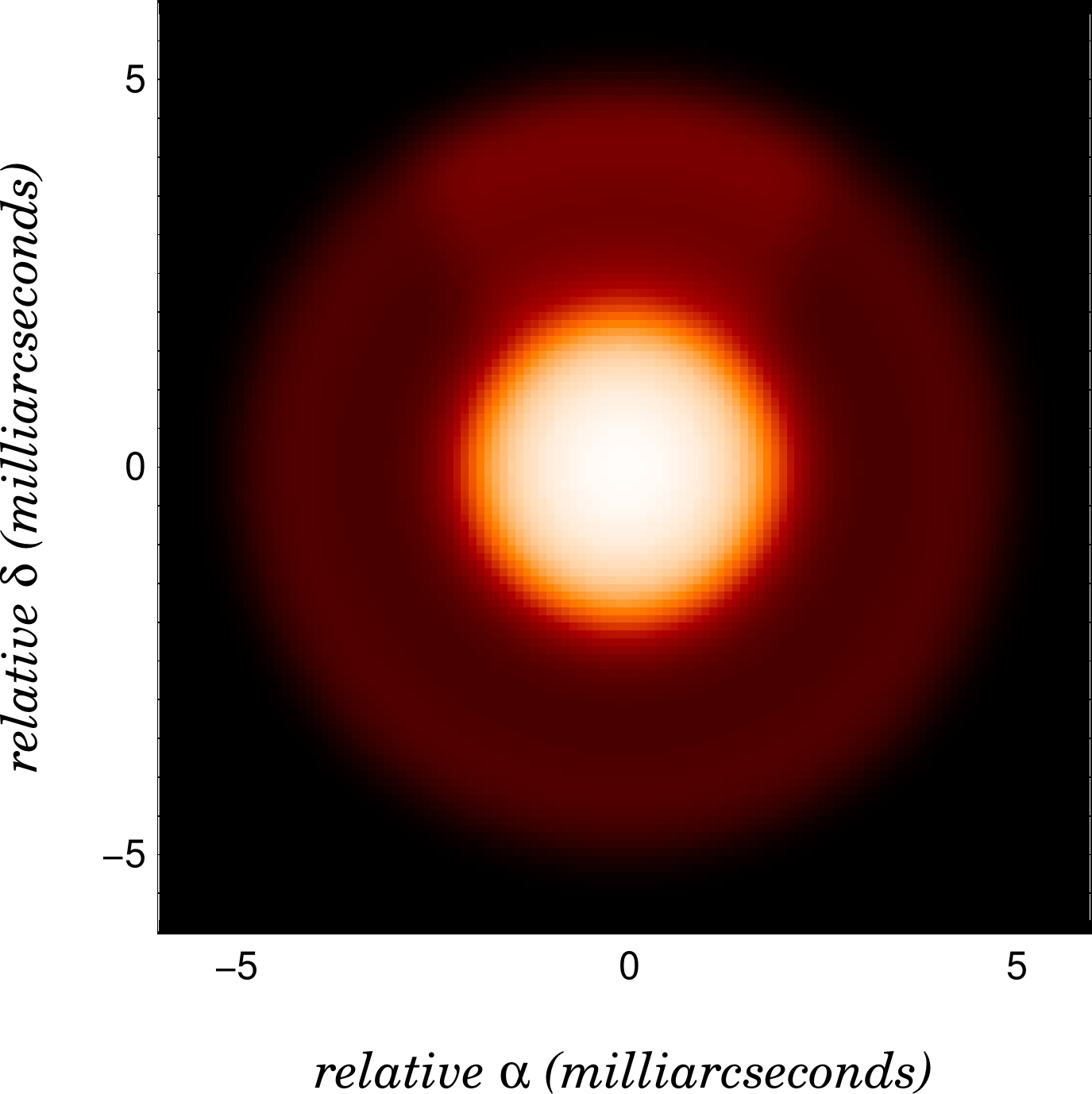}\\
   \includegraphics[height=0.42\textwidth]{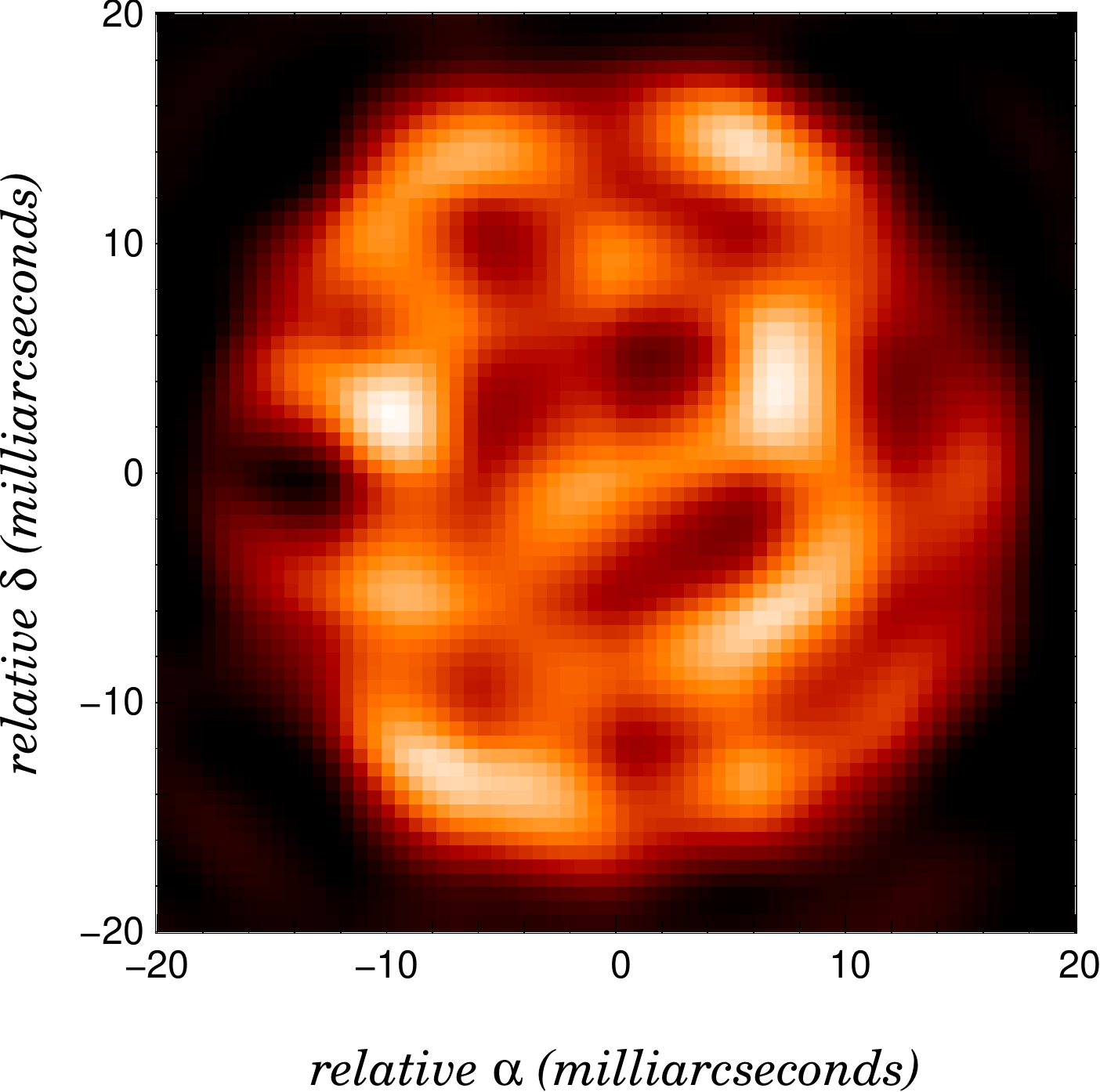} &
    \includegraphics[height=0.42\textwidth]{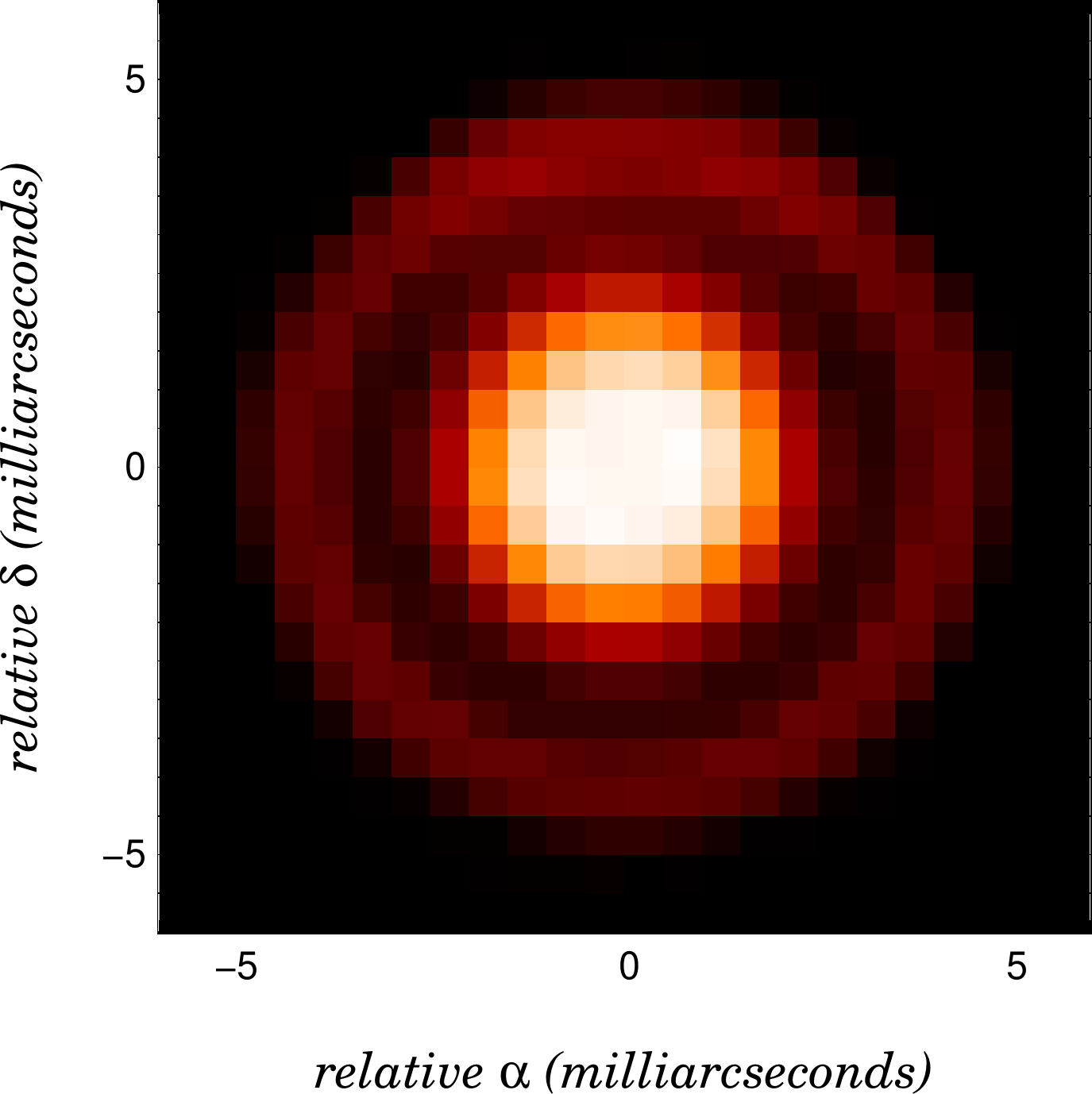}\\
\end{tabular}
\caption{Resulting images from the image reconstruction (bottom row)
  of the supergiant (left) and evolved star (right) cases compared to
  the initial images resulting from our modeling (top row) and
  compared also to the image that would be obtained by a single
  telescope whose diameter is the maximum interferometer resolution
  (central row).  The astrophysical cases are on the left column a
  supergiant star and on the right column a molecular layer around
  late type star. The color codes correspond to logarithmic scales and
  the image have been zoomed-in with respect to figure
  \ref{fig:imageswide} to concentrate on the meaningful reconstructed
  details.}
 \label{fig:images}
\end{figure*}

\begin{figure*}[p]
  \centering
  \begin{tabular}{c} 
  \includegraphics[height=0.42\textwidth]{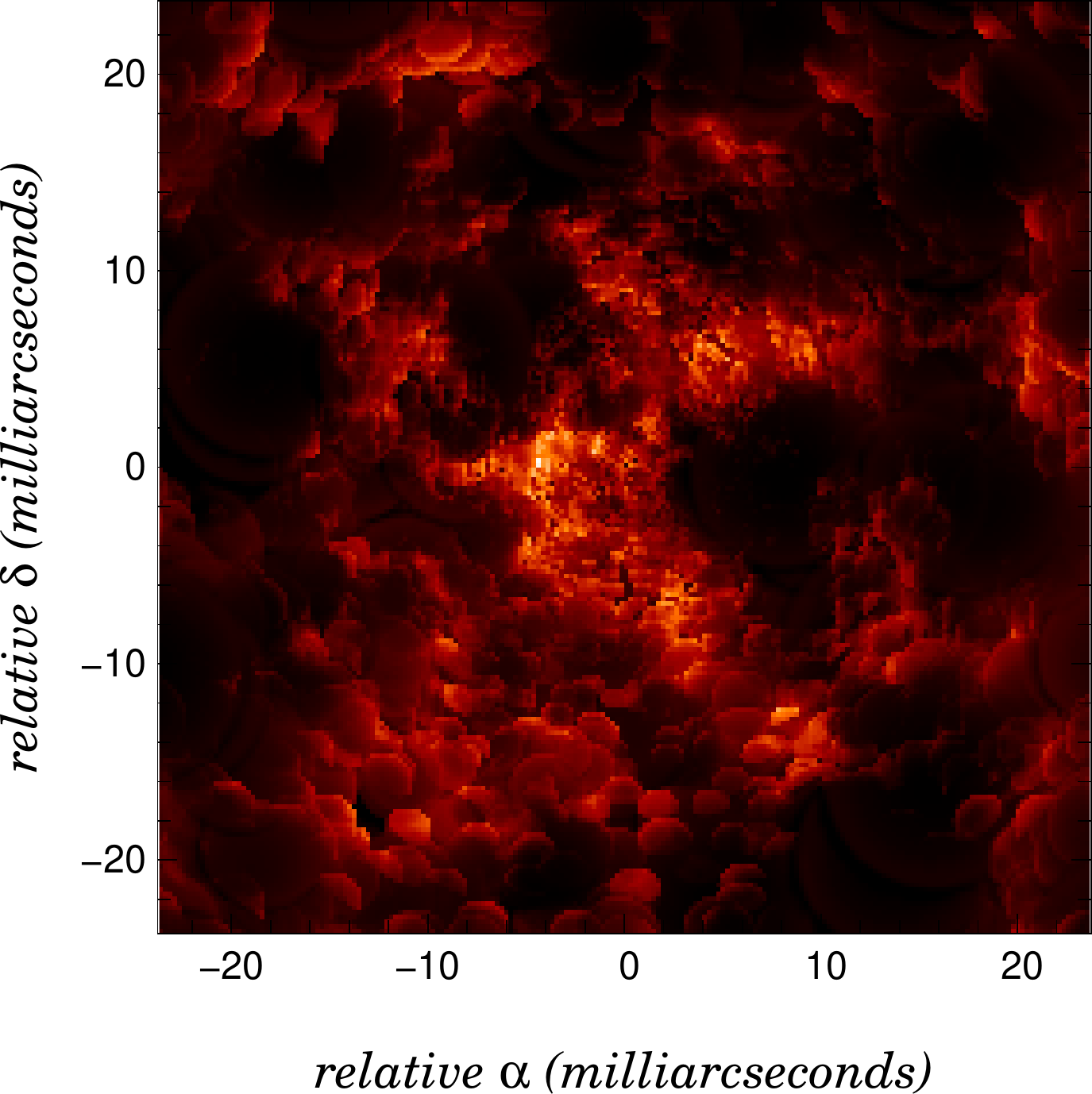} \\
  \includegraphics[height=0.42\textwidth]{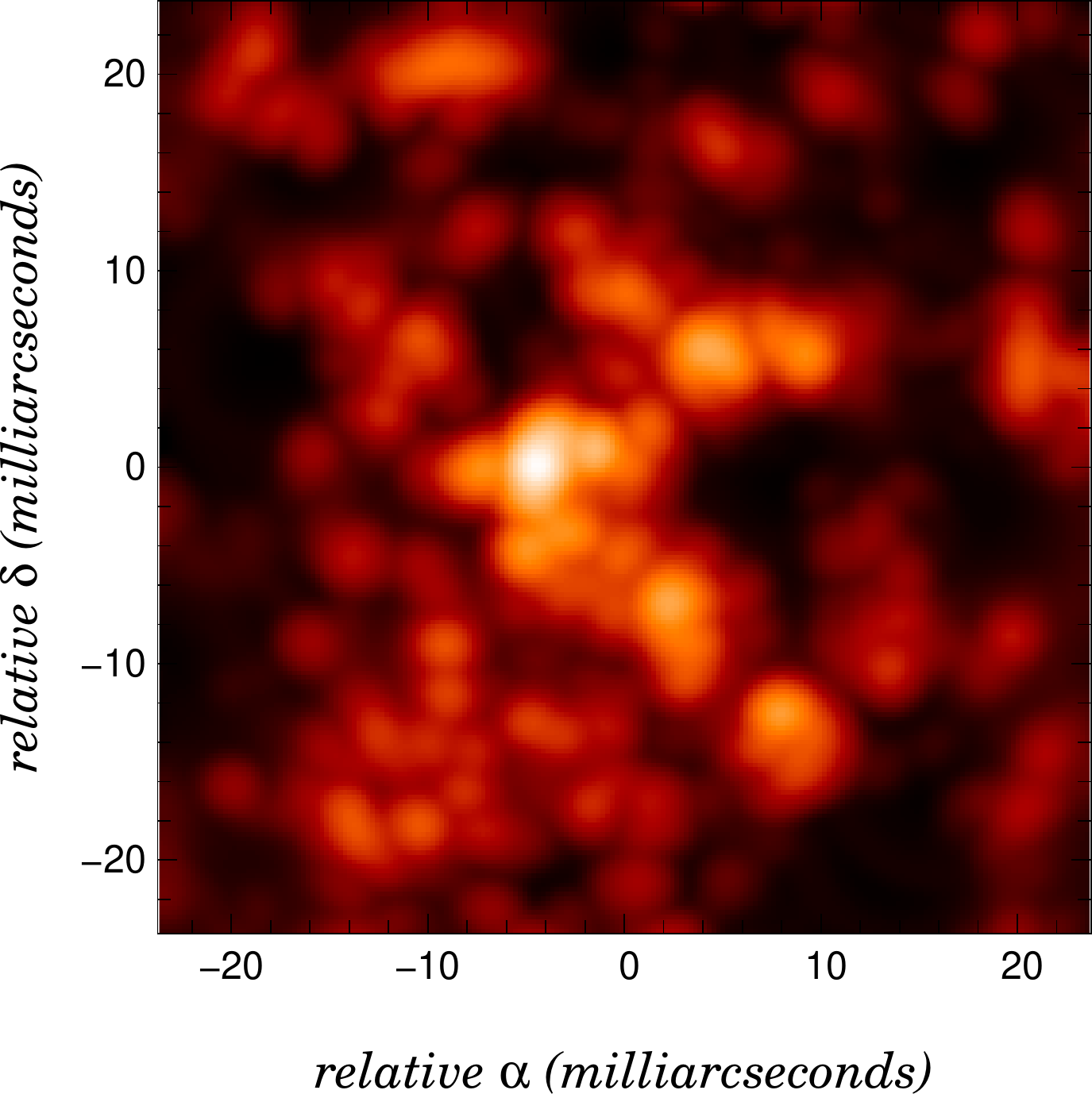} \\
  \includegraphics[height=0.42\textwidth]{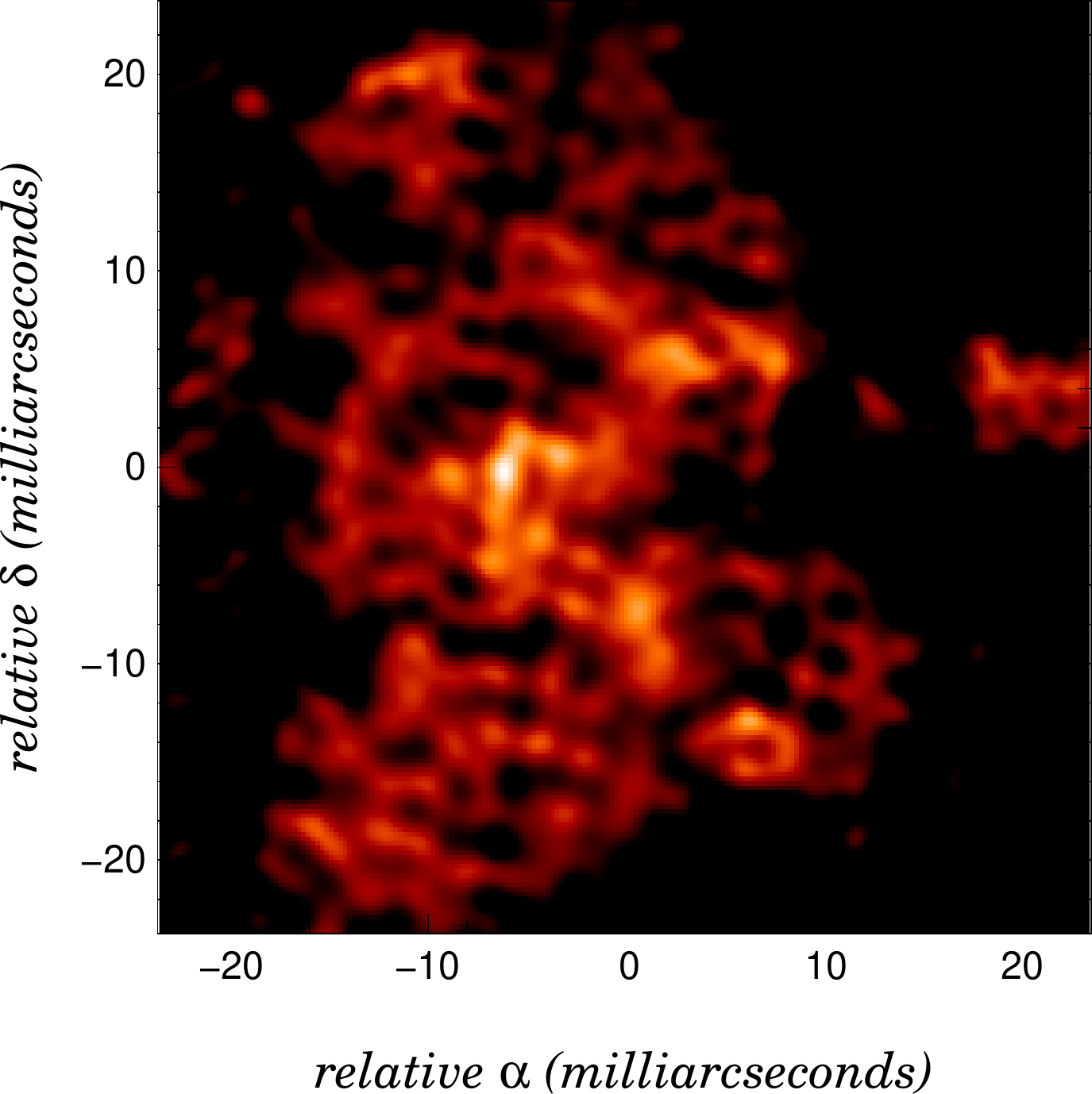}\\
 \end{tabular}
 \caption{Resulting images from the image reconstruction (bottom row)
   of the AGN images compared to the initial images
   resulting from our modeling (top row) and compared also to the
   image that would be obtained by a single telescope whose diameter
   is the maximum interferometer resolution (central row).  The
   astrophysical object is here a dusty tori in active galactic
   nuclei. The color codes correspond to logarithmic scales.}
 \label{fig:images2}
\end{figure*}

We have performed a series of simulations following the astrophysical
prescriptions given in Sect.~\ref{sec:astro-simus} and using the
different instrument configurations described above. We focus our attention on the four cases
listed in Table~\ref{tab:science_cases}, since we chose not to explore
all the possibilities, which would have led to more than several tens
of results. We have selected results obtained with the instruments
VSI/VLTI and I4/MROI in the $K$ band and Matisse/VLTI in the $N$ band.
The  $(u,v)$ coverage for each configuration is displayed in
Fig.~\ref{fig:uv}. The configurations were chosen visually to fill the
$(u,v)$ plane as uniformly as possible and therefore ensure the best
spatial frequency sampling. For the MATISSE case a fixed UT configuration
was used. For the VSI reconstruction two 6 AT telescope configurations
were chosen with the supergiant one pushed to the maximum possible
hour angle range (e.g -5h,5h). The MROI coverage corresponded to one of the rare AGN with
sufficient angular extension to permit imaging (NGC 1068).

The \emph{Planet signatures in protoplanetary disks} case (hereafter
called the protoplanetary disk case) was the only one for which we
were not able to retrieve a viable reconstructed image. The conclusion
applies both to the embedded-planets image and the disk+gap
images. The explanation for this lack of success is that the high contrast
between the planet induced gap or shadows and the disk/photosphere
emission clearly exceeded the dynamical range achievable with such a
technique. None of the cases computed in Sect.~\ref{sec:disks-simul}
was successful in imaging features beyond the central star, using
BSMEM or MACIM.  However, the example chosen here is voluntarily an
extreme case of imaging. Long-baseline interferometry has already begun to explore imaging of protoplanetary environments
\citep[e.g.][]{Benisty:2011} and it is expected that MATISSE will
have the ideal combination of sensitivity and angular resolution to
reveal the inner morphology of dust emitting inner disks.  Possible
improvements to the method would be to introduce a multi-wavelength
approach. The location of the gap being independent of wavelength
continuum observations at various bands might provide additional hints
to constrain the gap positioning. We note that recent gaps in
transition disks were indirectly revealed by visibility modeling
\citep[][]{Benisty:2010,Olofsson:2011}.

\begin{figure*}[p]
  \centering
  \begin{tabular}{cc} 
    \includegraphics[height=0.65\textwidth]{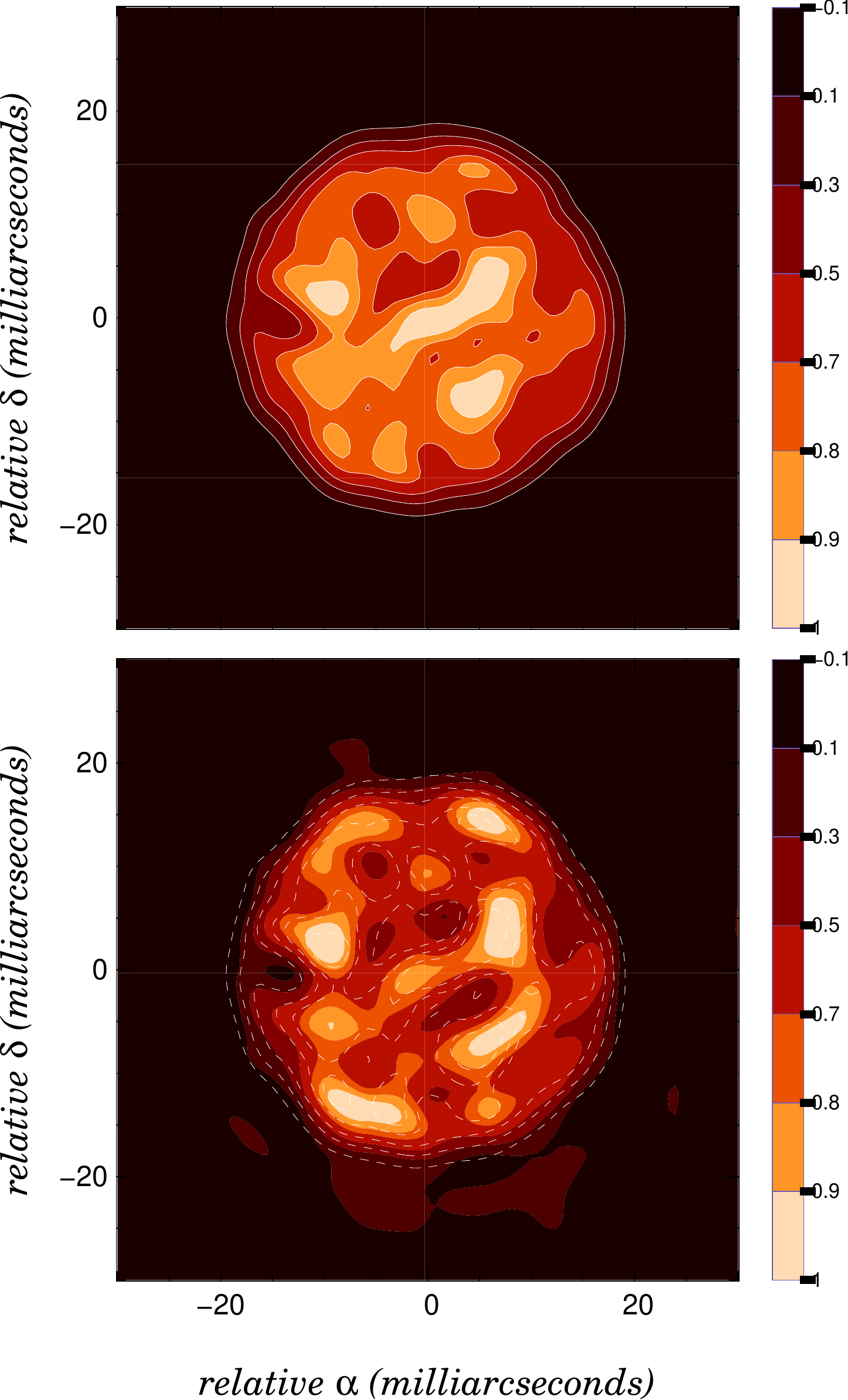} &
    \includegraphics[height=0.65\textwidth]{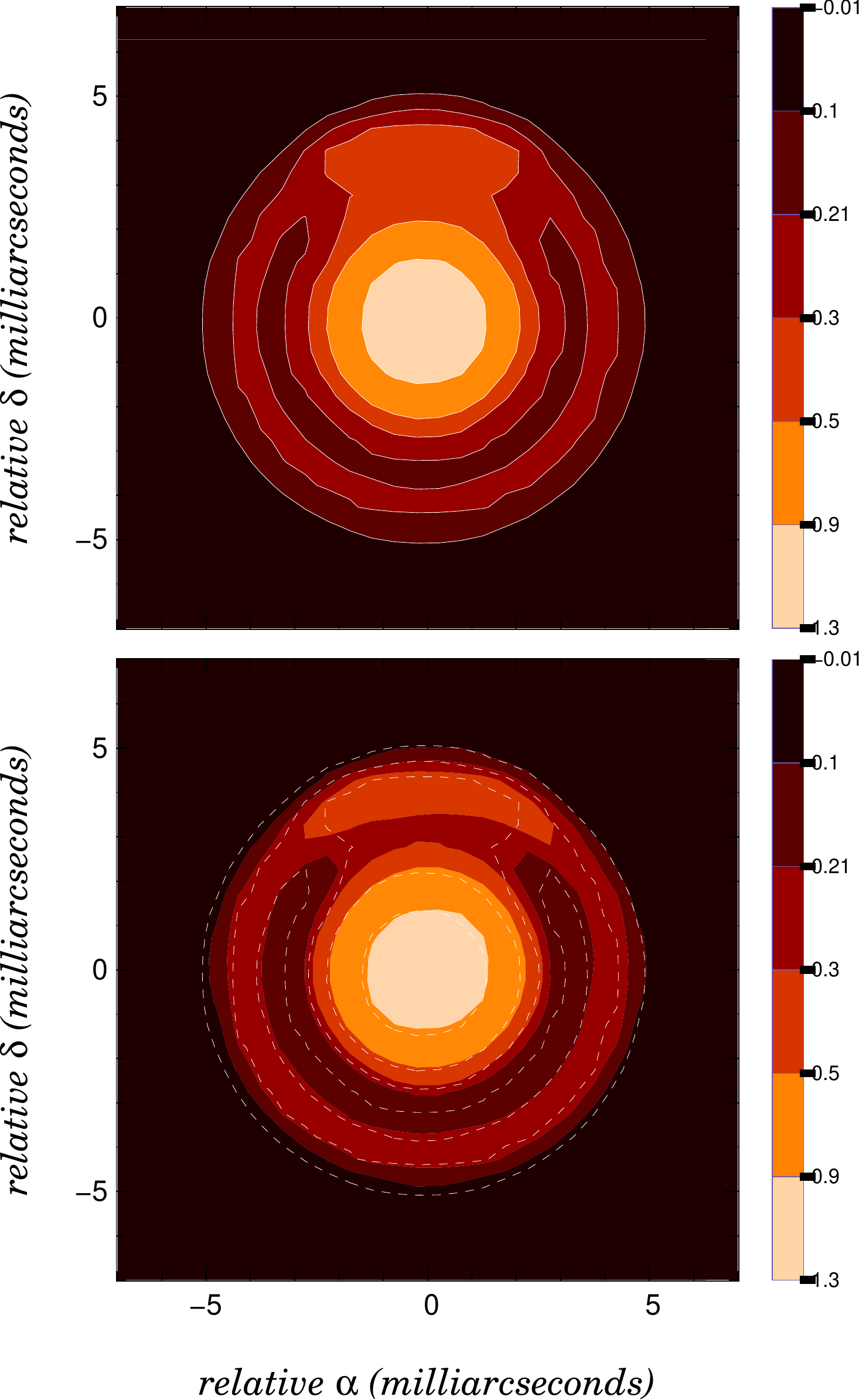}
\end{tabular}
\caption{Contour plots of both original (top row) and reconstructed
  image (bottom row) for the supergiant case (left) and evolved star
  with plume (right). Contour levels are expressed in fraction of
  maximum intensity. Contours of bottom row correspond to the model
  image for better comparison.Levels:  0.9,0.8,0.7,0.5,0.3,0.1); evolved star with plume (levels: 0.9,0.5,0.3,0.21,0.1).}
  \label{fig:contours}

\end{figure*}

\begin{figure*}[p]
  \centering
  \begin{tabular}{c} 
   \includegraphics[height=0.65\textwidth]{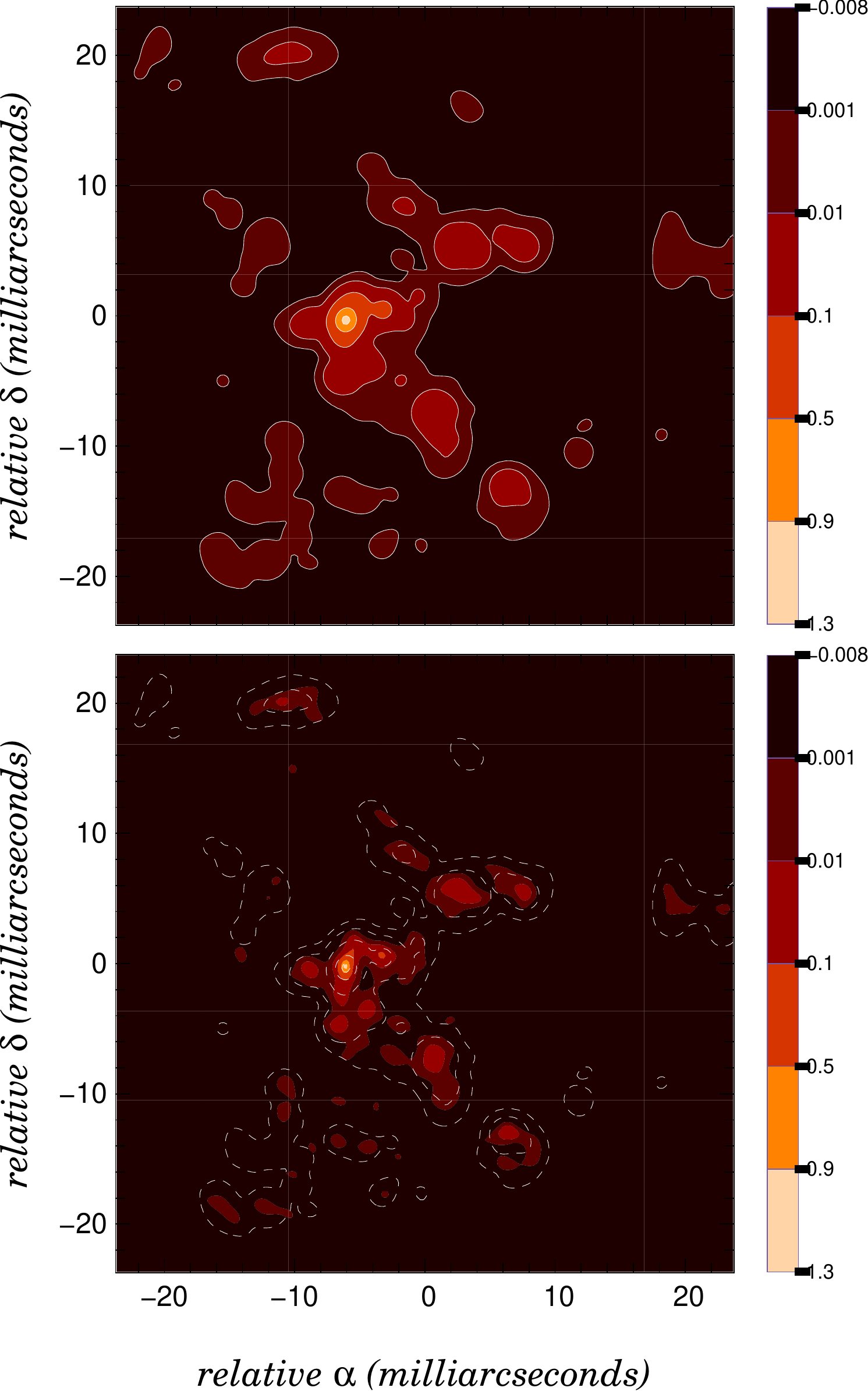}\\
\end{tabular}
\caption{Contour plots of both original (top row) and reconstructed
  image (bottom row) in the AGN science case. Contour levels are expressed in fraction of
  maximum intensity. Contours of bottom row correspond to the model
  image for better comparison.Levels: 0.9,0.5,0.1,0.01,0.001).}
  \label{fig:contours2}
\end{figure*}

For the remaining cases the results are summarized in
Fig.~\ref{fig:images} and Fig.~\ref{fig:images2} with the simulated images (top row), the
simulated images convolved by a Gaussian beam at the maximum
interferometer angular resolution (central row), and finally the reconstructed images
(bottom row). Figures \ref{fig:contours} and \ref{fig:contours2} show a contour comparison
between model and reconstructed image. These figures allow us to carry out the first
qualitative comparison between the reconstructed and original images
at the proper resolution. For the \emph{Molecular layer around
  late-type stars} case (hereafter called late-type envelope case) and
the \emph{Dusty torus around AGN} case (hereafter called the AGN
case), note that the images have been magnified in order to display the central
part of the images. In these cases, the image reconstruction was not
able to detect the less luminous structures present in a wider field
(Fig.~\ref{fig:imageswide}).

A first visual inspection shows that, 
\begin{enumerate}
\item for all cases, all the features with flux greater or equal to
  approximately $1/20$ of the peak intensity in the image are roughly
  correctly positioned and have the correct shape and size;
\item there is evidence that the image reconstruction process brings
  slightly better resolution than the fringe spacing $\lambda/B$ which
  is due to the super resolution allowed by the regularization process and
  the interferometer response function (this result is best
  illustrated in the AGN case in Fig.~\ref{fig:images2});
\item in the supergiant case, the most prominent "granular"
  atmospheric features are retrieved but the dark thin intergranular
  lanes of the underlying large-cell convective structure, visible in the model image, are not;
\item in the late-type envelope case the basis of the plume on top center of the image is marginally detected;
\item in the AGN case, the complexity of the scene is obvious and
  the emission morphology is globally well-reproduced in the
  reconstructed image. However as we will see later the 
quantitative interpretation of such an image requires caution.
\end{enumerate} 

A proper quantitative comparison cannot rely solely on this visual
inspection.  For the discussion below, we used one possible image
reconstruction quality estimator, called fidelity, described in
Appendix \ref{sec:fidelity} and inspired from ALMA work. There 
are also valid metrics like the variance of the difference between the two
images, reconstructed and original smoothed at the interferometer
resolution \citep{Renard:2011} or the figure of merit proposed by
\citet{2008SPIE.7013E..48C} but there is no single metric with  proven superiority over all other methods.


\section{Discussion}
\label{sec:discussion}

Except for the protoplanetary disk case, optical interferometric
imaging is quite successful in retrieving images similar to the
\emph{true} images. In this section, we discuss how new and useful
astrophysical information can be extracted from interferometric
imaging, the similarities and differences with
radio-interferometry synthesis imaging, and how to further improve image
reconstruction.   We conclude with a discussion of how to improve the
current facilities to obtain the best information possible using aperture synthesis imaging.

\subsection{Getting new and useful astrophysical information}

\subsubsection{Planetary signatures in protoplanetary disks}
\label{sec:discussion-disks}

High-resolution imaging of protoplanetary disks has the potential to
reveal where and when planets form as can be seen on the left image of
Fig.~\ref{fig:imageswide}. Growing cores of giant planets
gravitationally and tidally perturb the protoplanetary disks in which
they form to create indicators of their presence on scales much larger
than the planets themselves, including both localized dimples and
azimuthally symmetric gaps. Shadows and brightenings in scattered
light images from these perturbations will reveal their presence.

The difficulty in detecting these perturbations using optical
interferometry is that the stellar and disk brightness overwhelms the
emission from the features in the disk. At millimeter wavelengths, the
stellar emission is minimal, so the dynamic range is less of an issue
although these wavelengths probe a different regime of the disk as
well: the disk becomes optically thin and the images represent
vertically integrated thermal emission from the disk.  Because of the
nature of shadowing and illumination at the surface of the disk, the
contrast of the shadows and brightenings is higher at visible and
infrared wavelengths as compared to radio wavelengths.

However, these objects can still bring some interesting
surprises. In the past, it was believed that disks around young
stars would be only marginally resolved by long-baseline optical
interferometers in the infrared, but in fact the structures which have
been measured are indeed located at larger scales than expected
\citep[see review by][]{2007prpl.conf..539M}. There might be several
other signatures of planets and their formation in the potential
planet forming region of circumstellar disks (inner $0.1"$ in nearby
star-forming regions): overdense vortices \citep{2002ApJ...578L..79W};
gaps \citep{2002ApJ...566L..97W, 2004ApJ...612.1152V}; accreting giant
planets \citep{2005ApJ...619.1114W, 2006SPIE.6268E.112H,
  2007P&SS...55..569W} and other prominent features. Indeed
\citet{2005A&A...437..627M} and \citet{2006ApJ...645L..77M} have
reported oscillations in the visibilities that could trace the
presence of off-centered hot spots in the disk without identifying its
nature. If confirmed, imaging of these relatively bright spots should
be relatively easy since the dynamical range quoted in these works ranges
between 15 and 50. The failure of our tests reveals that the image reconstructions
are limited to dynamic ranges significantly smaller than $10^3$. The
reader is referred to \cite{Wolf:2012} for a detailed review of the
potential of long-baseline interferometers for the study of
circumstellar disks and planets.

\subsubsection{Granulation in late-type stars due to large convective cells}
\label{sec:discussion-supergiants}

The image we present here was computed in the $K$ band.
\cite{2009A&A...506.1351C} show that in this spectral region the
continuum-forming layers are more visible and the granulation pattern
is characterized by large-scale granules of about 400-500 $R_{\odot}$
\citep[$\approx$60$\%$ of the stellar radius in Fig.~4
of][]{2009A&A...506.1351C}. On top of these cells, there is small
scale structure with characteristic size of 50-100\,R$_{\odot}$,
$\approx$10-15\% of the stellar radius. This pattern is clearly
visible also in the top left panel of Fig.~\ref{fig:images}.

There are three main characteristics of the granulation pattern that
we may try to recover from image
reconstruction:
\begin{itemize}
\item the size of the granulation cells from the large scale granule
  to the smaller scale structures;
\item the intensity contrast;
\item the timescale of the convective motions. 
\end{itemize}
Images of stellar surfaces at various wavelengths will allow these
three questions to be answered, and will therefore be a very valuable tool to
(in)validate our simulations.  This is crucial in order to understand
the dynamics of the atmosphere of red supergiant stars, and its
importance in, e.g., the still unknown process at the origin of
mass-loss \citep{2007A&A...469..671J}, or in order to determine accurate
abundances from spectra using proper model atmospheres.

We compared the original image, smoothed to the interferometer
resolution (top left panel of Fig.~\ref{fig:images}), to the
reconstructed image (bottom left panel). The reconstructed image shows
medium sized granules, similar to the original smoothed image. There
is, however, much more information in the original image, and key
features like the large central granule are very hard, if not
impossible to see in the smoothed and reconstructed image. This is due
to the fact that the emission coming from the inter-granular lanes is too
weak compared to the surrounding regions.

We also compared the intensity contrast of the smoothed and
reconstructed images by using normalized intensities. For this purpose
we normalized the intensity to the peak intensity. The top images of
Figures~\ref{fig:contours} and \ref{fig:contours2} show for each
science case a contour display of the initial images, the bottom
ones correspond to contour displays of the reconstructed images.
Comparison can be quantified by measuring the fidelity of the image (Appendix
\ref{sec:fidelity}). We can see that the fidelity of the
reconstructed pixels with respect to the original ones can be fairly
high ($\geq 10$) on a significant fraction of pixels (Table
\ref{tab:fidelity}) but this is unrelated to the interesting stellar
surface structures.

Several issues need to be discussed:
\begin{itemize}
\item In the K band, it is possible to recover small-to-medium scale granules, while the information about large convective
  cells is lost. While near-infrared wavelengths have been used in this study, one could
consider  the interest of studying Red Supergiants in the
visible. In the visible,  the angular resolution is higher (e.g almost 4
  times between the V band than in the K band) and the intensity
  contrast of granulation is higher. At shorter wavelengths, corresponding to strong
  molecular absorption, e.g. TiO bands, synthetic
  intensity maps show more details and much larger contrast between bright
and dark areas than in the K band, and larger inter-granular lanes. 
While pioneering imaging results were carried out using visibile
instruments \citep[][]{Baldwin:1996,Benson:1997,2000MNRAS.315..635Y}
near infrared wavelengths have been the dominant source of
observations because of their lowered sensitivity to atmospheric
conditions.  However recent developments at CHARA with the VEGA
instrument have re-opened the way to such projects \citep[e.g.][]{Mourard:2009}.
\item Test image reconstructions should be made with interferometers 
containing more telescopes, or larger baselines, in order to investigate 
the best affordable way to recover the granulation properties outlined
above. As is suggested from comparing simulated images in the visible and
near-IR, the right choice of the wavelength may well be a key, 
despite technical difficulties growing at shorter wavelength.
\item Simultaneously to the measurement of the granulation size at
  different wavelengths, the temporal evolution of the convective
  pattern should also be compared to the simulations. Extracting typical 
convection turnover times from series of images should be possible
for the medium scales, as soon as the image reconstruction can be trusted,
which seems to be the case, even with a small number of telescopes
(Fig.~\ref{fig:images}).
\end{itemize}

\subsubsection{Molecular layers around late-type stars}
\label{sec:discussion-mollayers}

On a qualitative level, the image reconstruction was successful at
detecting the outer molecular shell as a ring-like structure. The base
of the plume is also visible, and the wavelength-dependent size of the
central object corresponds to the effect of the inner molecular layer.
Pixel fidelity computed in Appendix \ref{sec:fidelity} shows indeed
that the stellar-surface pixels are retrieved with a good fidelity (on
average 10, while the surrounding layer is less well determined
(average fidelity $\approx5$).  Our simulation did not include any
clumpiness in the circumstellar environment, but the present results
indicate that isolated and/or large enough clumps with a contrast of
$I_\mathrm{clump}/I_\mathrm{background} \ge 0.1$ would be detectable. However, the
fully resolved extended dust structure could not be detected with the
reconstruction algorithms, and the reconstructed limb darkening of the
central object is much stronger than that in the input image, making
diameter estimates for the central star unreliable and necessitating
rather detailed visibility model fitting. Although this has
not been investigated at length, we think this is due to the lack of
short baselines. Using an image of the star obtained at the
diffraction limit of a large telescope when performing the image
reconstruction would probably solve this issue like single dish
observation in radio astronomy. The latter problem is partly due to
an insufficient angular resolution in this particular simulation:
\cite{2008A&A...485..561L} show that the limb darkening can be
reconstructed if the data sample the 2nd and 3rd lobe of the
visibility curve, and a limb-darkened prior is chosen for the image
reconstruction. Given the highly variable nature of many evolved
stars, it is important that real observations come as close as
possible to snapshot images.

An evolved star and its circumstellar environments span a huge range
in sizes, from convective cells on the surface to the dusty shell at
hundreds of stellar radii. Although these different components lead to
comparable total flux contributions at IR wavelengths, the difference
in size of the radiating surfaces immediately indicates a huge range
in intensities. In fact, in our simulated image, the contrast between
central star and inner dust shell intensity is $I_{\star}/I_\mathrm{dust
  shell}\sim 10^{5}$, which is clearly beyond what can be retrieved
from the reconstructed image.

\subsubsection{Dusty torus in an active galactic nucleus}
\label{sec:discussion-agn}

One of the most interesting tasks for interferometric imaging of AGN
dust tori is the characterization of the substructure. At the moment,
the actual size of the dust clouds and their properties is closely
constrained by SED fitting (Sect.~\ref{sect:AGN}). However,
\citet{2004A&A...418L..39W} reported on $K$-band long-baseline
VLTI/VINCI interferometry of NGC~1068 where they found that $\sim$40\%
of the emission at 46\,m baseline is not coming from the larger-scale
emission region but confined to scales $\leq$5\,mas. Without knowledge
of the real distribution of the small-scale flux, this can be
interpreted either as emission from the dust-sublimation region or
accretion disk, or emission from dust clumps (or ``clumping of
clumps'') seen through the obscured screen of other clouds
\citep{2008A&A...485...33H}.

The reconstructed model images of the $K$-band emission of NGC~1068
show several small emission regions in the central region of the
torus. They reflect the parts of the emission from the inner torus
region seen through holes in the distribution of clouds at larger
radii. The total emission contained in these regions and their sizes
are consistent with the constraints from VINCI on NGC~1068. In this
way, interferometric imaging allows us to directly detect substructure
in the torus and even give an estimate of its contribution to the
overall flux of the torus. Interestingly, the analysis of our model images
shows that the holes in the obscuring screen of more distant clouds
have similar sizes as the typical cloud sizes. If that holds for other
model parameters as well, measuring the sizes of the small-scale
emission regions might even provide constraints for the typical size
of dust clouds in the torus.  

When trying to evaluate the fidelity of the pixel intensities with
respect to the model images (Appendix \ref{sec:fidelity}), the
fidelity is generally low and while the global morphology of the
emitting regions is nicely restored, the intensity of the emitting
regions is largely deviating from the original one.  This was expected
as this image was by far the most complex of our set.  Improving
intensity accuracy is essential in order to exploit color
maps (e.g $H-K$, $K-L$ or $N-H$) to bring constraints on dust
obscuration and the temperature distribution.

An additional issue that was not taken into account is the probable
contribution of incoherent flux by the host galaxy. As recognized by
\citet{Kishimoto:2011}, particular care will have to be taken in
calibrating this flux through  SED fitting and, when possible,
through single dish measurements such as those provided by aperture
masking \citep[e.g.][]{Monnier:2004}.

\subsection{Similarities and differences with mm-wavelength
  synthesis imaging} 
\label{sec:comparison-radio}

While early VLBI, with its poorly calibrated phases, confronted 
difficulties very similar to optical aperture synthesis,  we thought it
would be interesting to compare optical long baseline interferometry
with the millimeter one. mm-interferometry development
offers an interesting operational parallel to what could be the future of optical
interferometry.

Interferometry and aperture synthesis in the radio domain has been
routinely achieved since the 1970s.  The last 20 years have witnessed maturation in the 
millimeter domain as well, from the construction of large
facilities such as the IRAM Plateau de Bure interferometer \citep[six
15-m antennas,][]{1992A&A...262..624G} to the Atacama Large
Millimeter/submillimeter Array (ALMA, sixty-four 12-m antennas). Compared to
the infrared domain, two major differences make millimeter
interferometry easier to implement: (1) the coherence time of the
atmosphere is longer, making it possible to reference the phase of the
visibility to avoid relying on phase closure only, and (2) the millimeter
receivers are directly sensitive to the incoming electric field, which
is then down-converted to a frequency that can be processed with
electronic equipments; as a consequence, the delay between two
antennas can be corrected for after detection of each signal (before
any interference) and the signal can be amplified to allow to combine
many baselines simultaneously. Unlike in optical interferometry the
phase calibration can be done by observing, non simultaneously, a
point-source calibrator (within a much larger isoplanatic patch)
within one coherence time. Therefore the complex visibilities,
including both amplitudes and phases, can be {\em directly}
computed. Images are then obtained by Fourier transforming the
visibilities, and a deconvolution is later applied to correct for the
artifacts introduced by the $(u,v)$ plane sampling. Numerous examples
of scientific results obtained with millimeter or submillimeter
interferometry can be found in the literature (e.g. the special
edition of Astronomy and Astrophysics, volume 468, 2007).

To measure the source visibility on long (larger than a few hundreds
meters) baselines, special techniques must be deployed in order to
correct for the atmospheric phase errors. These have to be tracked on
timescales of $\sim$1 second and water vapor radiometers have proved
to yield very efficient corrections by estimating the atmospheric
contribution to the phase error (water content). This cophasing
process is much more painful in the optical domain where
fringe-trackers must actually form the fringes in order to maintain
the global wavefront as planar as possible \citep[see
e.g][]{Blind:2011}. In order to obtain the best possible images, a
special attention has then to be paid to the $(u,v)$ plane coverage
that is obtained in such millimeter observations. Indeed, the poor
sampling of the $(u,v)$ plane has been identified as one of the major
limitations of the imaging process: the point-spread-function (PSF) of
the synthesized instrument (the so-called {\em dirty beam}) is the
Fourier Transform of the $(u,v)$ coverage and may thus include strong
sidelobes if the $(u,v)$ sampling is poor, making the astronomical
interpretation of the resulting images quite difficult. Although this
can be partly corrected by the deconvolution process (CLEAN being the
most widely algorithm used), obtaining directly a high-quality image
is a central goal of the current instruments design and
operations. The Plateau de Bure interferometer configurations have
been optimized for that purpose \citep{2006ApJS..164..552K}. The
positions of the 64 ALMA antennas have also been optimized (ALMA memo
\#400), the large number of antennas making it possible to achieve
almost perfect beam shapes. In such optimization, the goal is to
obtain a Gaussian distribution in the $(u,v)$ plane (which means
higher weights on the short baselines), which results in a Gaussian
PSF. The behavior of a single-dish telescope with Gaussian beam taper
is thus reproduced as accurately as possible.

\subsection{Improving image reconstruction}
\label{sec:improving-reconstruction}

\subsubsection{What is a good image reconstruction?}
\label{sec:good-reconstruction}


In this work, we wondered whether a reconstruction is a good
representation of the model. Although visual inspection can do a good
job, it is difficult to go further. 
It is useful to compare
different results from different algorithms like in the Beauty
Contests \citep{2004SPIE.5491..886L, 2006SPIE.6268E..59L,
  2008SPIE.7013E..48C}, but not really useful when assessing the
quality of the image reconstruction by itself. Developing tools to
compare relative positions or relative fluxes of the image features
appears to be quite important but it is beyond the scope of this work.
A systematic testing of image
reconstruction ``fidelity'', such as the one mentioned in Appendix
\ref{sec:fidelity} is an essential step to build the trust in image
reconstruction. The reader is referred to \citet{Renard:2011} for a
first attempt to test systematically the quality of image reconstruction.

The answer to the question of whether model fitting should be performed
at the same time is certainly positive since this is the
only way to deduce the quantitative parameters of the models. However,
image reconstruction helps to recognize the general structure of the
image which can be quite complex as illustrated in the supergiant and
AGN cases. The modeler is then guided in his work by the general
architecture revealed by the reconstructed image.


\subsubsection{Improving current facilities}
\label{sec:current-facilities}

For this work, we used expected features of 
interferometers and instruments either in existence or under active development. 
In these cases, an ideal observing sequence is often impossible due to technical
limits: finite range of delay lines, telescope shadowing, bad telescope tracking near zenith, etc.
Indeed, to optimally use VLTI and CHARA, we must identify the current limitations of instruments at VLTI
and CHARA.

VLTI. Some general remarks concerning image reconstruction at VLTI can be
drawn: {\it (i)} so far simple objects are the most commonly imaged:
binaries, simple circumstellar environments, or images with small
field-of-view (perhaps 5x5 pixels at maximum); {\it (ii)} the main results
have been obtained in visitor-mode with at least three configurations
of 3 ATs; {\it (iii)} a sampling speed of the $(u,v)$ plane at
$\approx25$min per calibrated observation has been shown to be sufficient; {\it
  (iv)} the poor quality of the calibrated visibilities is an
important limitation leading to poor dynamic range; {\it (v)} the
impossibility to track on resolved fringes ($<10$\%) with the VLTI
fringe tracking instrument FINITO is a
limitation for imaging stellar surfaces; {\it
  (vi)} the observations are limited in hour angle because
of the delay line stroke reducing the possible $(u,v)$ coverage; {\it
  (vii)} finally all existing data sets suffer so far from an important hole
in the $(u,v)$ plane due to the current limitations in baseline configuration.
In the coming years, the VLTI will improve imaging by prioritizing improved
data quality as well as an increase of the
number of offered interferometric baselines. Background work is
on-going, focusing on  several open questions (such as the
trade-off between maximum baseline length, $(u,v)$ plane density and
sky-coverage). 

CHARA. The fixed locations of the CHARA telescopes mean that the
array can only image objects of a certain size range. For
near-infrared, the ideal range for imaging is between 1.5 and 3.5
milli-arcseconds, since this allows all the 15 baselines of CHARA to
make a useful contribution to the image reconstruction. If an object
is smaller than this, then there is not enough resolution; if the
object is bigger, then most of the short baselines are sampling beyond the
first null and this means too much information is missing to reliably
invert the Fourier data. One way to improve imaging on CHARA
substantially would be to add a telescope to the center of the
array\footnote{This addition is feasible since there is space on the
  mountain top for the telescope and space in the lab for the delay
  line.} to introduce more small baselines and to aid the baseline
bootstrapping. A simpler short-term solution to improving CHARA
imaging is to use all 6 telescopes simultaneously with new
combiners\footnote{The MIRC 6-T upgrade was successfully completed in summer 2012.}.


\subsection{How to obtain high-quality images}
\label{sec:future-facilities}

For future facilities, either new interferometers like MROI or new
instruments on existing facilities, it is important to identify the
most important criteria for high-quality imaging.
The main practical limitations are related to the complexity, dynamic
range, and magnitude of the source to be imaged.

As pointed out in Sect.~\ref{sec:comparison-radio}, as in the radio
domain the dominant source of artifacts is an inadequate $(u,v)$
sampling. The causes may come from the number of positions of the telescopes
(see the discussions above about current limitations at VLTI and
CHARA), and the reduced stroke of the delay lines preventing full
observation of the source through the sky. Most specialists agree that
this is the major issue by a large margin. In our simulations,
the $(u,v)$ coverage does not seem to be a limitation, but we have
been quite generous in $(u,v)$ points. Experience with the MIRC/CHARA
instrument shows that snapshot coverage with only 4 telescopes has
proven adequate for imaging objects such as stars with simple spots
and resolved binaries if they are of the "right" size. However, more
complicated objects such as multiple components objects, crowded stellar
fields or heavily spotted stars need more $(u,v)$ coverage. In
addition to that, short-baseline $(u,v)$ sampling is an essential part
when observing extended objects (such as the evolved star
case \cite[see also][]{Monnier:2004}). Diffraction-limited imaging
with a  conventional single-pupil
telescope, or, preferably single dish aperture masking, like in millimeter-wave interferometry, might be an
efficient way to cover that issue fore relatively compact objects.

How many telescopes are needed? Millimeter-wave interferometers have
found that 6-7 telescopes are adequate for many imaging applications
and experienced observers with the CHARA/MIRC instrument are coming to
similar conclusions. In the optical domain there is one crucial point
that has to be integrated in this discussion: the ability to cophase
the array. The so-called ``fringe-trackers'' of future generation
instruments will have to cope with the difficulty that highly detailed
images require high spatial frequency measurements. This translates
into a strong requirement on these cophasing instruments and on the
array configuration: they need to be able to maintain the array
phasing while measuring small visibilities. While this consideration was
integrated in the design of NPOI \citep[e.g.][]{Armstrong:1998b} and
MROI through the use of array configurations allowing for baseline
bootstrapping, this was not the case for VLTI. The ability to
re-configure telescope locations will be another key to making images
of high spatial dynamic range, although many interesting objects such
as spotted rotating stars and binary systems generally change in time
too quickly to make re-configuring practical to schedule.

Spectral dispersion has been quite successful to increase the $(u,v)$
coverage by wavelength synthesis. Line maps have been obtained at NPOI
for a long time \cite[e.g.][]{Schmitt:2009} while this has only
become possible recently at VLTI with the advent of the AMBER high
resolution spectroscopy capability \cite{Millour:2011}.  Observations
in an absorption line and in the continuum probe different atmospheric
depths and thus layers at different temperatures.  Such photosphere
observations provide important information on the wavelength
dependence of limb darkening and therefore possibly a strong
constraint to the supergiant case (and all stellar-surface
imaging). Another interesting application is the comparison  between a
central star emitting in the continuum and its surrounding hot 
line-emitting disk. Specific algorithms could be developed to image
simultaneously all channels within a line (e.g.\ taking into account
the fact that a smooth transition from one channel to the next one is
expected) although this has never been done in the millimeter wave
domain where each channel is imaged separately, most probably because
the $(u,v)$ coverage is sufficient for such mappings.

Finally, flux sensitivity and measurement accuracies appear to impose a
quite stringent limitation. The dynamic range $DN$ is not only a
direct function of the number $n$ of $(u,v)$ points but also of the
errors on each measurements $\delta V$ and $\delta \phi$ as pointed
out by \citet{2002RSPTA.360..969B} who propose a rule-of-thumb estimator: $DN
= \sqrt{n/[ (\delta V/V)^2 + (\delta \phi)^2]}$. This ideal metric
relies on the independence of the measurements whose calibration
should remain uncorrelated.  Instruments should therefore be conceived
to allow for precisely calibrated measurements.

The temporal fluctuations of the sources might be a severe limitation
especially with interferometers having a limited number of telescope
combinations. For example, the timescale of convection is crucial to
constrain the simulations of red-supergiant stars. We probably need a
time series of observations (at least 4-5 different epochs) about 1-2 months
apart in the $H$ and $K$ bands where only small differences are predicted and
at maximum 1 month apart in the optical where things change faster.
Increasingly, we do find our prime targets changing on timescales of
days or even hours and imaging snapshots will require as many
simultaneous measurements as possible\footnote{A warning to researchers that
believe building up many 3-telescope observations is the best
approach: it takes 20x longer to build up full closure phase coverage one
triplet at a time for a 6-telescope interferometer compared to schemes
which measure all baselines!.}.

\section{Conclusions}
\label{sec:conclusions}

We have reviewed past and present efforts to generate aperture
synthesis maps from optical long baseline interferometers showing the
tremendous progresses obtained since 1996.
We have presented in this paper  end-to-end simulations of 4 different
astrophysical cases relevant for image reconstruction with optical
interferometers. The purpose of these simulations was to assess if
reconstructed images would be faithful enough to be trusted when
interpreting the results and to identify ways to increase the quality
of the reconstructed images. This was not a trivial process and it
required the contribution of a large range of competences from
astrophysicists able to create sensible images of the objects,
interferometry experts to mock realistic data, and specialists of
image reconstruction to reach the best reconstructed images. We
addressed four different topics of interest for optical long-baseline
interferometry and found the following results:
\begin{itemize}
\item Detection of planet signatures in disk structures like shadows
  are challenged by dynamic range and might be out of reach of current
  optical interferometry facilities. However other larger-scale structures, with
  lower contrasts such as disk gravity waves, gaps, hot
  accreting bubbles around a protoplanet etc. might be revealed. Future
  efforts in that direction should address both instrumental issues
  (precision and accuracy of measurements, data reduction) and
  image-reconstruction dynamic range limits.
\item Interferometric imaging allows sizes of convection cells and the
  contrast of the surface intensity to be measured and allows us to
  better understand the photosphere and dynamics of evolved massive
  stars. These results are crucial for understanding their unknown mass
  loss mechanism that contributes heavily to the chemical enrichment of
  the Galaxy. The
  complex structure requires a rich $(u,v)$ plane sampling and
  therefore the combination of the highest possible number of telescopes.
\item The capabilities offered by current interferometers make it
  possible to reveal asymmetries in the molecular layers around
  evolved stars. Further imaging of molecular layers appearing in
  spectral lines is the path to understand these dust and
  molecule factories of the Galaxy.
\item Our simulations show that unraveling the clumpy structure of the
  dusty torus obscuration region in AGN is within reach and will
  certainly be a necessary step to progress in the understanding of
  the AGN unified scheme.
\end{itemize}
In general, we found that  image-reconstruction techniques have been successful
at retrieving global morphologies of all our objects. Their current
limitations lie (i) in their ability to match exact intensities and
(ii) to detect highly contrasted features.

The $(u,v)$ coverage is of crucial and primary importance and therefore
we look forward to results from the 6-telescope mode of CHARA/MIRC,  the
implementation of at least 6 (up to 8) telescopes at the VLTI
with a fully operational set of delay lines, and the construction of
MROI with full capacity. We also recall the importance of imaging
instruments (Matisse, Gravity, VSI at VLTI, MIRC and MIRC6T at CHARA).
The spectral resolution is an asset whose full exploitation has only started to be fully exploited.
Of course, like in millimeter astronomy, the future of interferometry
has to be prepared and will probably aim at a facility similar to
the \emph{Very Large Array (VLA)} or ALMA.

Our study also shows that, in the context of image reconstruction,
the tools have reached a
certain level of maturity to produce on-the-fly results and with a reliability and reproducibility that allows 
productive results and discussion. Optimization of the ``\emph{theory $\rightarrow$ models $\rightarrow$ simulated
 observations $\rightarrow$ reconstructed images}'' cycle is never finished, 
  but a first level of quantitative analysis shows that
optical interferometry is capable of producing meaningful images. We have
now tools to improve the comprehension of astrophysical phenomena and to
reach ambitious research topics.

We believe that imaging interferometry will provide to the broad astronomical
community a new window to the universe if we are able to gather
resources (expertise, manpower and relatively limited investments).

\begin{acknowledgements}
We are very grateful to A. Quirrenbach, the referee, for his in-depth
reading and his many suggestions that helped improve the article.
  This work is the result of a workshop on \emph{Interferometry
    Imaging} held in Ch\^ateau de Goutelas from 26 May to 29 May 2009
  and organized by F.~Malbet and J.-P.~Berger following an idea of
  J.-L.\ Monin. We would like to thanks the members of the
  \emph{Science Organizing Committee} O.\ Chesneau, T.\ Driebe, A.\
  Marconi, J.\ Monnier, B.\ Plez, L.\ Testi, S.\ Wolf and J.\ Young,
  as well as the \emph{Local Organizing Committee}.  This workshop has
  been possible thanks to the financial participation of the
  \emph{Laboratoire d'Astrophysique de Grenoble} (LAOG), of the
  \emph{Programme National of Physique Stellaire} (PNPS) from CNRS,
  the \emph{Jean-Marie Mariotti Center} (JMMC) and the
  \emph{Universit\'e Joseph Fourier}. The web page for the workshop is
  at \texttt{http://wii09.obs.ujf-grenoble.fr}.  M.~Elitzur
  acknowledges the support of NSF (AST-0807417) and NASA
  (SSC-40095). S.~H\"onig acknowledges support by
  DFG. T.V. acknowledges support from the Fund for Scientific
  Research, Flanders as Postdoctoral Fellow". "B.F. acknowledges financial support from ANR and the PNPS of CNRS/INSU, France." This research has made use of the Jean-Marie  Mariotti Center \texttt{SearchCal} service \footnote{Available at
    http://www.jmmc.fr/aspro}. We have made use of the SAO/NASA
  Astrophysics Data System. Figures were generated using the free
  Yorick software, under BSD license \footnote{Available at
    http://www.maumae.net/yorick/doc/index.php}.
\end{acknowledgements}

\appendix

\section{Noise model for ASPRO simulations}
\label{sec:noise-model}

This Appendix is aimed at describing the noise model used in ASPRO for
creating the simulations used in this paper (see
Sect.~\ref{sec:simulation_methodology}).

The noise model is based on a general scheme valid for spatially
filtered recombiners where the detection of fringes is made on a
detector with ``pixels''. This scheme is valid for image-plane
recombination, with fringes covering a surface of a pixel camera, and
for pupil recombination where fringes are obtained on a few pixels
detector by scanning in optical path difference.

The flux $\overline{N_{\rm tot}}$ from the object of magnitude $m_b$
in a given bandwidth $\Delta\lambda$ of a photometric band $b$ is
collected by $N_{\rm tel}$ telescopes of diameter $D$, transmitted
with some instrumental transmission $T_{\rm inst}$, and injected with
some Strehl factor $s$ due to incomplete correction of wavefront
aberrations due to seeing into a
spatial filter like an optical single mode fiber for example, during a
time $t_{\rm int}$. Thus:
\begin{equation}
  \label{eq:Nphot}
  \overline{N_{\rm tot}} = \eta \, F_0 \, 10^{-0.4 m_b} \,
  T_{\rm inst} \, N_{\rm tel} \, s \, \frac{\pi D^2}{4} \, \Delta\lambda,    
\end{equation}
where $F_0$ is the zero-magnitude flux in band $b$, expressed in
$\mbox{ph}\,\mbox{s}^{-1}\,\mbox{m}^{-2}\,\mbox{$\mu$m}^{-1}$ 
transmitted through the atmosphere with an absorption $\eta$.

This flux is divided between the photometric flux and interferometric
flux with a branching value $b_i$, where $b_i$ equals $1$ for
recombiners which do not have simultaneous flux monitoring.

The $N_{\rm tel}$ photometric fluxes $\overline{N_p}=(1-b_i) \,
\overline{N_{\rm tot}}/N_{\rm tel}$ are distributed on $N_{\rm pix}$
pixels. The interferometric flux $\overline{N_i}=b_i \,
\overline{N_{\rm tot}}/N_{\rm tel}$ consists in $N_f= N_{\rm tel} \,
(N_{\rm tel}-1) / 2 $ fringes that cover $N^i_{\rm pix}$ pixels. There
are thus $N_{\rm ppf}=N^i_{\rm pix}/N_f$ pixels per fringe.

These fringes code the intrinsic visibility $V(u,v,\lambda)$ degraded
by the interferometer instrumental contrast and the atmosphere
(through the jitter associated with the temporal coherence of the
seeing). $V(u,v,\lambda)$ and the derived interferometric observables
are thus affected by the sum of the variance of the flux used to code
the corresponding fringe in the interferometric flux and of the
associated two photometric fluxes. For example, since the squared
visibility estimator of a correlated flux $F_c^{ij}$ measured
alongside with photometries $F_i$ and $F_j$ is
$V^2=\frac{1}{4}<|F_c^{ij}|^2>/<F_iF_j>$, the associated variance is 
\begin{equation}
  \label{eq:var}
\sigma^2(V^2) = \overline{N_i}  V(u,v,\lambda) 
     + N_{\rm ppf} \sigma^2_{\rm det}
     + 2 \left( \overline{N_p}+N^p_{\rm pix} \sigma_{\rm det}^2 \right),
\end{equation}
where $\sigma_{\rm det}$ is the readout noise of the detector.

The noise model used in \texttt{ASPRO} takes also into account the
possibility of increasing the integration time to keep observations in
a photon-dominated regime, when a fringe tracker is present.

Finally, no detailed calibration error was computed. We took instead  an
additional visibility and closure phase threshold error set to 0.002 in visibility.
and 0.1 degree in closure phase.

\section{Computing fidelities}
\label{sec:fidelity}

\begin{table*}[t]
  \centering
  \caption{For a given fraction of the brightest pixels (col 1) the
    corresponding number of pixels, median and average fidelity is
    given for each object.}
  \label{tab:fidelity}
  \begin{tabular}[c]{cccccccccc}
\hline
\multicolumn{1}{c@{\hspace*{2em}}}{}
&\multicolumn{3}{c@{\hspace*{2em}}}{Supergiant}
&\multicolumn{3}{c@{\hspace*{2em}}}{Evolved Star}
&\multicolumn{3}{c}{AGN} \\
&\multicolumn{3}{c@{\hspace*{2em}}}{($100\times100$ pixels)}
&\multicolumn{3}{c@{\hspace*{2em}}}{($200\times200$ pixels)}
&\multicolumn{3}{c}{($200\times200$ pixels)} \\
\hline
\hline 
Fraction of&$N_{\rm pixel}$&Med&Avg&$N_{\rm pixel}$&Med&Avg&$N_{\rm pixel}$&Med&Avg\\
 brightest pixels \\*[1ex]
0.3\% & 30 & 2.9 & 3.2  & 119 & 13.7 & 30.7    &   122 & 1.4 & 2.9 \\
1\% &  100 & 3.6 & 4.8  & 398 &4.4 & 14.0       &  408
& 1.4 & 4.5 \\
10\% &  1000 & 5.7 & 11.5  & 3980 & 0.0 & 1.4 &   4080
& 1.5 & 4.5\\
\hline
\end{tabular}
\end{table*}

\begin{figure*}[p]
  \centering
  \begin{tabular}{cc} 
    \includegraphics[height=0.35\textwidth]{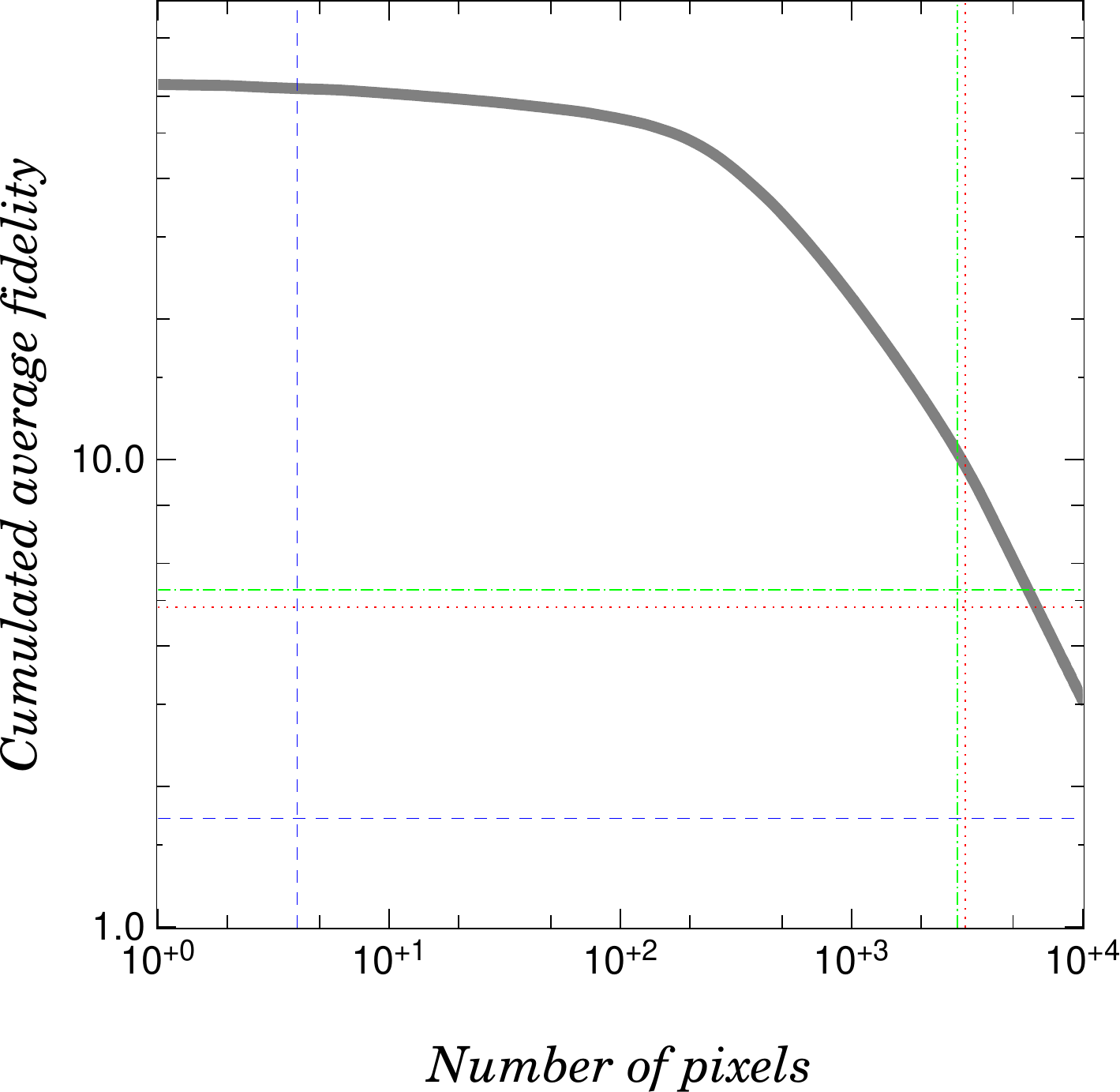} &
    \includegraphics[height=0.35\textwidth]{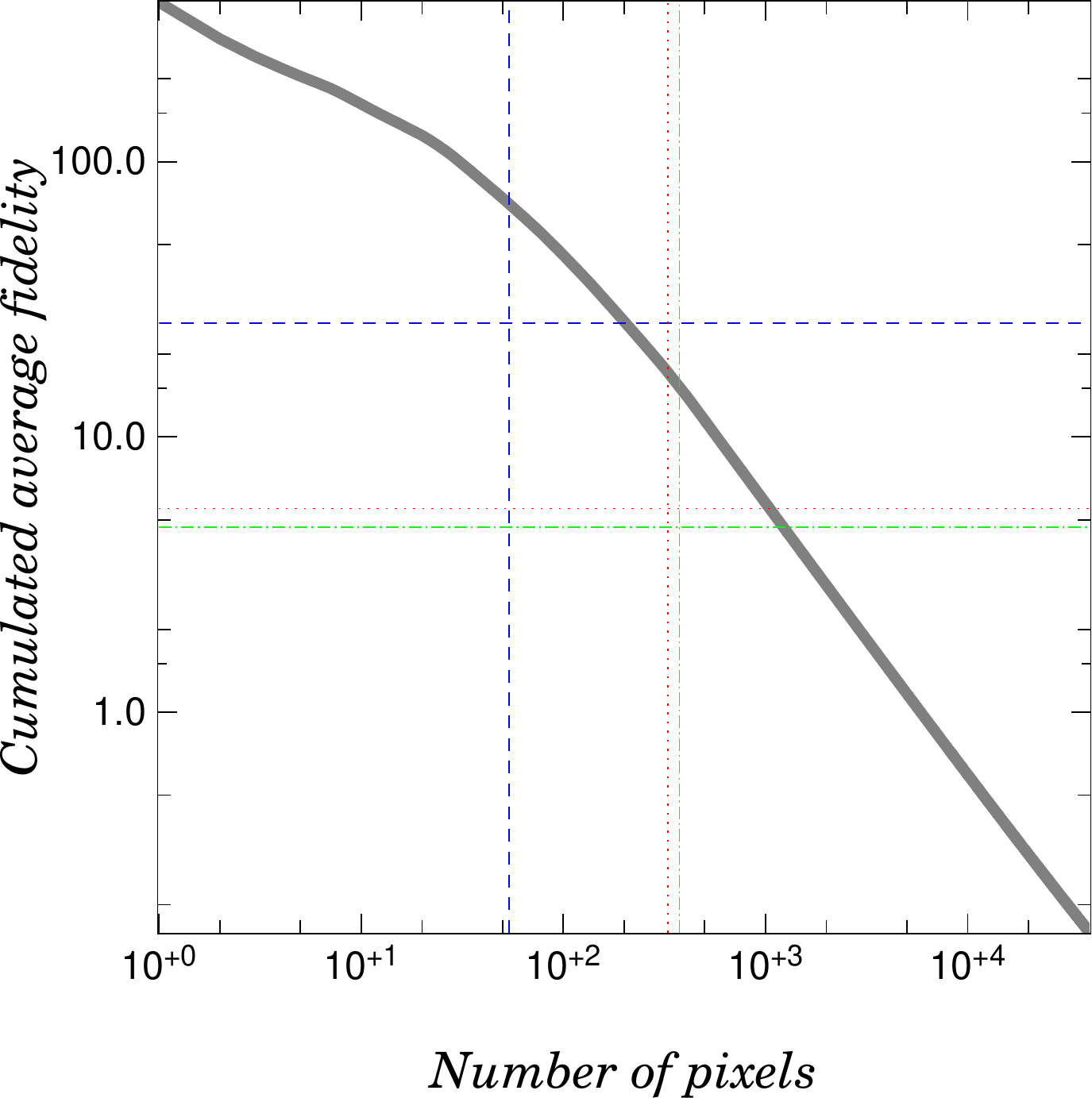}\\
    \includegraphics[height=0.70\textwidth]{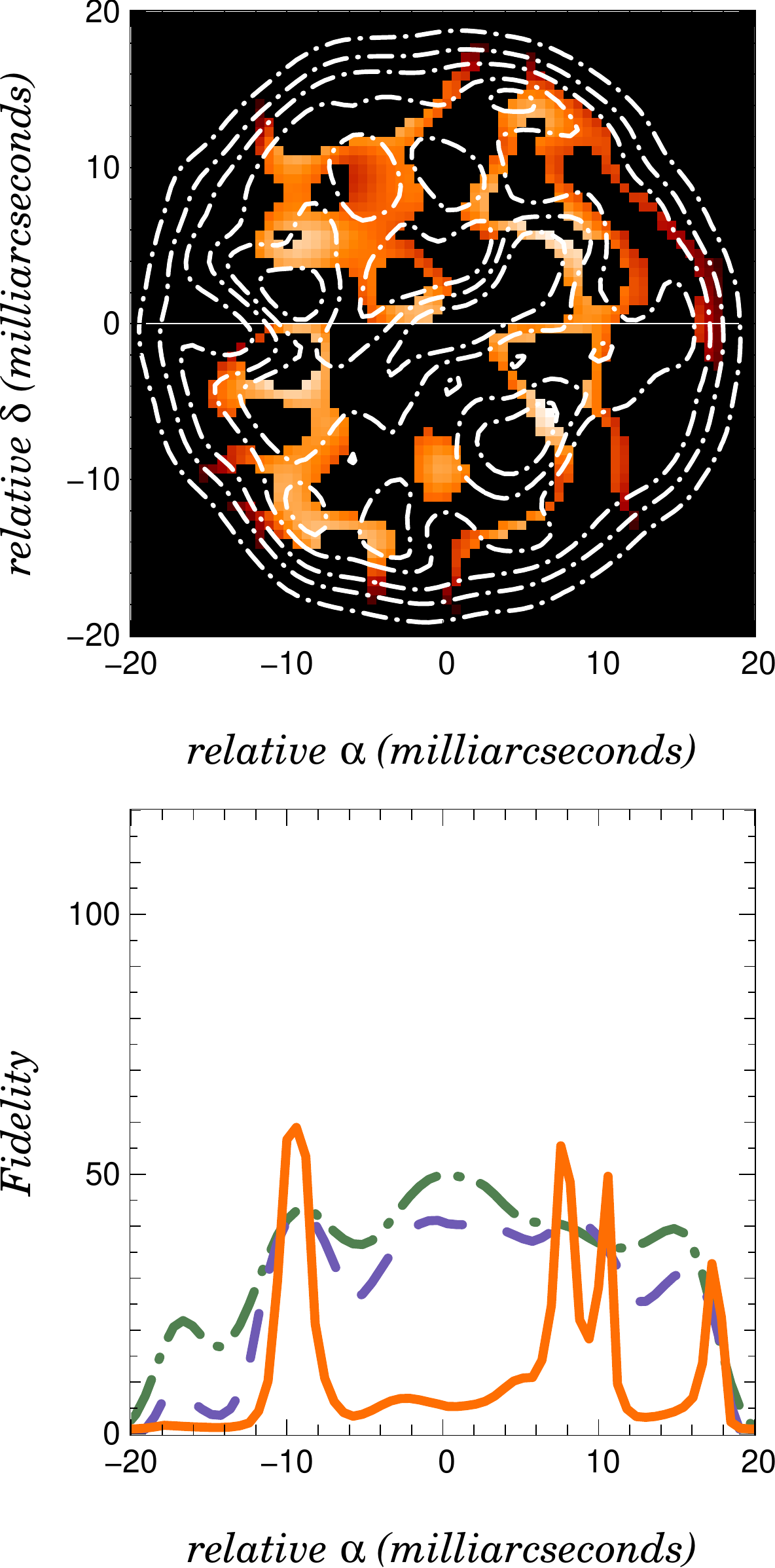} &
    \includegraphics[height=0.70\textwidth]{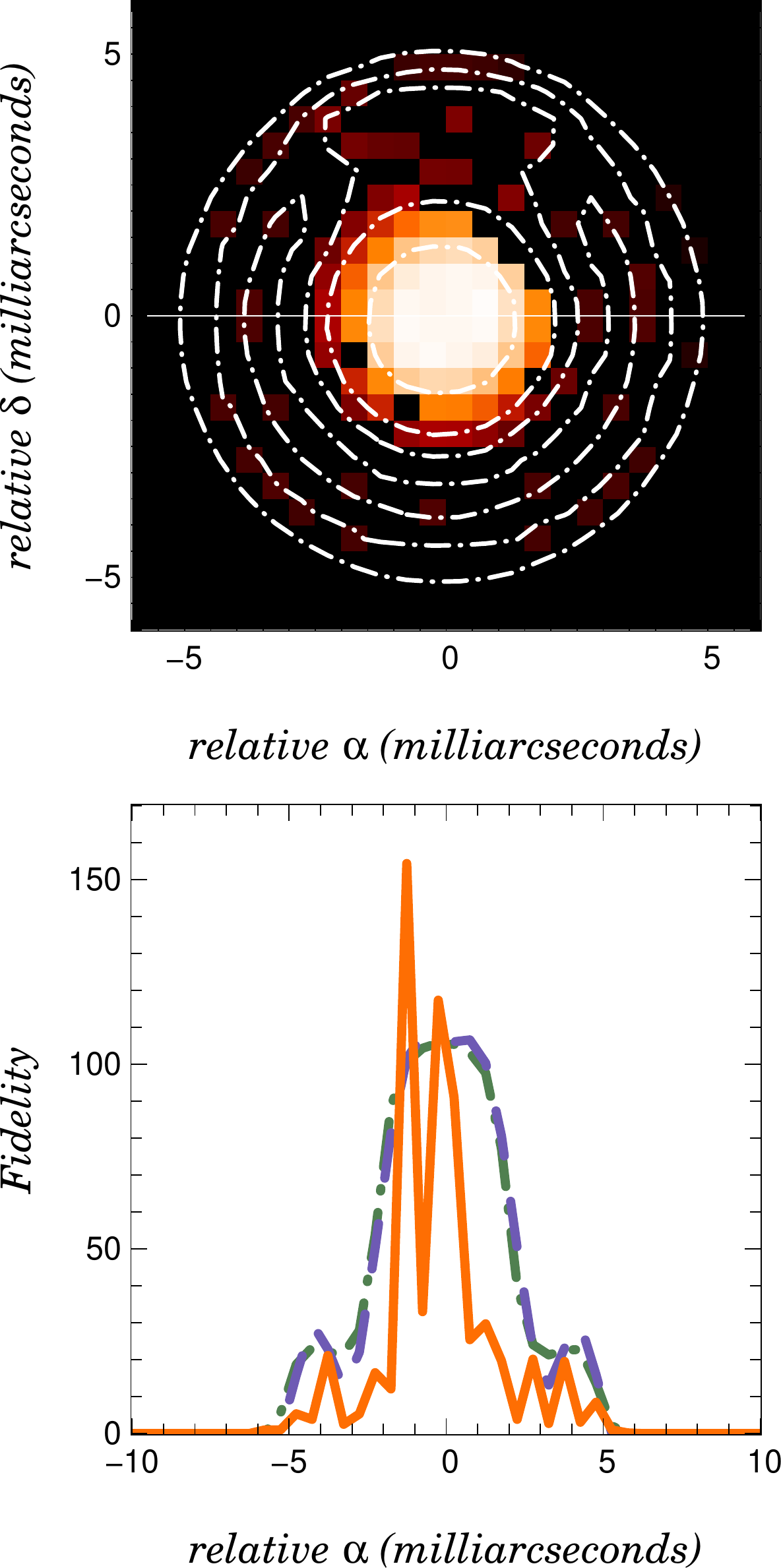} \\
\end{tabular}
\caption{\textbf{Top row}: cumulated average fidelity for
    the supergiant and evolved star cases. Blue dashed, green
    dashed-dotted, red dotted vertical lines: number of pixels with
    intensity greater than a given fraction of the total image
    intensity, respectively: $4.10^{-4}, 1.10^{-4}, 4.10^{-5}$ for the
    supergiant (left), $5.10^{-3}, 1.10^{-3}, 5.10^{-4}$ for the
    evolved star (right). Horizontal lines share the same color code: median
    fidelity for pixels having intensity greater than a given fraction
    of image total intensity.  \textbf{Central row}: original model
    image filtered with a fidelity threshold (supergiant: 10, evolved
    star: 10).  \textbf{Bottom row}: fidelity profile along
    the solid cut line in central row.  Orange solid line: fidelity;
    green dashed-dotted line: model image; dashed blue: reconstructed
    image. Units are given in fidelity, image profiles have been
    scaled to fit the figure.}  
  \label{fig:fidelity}
\end{figure*}

\begin{figure*}[p]
  \centering
  \begin{tabular}{c} 
    \includegraphics[height=0.35\textwidth]{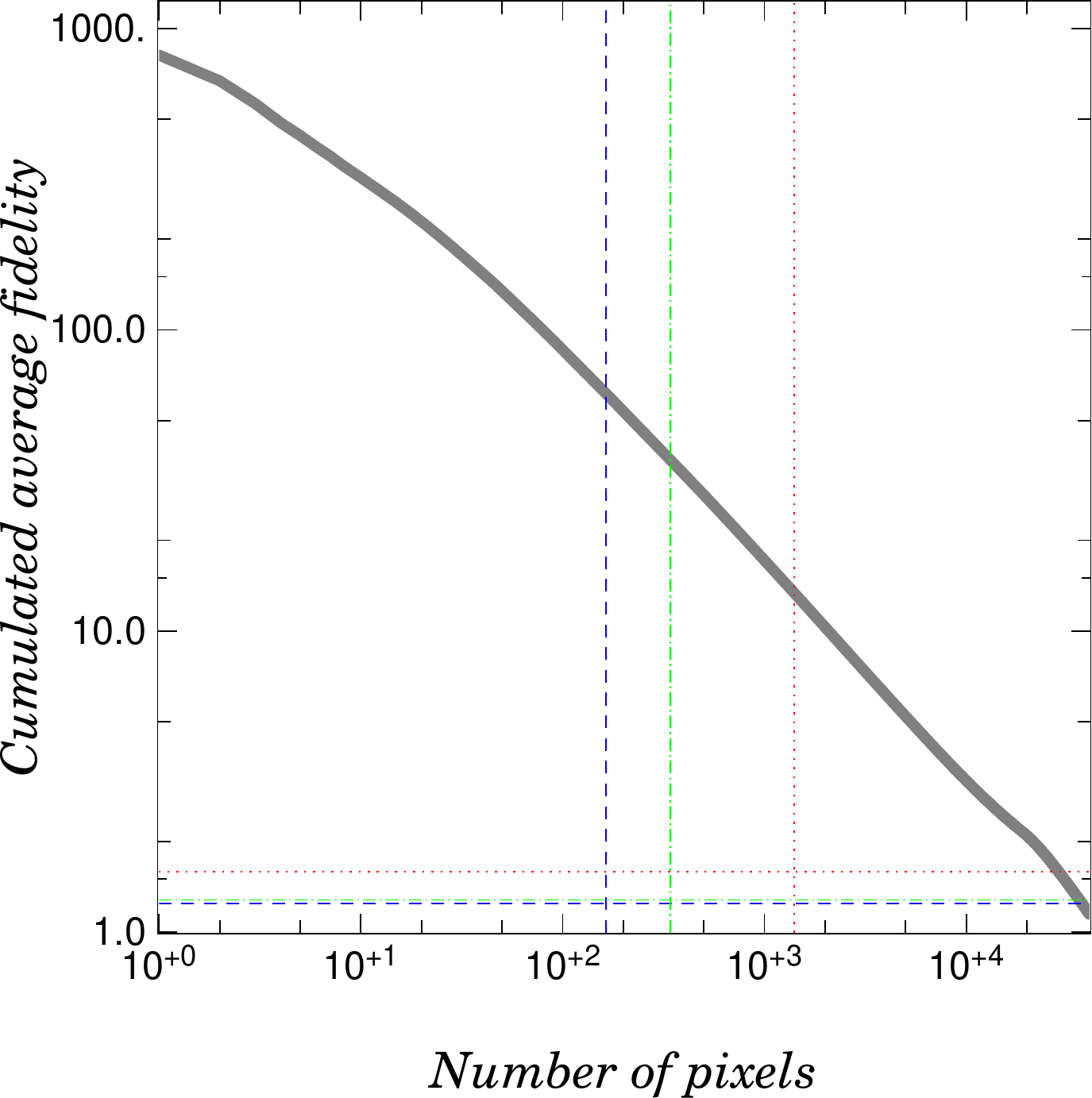}\\   
  \includegraphics[height=0.70\textwidth]{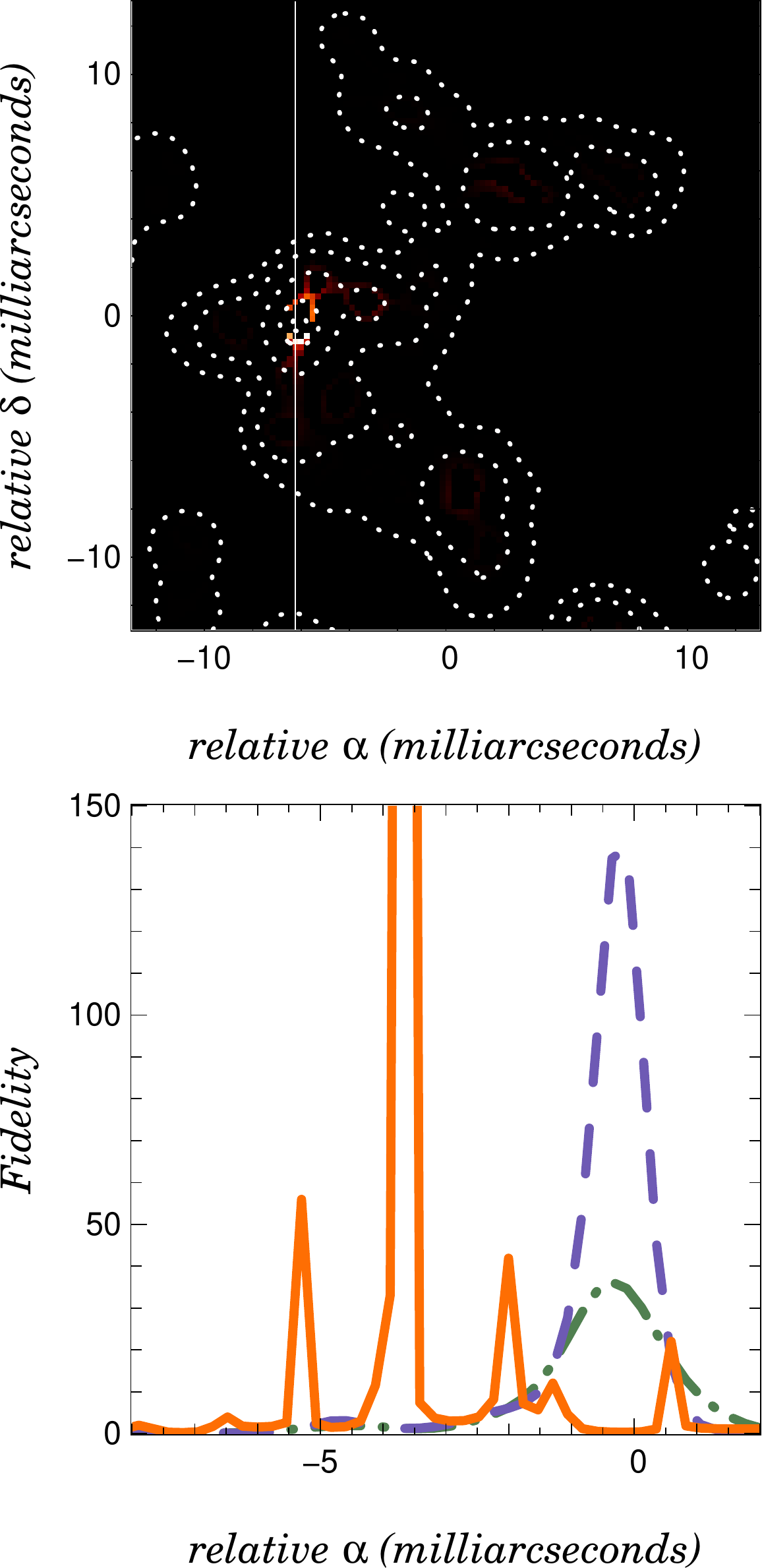} 
\end{tabular}
\caption{\textbf{Top}: cumulated average fidelity for
    the AGN case. Blue dashed, green
    dashed-dotted, red dotted vertical lines: number of pixels with
    intensity greater than a given fraction of the total image
    intensity i.e. $1.10^{-3}, 5.10^{-4}, 1.10^{-4}$. Horizontal lines share the same color code: median
    fidelity for pixels having intensity greater than a given fraction
    of image total intensity.  \textbf{Central}: original model
    image filtered with a fidelity threshold (AGN: 3).  \textbf{Bottom}: fidelity profile along
    the solid cut line in central row.  Orange solid line: fidelity;
    green dashed-dotted line: model image; dashed blue: reconstructed
    image. Units are given in fidelity, image profiles have been
    scaled to fit the figure.}  
  \label{fig:fidelity2}
\end{figure*}

Testing image reconstruction software is out of scope of this
paper. In this Appendix, we follow the approach described in ALMA
Memo~398 (F.\ Gueth private communication) to evaluate the quality of
the reconstructed images presented in this paper. One of the possible
methods to compare original (convolved to the interferometer
resolution) and reconstructed image is to compute the {\it fidelity}
of the image. This can be done either in the direct image plane or in
the spatial frequency $(u,v)$ plane.  Such a pixel to pixel comparison
requires subpixel alignment to limit the effect of sharp transitions.

In the image plane this fidelity can be expressed as
\begin{equation}
  \label{eq:fidel1}
\mathcal{F}(x,y)= \frac{\mbox{abs}(\mathit{Model} (x, y))}{\mbox{max}(\mathit{Diff}(x, y), \mathit{Threshold})},   
\end{equation}
where $\mathit{Model}(x,y)$ describes the object ``true'' brightness
distribution and
\begin{equation}
  \label{eq:fidel2}
\mathit{Diff}(x,y)=\mathit{Model}(x,y)-\mathit{Reconstructed}(x-\Delta
x,y-\Delta y)
\end{equation}
describes the difference between the model and the reconstructed image
shifted by the offset $(\Delta x,\Delta y)$ to have the images
centered. For a proper comparison images are normalized to the total
intensity contain in the image. $\mathit{Threshold}$ is defined here
as $0.7\,\mbox{rms}[\mathit{Diff}(x,y)]$ which provides an
estimation of the threshold noise of this difference.  For example a
pixel fidelity of 100 corresponds to a difference of 1\% between the
model and reconstructed image. We have computed such a fidelity for
three of our objects: the supergiant, the evolved star and the active
galactic nuclei. Figures \ref{fig:fidelity} and \ref{fig:fidelity2} offer different ways
to visualize fidelity. The first row shows the cumulated average
fidelity over the image for the three different objects.  The second
row displays the original model image filtered by fidelity level, i.e.\
only the pixels with fidelity above a certain level are displayed.
Finally Table~\ref{tab:fidelity} shows the average fidelity of the
pixels whose intensity is above a certain fraction of the total image
intensity.

\bibliographystyle{spbasic}       
\bibliography{interfIMagingAAR2011}

\end{document}